\documentclass[11pt]{extarticle}
\usepackage{longtable}
\usepackage[margin=14mm]{geometry}
\usepackage{multirow}
\usepackage[table]{xcolor}
\usepackage{diagbox} 
\usepackage[normalem]{ulem} 
\usepackage[dvipsnames]{xcolor}
\usepackage{amsmath,amsthm,amssymb,amsbsy,amsopn,mathtools,comment,bbm}
\usepackage{graphicx,color,float,subcaption,caption}
\usepackage[bookmarks=true,bookmarksnumbered=true,bookmarkstype=toc]{hyperref}
\usepackage{mathptmx,bm,bbm}
\usepackage{enumitem}

\usepackage[most]{tcolorbox}
  \tcbset{
    colback=white,
    colframe=black!10,
    coltitle=black,
    boxsep=1.0ex,
    left=1.2ex,right=1.2ex,top=1.0ex,bottom=1.0ex,
    fonttitle=\bfseries
  }
\tcbuselibrary{breakable}

\usepackage{lineno}

\allowdisplaybreaks

\usepackage{tikz}
\usetikzlibrary{arrows,automata}
\usetikzlibrary{positioning}

\usepackage[titletoc]{appendix}

\graphicspath{{../fig/}{figures/}}

\theoremstyle{definition}

\makeatletter
\@addtoreset{equation}{section}
\makeatother

\title{Stopping rules for Monte Carlo methods: A review\footnotetext[0]{This version: \today.}}
\author{\sc Jiezhong Wu\footnote{Email: jiezhongwu2021@u.northwestern.edu. Affiliation: Department of Industrial Engineering \& Management Sciences, Northwestern University, USA.} \,
and Reiichiro Kawai\footnote{Corresponding author. Email: raykawai@g.ecc.u-tokyo.ac.jp. Affiliation: Graduate School of Arts and Sciences / Mathematics and Informatics Center, The University of Tokyo, Japan.
This work was partially supported by JSPS Grants-in-Aid for Scientific Research 21K03347 and 24K06844.}}
\date{}
\begin{document} 

\maketitle

\begin{abstract}
\noindent 
Sequential analysis encompasses simulation theories and methods where the sample size is determined dynamically based on accumulating data.
Since the conceptual inception, numerous sequential stopping rules have been introduced, and many more are currently being refined and developed.
This article aims to deliver an up-to-date review of recent developments in sequential stopping rules, 
\textcolor{black}{with a deliberate emphasis on Monte Carlo methods for estimating an unknown expectation, including binomial proportions, primarily under standard iid sampling and also under certain lightly generalized settings.}
These methodologies have long served and likely will continue to serve, as fundamental bases for both theoretical and practical developments in stopping rules for general statistical inference, advanced Monte Carlo techniques and their modern applications.
{Building upon over a hundred references and empirical studies, we explore the essential aspects of these methods, such as core assumptions, numerical algorithms, convergence properties, and practical trade-offs to guide further developments, particularly at the intersection of sequential stopping rules and related areas of research.}

\vspace{0.3em}

\end{abstract}

\tableofcontents

\section{Introduction}
\label{section introduction}

Monte Carlo methods are widely recognized as versatile computational tools based on the principle of repeated random sampling to achieve numerical results that are often difficult or even infeasible to obtain through other techniques.
Their remarkable adaptability has enabled their integration into a wide array of fields, including actuarial science, 
biology, 
chemistry, 
materials science, 
medical research, 
quantum mechanics, 
social sciences, 
physics, 
and structural reliability analysis,
such as \cite{HE20231, nightingale1998quantum, d2019monte, SONG2023103479}, among numerous relevant disciplines.

In Monte Carlo methods, sequential stopping rules are strategies designed to determine when to stop the procedure based on real-time observations, rather than a fixed endpoint set before the simulation begins.
By monitoring suitable metrics, these stopping rules aim to ensure that simulations terminate once the desired accuracy is achieved, thereby optimizing computational effort which is a critical consideration for resource-constrained or expensive simulations.
An effective stopping rule needs to strike a balance between efficiency and accuracy by wisely adjusting the sample size based on intermediate results. 
For instance, excessively long simulations can waste computational resources, whereas arbitrary termination increases the risk of slow convergence to steady-state solutions.
Such arbitrary stopping also falls under the scope of optional stopping, particularly when the experimenter retains the ability to resume the procedure retrospectively \cite{OHKUBO2021}.
{This, however, violates the so-called stopping rule principle: ``{\it the evidential relationship between the final data from a sequential experiment and a hypothesis under consideration does not depend on the stopping rule: the same data should yield the same evidence, regardless of which stopping rule was used}'' \cite{fletcher2024}.}

More broadly than Monte Carlo methods, the concept of sequential stopping rules originates from the classical gambler's ruin problem in the 17th century and gained practical significance in the early 20th century through a variety of applications, such as the sequential sampling inspection procedure, the quality control charts and the two-stage designs \cite{handbook}, often interpreted in various equivalent ways over years, such as the perspective of testing by betting \cite{10.1093/jrsssb/qkad009}.
The framework of sequential stopping rules was formalized in the 1940s and gained further developments in the 1950s, {with pioneering studies such as \cite{Anscombe_1949, Anscombe_1952, Anscombe_1953, Anscombe_1954, RL1956, Ray_1957, Stein_Wald_1947, 10.1214/aoms/1177730439}.
These works highlighted the fundamental challenge of balancing sample size and computational effort against statistical accuracy and theoretical justification in stochastic simulation and sequential inference \cite{dantzig1940, 10.1214/aoms/1177731088}.}

\textcolor{black}{Given the broad significance of Monte Carlo methods, the present article aims to deliver \textcolor{black}{an up-to-date review} of fundamental concepts and emerging developments in sequential stopping rules, with a deliberate emphasis on their application to Monte Carlo methods for estimating an unknown expectation (including binomial proportions), where the empirical mean serves as the canonical estimator.
In particular, we focus primarily on the standard setting of independent and identically distributed (iid) sampling, while also considering certain extensions beyond the iid framework, referred to here as lightly generalized Monte Carlo methods, such as those involving martingale difference sequences, mixing processes, and quasi-Monte Carlo.
Throughout, we explore the underlying assumptions, numerical algorithms, convergence behavior, and practical trade-offs involved in implementation.}


Drawing from a wide array of references, this review article is designed to provide a structured overview to highlight recent advancements, categorize various approaches, provide comparisons, and guide future research, particularly at the intersection of sequential stopping rules and related areas of research.
Although sequential stopping rules for such standard and lightly generalized Monte Carlo methods have only appeared somewhat sporadically in the recent literature, those have long served, and are expected to continue to play a pivotal role, as fundamental bases for theoretical insights, practical findings, and advancements in stopping rules for advanced Monte Carlo techniques in light of the growing interest in modern applications (such as \cite{dammertz09wscg}).
\textcolor{black}{To our knowledge, few studies have so far jointly considered (i) a structured classification of method families, (ii) practitioner-oriented selection guidelines, and (iii) empirical comparisons across representative simulation regimes.}

\textcolor{black}{The contribution of this review is threefold.
First, we organize a broad and often fragmented literature on sequential stopping rules into a structured taxonomy, highlighting the relationships among different methodological approaches.
Second, we provide a comparative perspective on major families of stopping rules, examining their underlying assumptions, theoretical guarantees, and practical trade-offs.
Third, we complement this synthesis with illustrative numerical studies that demonstrate how these trade-offs manifest across representative simulation settings.
Rather than proposing new stopping rules, the goal of this article is to present a cohesive and practice-oriented overview that clarifies existing methodologies and helps guide their application in different contexts.}

\textcolor{black}{We remark that the term ``lightly generalized Monte Carlo methods'' is employed throughout this review as an informal but descriptive label for modest extensions of iid Monte Carlo methods for estimating an unknown expectation, such as martingale difference sequences, mixing models, and quasi-Monte Carlo, that retain the empirical mean as the primary estimator and continue to satisfy variants of classical limit results.
In particular, the qualifier ``lightly'' is intended to convey that these modifications introduce only moderate structural deviations from the iid setting, so that much of the conventional stopping-rule methodology can still be applied with relatively minor adjustments.}

\color{black}
\subsection{Existing reviews and empirical studies}
\label{Existing reviews and empirical studies}

Due to significant attention received over time, the topic of sequential stopping rules has been reviewed from time to time, both comprehensively and selectively, with seminal works dating back to 1953 \cite{Anscombe_1953}, 1968 \cite{M1968} and then 1973 \cite{doi:10.1287/opre.21.6.1309}, such as surveys focusing on steady-state simulation in the early 1980s \cite{AJ1982, survey_SteadyStateI}.
In the mid-1980s, sequential analysis was extensively reviewed in numerous monographs and articles, such as \cite{PBL1987, Z1987, D1985, GK1986, WOODROOFE198670}, some covering much broader topics beyond sequential stopping rules.

In the 1990s, this topic continued to receive thorough reviews, such as those found in the handbooks \cite{handbook, 10.5555/554952} and the invited article \cite{e0ef43b4-f485-3a19-b019-5fb86fa01603}.
In particular, the latter provides a comprehensive review with a broad scope, organized with respect to representative applications, such as 
sequential tests of hypotheses \cite{frick1998, wald1947}, sequential detection \cite{https://doi.org/10.1002/j.1538-7305.1930.tb00373.x}, sequential estimation \cite{10.1093/biomet/33.3.222} and stochastic approximation \cite{10.1214/aoms/1177729586}. 
Taking a very broad perspective that extends well beyond stopping rules and relevant techniques, a comprehensive survey is provided in 1994 \cite{O1994} on validation, verification and testing, encompassing virtually all aspects involved throughout the life cycle of simulation studies, from initial planning and software development to programming and presentation.
Although somewhat tangential to the primary topic of the present article, the issue of sample size is comprehensively reviewed in \cite{jennison1999group}, with a particular focus on applications in clinical trials.

In the 21st century, the seminal research of Herbert Robbins in sequential analysis (from 1952 until roughly 1980) is comprehensively reviewed in \cite{10.1214/aos/1051027870}.
This topic is further explored in the monograph \cite{Mukhopadhyaybook}, as well as in \cite{Schoen2009, doi:https://doi-org.utokyo.idm.oclc.org/10.1002/9781118445112.stat08283} on related advanced directions, such as Markov chain Monte Carlo \cite{https://doi.org/10.1111/bmsp.12357}, stochastic programming \cite{df3b10f8-26b4-38aa-9eb3-41173d4233f0, Schoen2009}, machine learning related techniques \cite{Prechelt2012EarlyStopping}, and Bayesian cubature \cite{jagadeeswaran_hickernell_2019, mahsereci2026bayesianquadraturegaussianprocesses}, which have been evolving more systematically. 

Empirical case studies on stopping rules have rarely been presented in the literature, with some exceptions for those on discriminant analysis \cite{Costanza01121979}, clinical trials \cite{7843498}, a cellular automata model for simulation of a susceptible-infectious-recovered epidemic \cite{bicher}, and the lightning performance of transmission lines \cite{ALMEIDA2023108797}, among only a few others.
{For stimulating research in diverse directions, more application-specific empirical case studies would be of particular value in the study of sequential stopping rules.
Note that, although the present article provides a practical guide and empirical studies (Section \ref{section practical guide and comparison}), those are organized in a general (rather than application-specific) manner, reflecting the foundational nature of the present review.}

\color{black}
\subsection{Notation}\label{subsection notation}

We begin with some general notation that will be used throughout this review article.
We use the notation $\mathbb{N}:=\{1,2,\cdots\}$ and $\mathbb{N}_0:=\{0\}\cup \mathbb{N}$.
Let $\mathbb{P}$ and $\mathbb{E}$ denote the underlying probability measure and corresponding expectation which we work with.
We let $\stackrel{\mathcal{L}}{\rightarrow}$, $\stackrel{\mathbb{P}}{\rightarrow}$ and $\stackrel{a.s.}{\rightarrow}$ denote the convergences in law, in probability and almost surely, respectively.
We reserve the notation $\Phi$ for the standard normal cumulative distribution function and let $z_{\delta/2}$ represent the $100(1-\delta/2)\%$-quantile of the standard normal distribution for $\delta\in (0,1)$, that is, $z_{\delta/2}=\Phi^{-1}(1-\delta/2)$, for instance, $z_{0.025}\approx 1.96$ and $z_{0.005}\approx 2.58$, respectively, for constructing $95\%$- and $99\%$-confidence intervals.

Unless stated otherwise, we are throughout concerned with the estimation of the expectation $\mu$ by the empirical mean $\mu_n:=n^{-1}\sum_{k=1}^{n} X_k$, where $\{X_k\}_{k\in\mathbb{N}}$ denotes a sequence of random elements, which can be univariate, multivariate, iid or not iid, depending on the context.
{We denote by $(\mathcal{F}_k)_{k\in\mathbb{N}}$ the natural filtration generated by the sequence of random variables $\{X_k\}_{k\in\mathbb{N}}$.  
In other words, $\mathcal{F}_k$ represents the information available up to time $k$, as encoded by the observations $(X_1, X_2, \cdots, X_k)$.}
When univariate, we often write $\sigma^2$ and $s_n^2$ for the theoretical and empirical variances, that is, $s_n^2=(n-1)^{-1}\sum_{k=1}^n (X_k-\mu_n)^2$, where the factor $n-1$ should be adjusted as appropriate (for instance, to $n$), depending on the nature of the problem.

{Let us stress our primary focus on Monte Carlo methods for estimating expectations, for which the empirical mean $n^{-1}\sum_{k=1}^{n} X_k$ is the estimator naturally employed.
This contrasts with parametric statistical inference (which is not our central interest), where unbiased estimators may differ from the empirical mean, for instance, $(n-1)^{-1}(\sum_{k=1}^nX_k-1)$ in inverse binomial sampling (Section \ref{section confidence intervals for inverse binomial sampling}).}

\subsection{Structure of the present review}

This introduction concludes with a brief overview of the structure of the present review.
In pursuit of our primary goal of providing a \textcolor{black}{structured and practice-oriented overview}, we focus on overviewing sequential stopping rules for standard and lightly generalized Monte Carlo methods, while complementing the theoretical discussion with illustrative numerical comparisons.
Building on this perspective, in Section \ref{section fundamental concepts}, we set out the fundamental concepts and key components of sequential stopping rules for Monte Carlo methods, including error tolerances, confidence and significance levels, non-iid random elements, large sample theory, normal approximation and procedures for mean estimation.
We then, in Section \ref{section recent developments}, present a survey of various recent developments in sequential stopping rules for standard and lightly generalized Monte Carlo methods.
{In Section \ref{section practical guide and comparison}, we provide a practice-oriented guide for selecting stopping rules by context and a qualitative comparison of method families to bridge theory and practice along with representative empirical studies.}
Finally, Section \ref{section concluding remarks} offers a synthesis of both the theoretical and empirical findings and a concise outlook on possible avenues for future study.

\color{black}
\section{Fundamental concepts and key components}
\label{section fundamental concepts}

In this section, we present a systematic and minimally necessary overview of the essential concepts and main components of sequential stopping rules for {standard and lightly generalized Monte Carlo methods (for instance, those of martingale difference and mixing types, and quasi-Monte Carlo), employed primarily for estimating an unknown expectation, including binomial proportions}, along with the common assumptions in existing studies, to make the review of recent developments (Section \ref{section recent developments}) more accessible to a broad audience.

We begin by establishing the terminology and notation for further discussion.
As defined in Section \ref{subsection notation}, we denote by $\mu$ the theoretical mean to be estimated by Monte Carlo methods and by $\mu_n$ the estimator for the mean $\mu$ based on $n$ iterations.
{The primary objective of sequential stopping rules with a confidence interval, whose fixed half-width is $\epsilon$, is to determine or, more precisely, estimate the minimum number of iterations $\tau_{\epsilon,\delta}$ required to ensure that the error probability remains below the significance level $\delta$ during the simulation run.
Confidence intervals naturally complement estimation by quantifying the uncertainty of the point estimator, and in sequential procedures, their width $2\epsilon$ serves as the precision threshold that determines when the estimator is sufficiently accurate to terminate the run.}

For most absolute-precision sequential stopping rules, this condition can be formulated as
\begin{equation}\label{core requirement for stopping}
\tau_{\epsilon,\delta}\leftarrow \inf\left\{n\in \mathbb{N}:\,\mathbb{P}\left(|\mu_n-\mu| >\epsilon\right) \le \delta\right\},
\end{equation}
while for relative-precision sequential stopping rules, it often takes the form
$\tau_{\epsilon,\delta}\leftarrow \inf\{n\in \mathbb{N}:\,\mathbb{P}(|\mu_n-\mu| >\epsilon|\mu|) \le \delta\}$ with a few alternative frameworks reviewed in Section \ref{section outside the conventional framework}.
Importantly, the number of iterations $\tau_{\epsilon,\delta}$ defined as in \eqref{core requirement for stopping} is deterministic and infeasible to determine exactly in practice.
In what follows, we call the quantities $\epsilon$, $1-\delta$, $\delta$, $\mathbb{P}(|\mu_n-\mu| >\epsilon)$ and $\mathbb{P}(|\mu_n-\mu| \le \epsilon)$, respectively, the error tolerance (Section \ref{subsection error tolerance}), the confidence and significance levels (Section \ref{subsection confidence level}), and the error and coverage probabilities.
{In addition, the asymptotic regime (in the sense of $\epsilon\to 0$) is central not only to the theoretical development but also to the conceptual understanding of sequential procedures (Section \ref{subsection asymptotic validity}).}

\color{black}
\subsection{Error tolerances: absolute and relative}
\label{subsection error tolerance}

The term ``error tolerance,'' most commonly denoted by $\epsilon$, is widely employed to describe the permissible deviation between the estimate $\mu_n$ and the true mean $\mu$.
Other terms, such as precision constant and tolerance, have also appeared, while ``error tolerance'' seems to be the most common in the literature.

Error tolerances are often classified into two kinds: absolute and relative.
In most stopping rules, the absolute precision is the criterion in measuring errors, that is, $\tau_{\epsilon,\delta}\leftarrow \inf\{n\in \mathbb{N}:\,\mathbb{P}(|\mu_n-\mu| >\epsilon) \le \delta\}$ as of \eqref{core requirement for stopping}, whereas the relative precision (also referred to as the proportional accuracy \cite{A1969}) has often attracted great attention \cite{Law01011981}, such as 
\[
\tau_{\epsilon,\delta}\leftarrow \inf\left\{n\in \mathbb{N}:\,\mathbb{P}\left(\left|(\mu_n-\mu)/\mu\right| >\epsilon\right) \le \delta\right\}.
\]
In particular, the relative precision can be valuable when the appropriate level of precision is unclear due to the unknown scale of the sample mean.
Often, the error probability $\mathbb{P}(|(\mu_n-\mu)/\mu_n| >\epsilon)$ is instead adopted by slightly inflating the original relative error $|(\mu_n-\mu)/\mu|$ to, at most, $\epsilon/(1-\epsilon)$, relative to $\epsilon$ for the original relative error. 
A potential issue with relative-precision rules is that if the empirical mean is disproportionately large compared to the empirical variance, the procedure may stop too early, compromising the coverage probability.
Asymptotic validity has also been investigated for such relative-precision sequential stopping rules in \cite[Theorem 3]{PW1992}.
Later in Sections \ref{subsubsection absolute error}, \ref{subsubsection relative error gajek}, \ref{subsection Stopping rule for Bernoulli trials} and \ref{subsection several means}, we review various relative-precision stopping rules.


Error tolerances can also be dynamically refined during the simulation process \cite{D2014}, albeit deviating from the framework of fixed-width confidence intervals.
For instance, when the coverage probability is found insufficient, the error tolerance can be iteratively adjusted to improve the coverage.
In the present article, we do not delve into non-fixed-width approaches, such as confidence sequences
\cite{10.1093/jrsssb/qkad009}.

{In comparison with other error measures (such as mean-square, root-mean-square, or mean-absolute error), absolute and relative errors align more naturally to stopping rules for Monte Carlo methods, which depend on clear and interpretable thresholds for determining when simulation accuracy is adequate.
This contrasts sharply with parametric statistical inference, where the emphasis is on evaluating estimator performance under a parametric model, whether in fixed-sample or asymptotic regimes, rather than terminating a computation based on numerical accuracy.
In that setting, mean-square and root-mean-square errors are central because their bias-variance decomposition underlies efficiency, the Cram\'er-Rao bound, and other standard optimality results, all of which fundamentally depend on squared-error loss.}

\subsection{Confidence and significance levels}
\label{subsection confidence level}

The confidence level $1-\delta$, with the parameter $\delta\in (0,1)$ called the significance level, quantifies the coverage probability $\mathbb{P}(|\mu_n-\mu| \le \epsilon)$ of the estimate $\mu_n$ falling within the specified error tolerance $\epsilon$ of the theoretical mean $\mu$.
In asymptotic analyses (Section \ref{subsection asymptotic validity}), the confidence and significance levels are typically held constant to investigate the behavior of stopping rules when the error tolerance $\epsilon$ approaches zero \cite{FLYA2012, NS1996}.
{Despite the confidence and significance levels often being fixed} in sequential stopping rules in the non-asymptotic regime (Section \ref{subsection normal approximation}), it is widely recognized that fixed levels can result in poor coverage when sample sizes are small.
In contrast, adaptive methods have also been developed to dynamically adjust those levels \cite{D2014, DL2009}.
For instance, coverage functions (Section \ref{subsection coverage functions and related stopping rules}) can be employed to estimate parameters and adjust confidence levels during the simulation.
{It is worth noting that earlier literature often employed a different notation: the error tolerance was written as $\delta$ (or simply $d$) instead of $\epsilon$, while the significance level was denoted by $\alpha$ instead of $\delta$.
The use of this convention dates back at least to \cite{YH1965} and continues to appear in recent studies, such as \cite{Hickernel2018}.}

\color{black}
\subsection{Multivariate outputs or multiple means}
\label{subsection multivariate outputs}

In the most usual terms, sequential stopping rules are built on a sequence of univariate random variables for estimating a scalar mean. 
In the literature, there exist a few stopping rules tailored for estimating a vector mean.
An asymptotic analysis has been conducted in \cite{PW1992} for multivariate outputs, while a stopping rule is proposed for separately yet simultaneously estimating multiple means (as opposed to multivariate outputs) in \cite{Luck1993}, where the main difference lies in the role of the covariance matrix $\Sigma$ of the underlying random vector $X$.
That is, for multivariate outputs, the covariance matrix is typically incorporated through the quadratic quantity $\langle X-\mu,\Sigma^{-1}(X-\mu)\rangle$, leading to an ellipsoidal confidence region.
In contrast, the covariance matrix is disregarded in the case of multiple means, resulting in a rectangular confidence region.

In the former framework, on the one hand, the precision of individual means cannot be directly controlled.
Instead, only the volume of the ellipsoidal confidence region can be regulated, necessitating accurate estimation and approximation of the covariance matrices and confidence regions \cite{JA2006}.
On the other hand, in the latter framework, the precision of individual means needs to be predetermined.
Later in Section \ref{subsection several means}, we review a relative-precision stopping rule for multiple means \cite{Luck1993}. 

\subsection{Non-iid random elements}
\label{subsection non-iid random elements}

Sequential stopping rules were primarily designed for wisely stopping queueing systems \cite{ADLAKHA1982379, 5078cc27-f2a4-395f-864d-471f6693e06d} and steady-state simulations \cite{10.1145/2567907, AJ1, AJ1982}, in which the underlying sequence $\{X_k\}_{k\in \mathbb{N}}$ represents a discrete-time stochastic process
(Section \ref{subsection quasi-independent sequences}).
Their regenerative property can also play a central role in constructing sequential stopping rules \cite{GS1, SC1977}.
Aiming to handle such steady-state simulations, the asymptotic analysis of stopping rules 
(Section \ref{subsubsection a general framework for asymptotic analysis}) is founded on the framework of the so-called estimation process.
To estimate variances for stopping rules in steady-state simulations, spectral methods are employed in \cite{PP1981a} by analyzing the spectral density at zero frequency, involving a regression on the logarithm of the averaged periodogram.
Enhancements introduced in \cite{PP1981b} incorporate smoothing techniques that adapt to variations in the spectral shape, aiming to improve performance for both small and large sample sizes.
Subsequently, the methodology is further refined in \cite{PP1983} to address scenarios involving an initial transient phase, accommodating a stationarity test to be conducted during the simulation.

Many other types of non-iid random elements have been addressed in the context of sequential stopping rules, such as stochastic programming, Markov chain Monte Carlo, and self-normalized importance sampling \cite{Schoen2009, doi:https://doi-org.utokyo.idm.oclc.org/10.1002/9781118445112.stat08283}, often building upon advancements in stopping rules for standard Monte Carlo methods.
{In contrast, in the context of standard and lightly generalized Monte Carlo methods, sequential stopping rules have been developed and appeared in a scattered manner in the recent literature, each relying on different assumptions and conditions,} such as sequences of martingale difference type $\mathbb{E}[X_k|\,\mathcal{F}_{k-1}]=\mu$ 
(Sections \ref{subsubsection relative error gajek} and \ref{subsection martingale difference type}), and low-discrepancy sequences for quasi-Monte Carlo methods 
(Section \ref{subsection low discrepancy sequences for quasi-Monte Carlo methods}).
In this article, we do not explore resampling methods, such as those discussed in \cite{1737d468-f05a-3760-96b3-0b4b960811fd}, which are inherently important, particularly in contexts like clinical trials where sample sizes are often constrained.
Instead, we assume throughout that the sample can be generated at the discretion of the experimenter, as is typically the case in scenarios where Monte Carlo methods are employed.

\subsection{Preliminary and final sample sizes}
\label{subsection preliminary and final sample sizes}

{It is important to understand the circumstances under which absolute-precision fixed-width guarantees cannot be achieved.
In the fully nonparametric setting \cite{RL1956}, where the underlying distribution is allowed to range over a convex class with nothing more than a finite-mean assumption, no sequential stopping rule can provide a uniformly valid fixed-width confidence interval without additional prior information, such as bounds on the distribution, or knowledge of its variance or tail behavior.
In fact, the worst-case coverage can deteriorate all the way to zero in some instances.
This limitation arises because even very small changes in the distribution tail can dramatically shift the mean while leaving finite samples nearly unaffected.
Related impossibility phenomena in parametric models occur when distinct parameter values produce sampling distributions that can be made arbitrarily similar \cite{takada}.}

{In contrast, such impossibility results do not apply to more restricted models with inherent bounds or known scale. For example, in Bernoulli models (that is, with fixed support), Hoeffding-type methods yield valid fixed-width or relative-error procedures. 
Likewise, in the normal model with known variance, exact fixed-width confidence intervals are available using standard normal-based theory.
However, for the normal model with unknown variance, achieving exact fixed-width intervals with uniform coverage is impossible, as the width must depend on the unknown variance.
This limitation traces back to classical results concerning inference when the variance is uncertain \cite{dantzig1940, 10.1214/aoms/1177731088}.}

These facts motivate two-stage designs involving multiple sampling \cite{e1db1290-dea4-3684-b116-06f625239089, 10.1214/aos/1176345639, 4a785a9d-f0f6-3ad7-b6ff-a8b0190a3f52}: use an initial stage to estimate scale (or verify model constraints), then set the final run length to meet the desired $(\epsilon,\delta)$ target under those validated assumptions; the choice of the preliminary size is nontrivial and materially affects both efficiency and coverage.
The sample size in the initial stage is not a trivial quantity but requires careful consideration, as it largely impacts the efficiency and accuracy of sequential stopping rules \cite{10.1111/j.2517-6161.1983.tb01243.x}.
In practice, no universal criterion seems to exist for determining such preliminary sample sizes, often depending on the underlying assumptions and objectives.
Increasing the preliminary sample size is recognized as one of the simplest ways to enhance performance, albeit not always feasible, for instance, particularly in scenarios involving extremely costly simulation environments.
For instance, it is demonstrated in \cite{D2014} how much the preliminary sample size should be increased to avoid premature stopping and improve low coverage probabilities at termination (Section \ref{subsection assessment and selection of sequential stopping rules}).


\subsection{Large sample theory}
\label{subsection asymptotic validity}

{From a conceptual standpoint, it is essential to let the error tolerance $\epsilon$ tend to zero because only then does the stopping time grow large enough for central limit theorems to apply and for the estimator to stabilize.
In principle, small $\epsilon$ forces long run lengths, ensuring the procedure enters its asymptotic regime where confidence regions behave predictably and achieve the desired coverage.}

In the present context, the large sample theory has been studied for a long time \cite{Anscombe_1949, Anscombe_1952, Anscombe_1954, Ray_1957}, in which the fixed sample-size formulas remain approximately applicable in the sequential procedure for high required accuracy.
In the literature with more flavor of mathematical statistics and operations research, there have been a wide variety of studies on asymptotic properties of stopping rules, such as double-sampling \cite{e1db1290-dea4-3684-b116-06f625239089}, asymptotic expansions \cite{doi:10.1080/07474949108836228, W1977, 10.1214/aos/1176350049}, the convergence rate of fixed-width sequential confidence intervals \cite{2274e8f1-41d9-3449-bdd4-d9796a099dc0, c1980}, the issue of oversampling \cite{10.1111/j.2517-6161.1983.tb01243.x}, and the parameter estimation of specific distributions and statistics, such as normal \cite{Stein_Wald_1947, WOODROOFE198670}, exponential \cite{hirose_isogai_uni_1997, doi:10.1080/03610928208828311}, U-statistics \cite{https://doi-org.utokyo.idm.oclc.org/10.1111/j.1467-842X.1986.tb00587.x, DL1976}, linear regression \cite{10.1214/aoms/1177700157}, as well as the steady-state stochastic
simulation experiment \cite{doi:10.1287/mnsc.35.11.1341, doi:10.1287/opre.40.2.279}, among others, too numerous to list comprehensively.

{In the seminal work of Chow and Robbins \cite{YH1965}}, the key asymptotic analysis was established in this context.
To describe it in brief, let $\{X_k\}_{k\in\mathbb{N}}$ be a sequence of iid random variables with $\mu=\mathbb{E}[X_1]$ and $\sigma^2={\rm Var}(X_1)$. 
For $\epsilon>0$ and $\delta\in (0,1)$, let $\{a_{\delta,n}\}_{n\in \mathbb{N}}$ be a sequence of positive constants converging to $z_{\delta/2}$, and define 
$\tau_{\epsilon,\delta}:=\inf\{n\in \mathbb{N}:\,s_n\le \epsilon \sqrt{n}/a_{\delta,n}\}$, serving as a stopping rule, satisfying $\tau_{\epsilon,\delta}\stackrel{a.s.}{\to} +\infty$ as $\epsilon\to 0$ if $s_n^2/n\stackrel{a.s.}{\to} 0$.
That is, small choices of the precision constant $\epsilon$ need to be associated with correspondingly large simulation times.
Then, it is shown that for every $\delta\in (0,1)$,
\begin{equation}\label{results Chow Robbins}
 \frac{\epsilon^2 \tau_{\epsilon,\delta}}{z_{\delta/2}^2\sigma^2}\stackrel{a.s.}{\to} 1,\quad \mathbb{P}(|\mu_{\tau_{\epsilon,\delta}}-\mu|> \epsilon)\to \delta,\quad \frac{\epsilon^2 \mathbb{E}[\tau_{\epsilon,\delta}]}{z_{\delta/2}^2\sigma^2}\to 1,
\end{equation}
as $\epsilon\to 0$.
Note that the empirical mean $\mu_{\tau_{\epsilon,\delta}}$ here is constructed based on the original sample used for determining the stopping time $\tau_{\epsilon,\delta}$, due to the central limit theorem for random sums (Section \ref{subsection normal approximation}).
The second and third convergences are referred to as asymptotic consistency and asymptotic efficiency, respectively.
A thorough performance analysis is presented in \cite{N1966} under the assumption of iid normal data.

Those pioneering studies were further expanded upon in various studies, such as \cite{NS1996, mukhopadhyay1989, A1969, RD1976, W1977}.
For instance, the convergence rates of relevant quantities are further investigated, such as the expected run length $\mathbb{E}[\tau_{\epsilon,\delta}]$ and the error probability $\mathbb{P}(|\mu_{\tau_{\epsilon,\delta}}-\mu|> \epsilon)$.
Later, we devote Section \ref{subsubsection a general framework for asymptotic analysis} to reviewing a general framework for asymptotic analysis.

Broadly speaking, it is natural to anticipate that various issues in the non-asymptotic regime may be mitigated in line with the asymptotic results by choosing a sufficiently small value of the error tolerance $\epsilon$.
That being said, let us stress that the large sample theory is not fully feasible in practice, as the parameter pair $(\epsilon,\delta)$ needs to be fixed in advance.
Therefore, appropriate adjustments to stopping criteria are crucial \cite{D2014}, such as inflating the empirical variance and imposing the maximum allowable simulation length, particularly crucial when coding in array programming languages. 

\color{black}
\subsection{Normal approximation}
\label{subsection normal approximation}

In a broad sense, as implied via the asymptotic consistency in \eqref{results Chow Robbins}, sequential stopping rules often rely on the following normal approximation involving a non-negative sequence $s_n$ (most commonly, an empirical standard deviation):
\begin{align}
 \tau_{\epsilon,\delta}=\inf\left\{n\in\mathbb{N}:\,\mathbb{P}\left(\left|\mu_n -\mu\right|>\epsilon\right)\le \delta\right\}
 &=\inf\left\{n\in\mathbb{N}:\,\mathbb{P}\left(\sqrt{n}\frac{|\mu_n-\mu|}{s_n}>\sqrt{n}\frac{\epsilon}{s_n}\right)\le \delta\right\}\label{CoV is useless here}\\
 &\approx \inf\left\{n\in\mathbb{N}:\,2\left(1-\Phi\left(\sqrt{n}\epsilon/s_n\right)\right)\le \delta\right\}.\label{typical approximation of the criterion}
\end{align}
Here, an approximation of the error probability $\mathbb{P}(|\mu_n-\mu|>\epsilon)$ (such as done in \eqref{typical approximation of the criterion}, or a different approximation as appropriate) is necessary because the distribution of the estimator $\mu_n$ is generally unknown, as well as the parameter $\mu$ being estimated is inherently unknown.
The normal approximation \eqref{typical approximation of the criterion} clearly comes with a certain central limit theorem $\sqrt{n}(\mu_n-\mu)/s_n\stackrel{\mathcal{L}}{\to}\mathcal{N}(0,1)$, thus implicitly assuming that the resulting simulation length $\tau_{\epsilon,\delta}$ is long enough for the central limit theorem to hold effectively.
On the contrary, early terminations can thus pose a serious issue here, often due to an inaccurate estimate of the standard deviation occurring, for instance, when the early data points are positioned with atypically small spacing, resulting in making the empirical variance extremely small.
Hence, for instance, in \cite{GS1}, a normality test is incorporated into stopping rules.

Once the approximation \eqref{typical approximation of the criterion} is established, parameterizing the simulation length $\tau_{\epsilon,\delta}$ with the two separate parameters $(\epsilon,\delta)$ becomes no longer essential, as they are not independent.
{Clearly, the criterion within \eqref{typical approximation of the criterion} depends on the two parameters $\epsilon$ and $\delta$, but ultimately solely on the single quantity $\epsilon/z_{\delta/2}$, which is a jointly smooth function of the pair $(\epsilon,\delta)$.}
Thus, there exist infinitely many parameter pairs $(\epsilon,\delta)$ that yield the same value of the constant $\epsilon/z_{\delta/2}$.

\subsection{Single- vs. two-stage procedures for mean estimation}
\label{subsection single- or multi-stage procedures}

If the approximate formulation \eqref{typical approximation of the criterion} (in contrast to the exact one \eqref{CoV is useless here}) is adopted to determine the simulation length (that is, $\tau_{\epsilon,\delta}=\inf\{n\in\mathbb{N}:\,2(1-\Phi(\sqrt{n}\epsilon/s_n))\le \delta\}$), then the quantity $\tau_{\epsilon,\delta}$ is evidently defined in terms of the stopped empirical standard deviation $s_{\tau_{\epsilon,\delta}}$, resulting in the first convergence result in \eqref{results Chow Robbins}, due to  the inequalities $\epsilon\sqrt{\tau_{\epsilon,\delta}-1}/s_{\tau_{\epsilon,\delta}-1}<z_{\delta/2}\le \epsilon\sqrt{\tau_{\epsilon,\delta}}/s_{\tau_{\epsilon,\delta}}$ by the formulation \eqref{typical approximation of the criterion}.
In turn, the corresponding asymptotic consistency $\mathbb{P}(|\mu_{\tau_{\epsilon,\delta}}-\mu|>\epsilon)\to \delta$ presented in \eqref{results Chow Robbins} is based on the central limit theorem for random sums (see, for instance, \cite{Anscombe_1952, landers_rogge_1976} and \cite[Problem 27.14]{{billingsley1995probability}}):
\begin{equation}\label{a certain CLT}
\sqrt{\tau_{\epsilon,\delta}}\frac{\mu_{\tau_{\epsilon,\delta}}-\mu}{s_{\tau_{\epsilon,\delta}}}\stackrel{\mathcal{L}}{\to}\mathcal{N}(0,1),\quad \epsilon\to 0,
\end{equation}
where the triplet $(\tau_{\epsilon,\delta},\mu_{\tau_{\epsilon,\delta}},s_{\tau_{\epsilon,\delta}})$ is constructed based upon a single sample set (for instance, \cite{YH1965, PW1992}).
 
In sharp contrast, the stopped quantities $\mu_{\tau_{\epsilon,\delta}}$ and $s_{\tau_{\epsilon,\delta}}$ in the central limit theorem \eqref{a certain CLT} are often interpreted differently in the context of two-stage experiments.
During the first stage, a sample set is used exclusively to determine the simulation length $\tau_{\epsilon,\delta}$. In the second stage, an independent sample set is generated to construct the pair $(\mu_{\tau_{\epsilon,\delta}}, s_{\tau_{\epsilon,\delta}})$, conditioned on the fixed simulation length $\tau_{\epsilon,\delta}$ obtained from the first stage.
Thanks to this conditioning, the central limit theorem \eqref{a certain CLT} effectively becomes identical to the standard form $\sqrt{n}(\mu_n - \mu)/s_n \stackrel{\mathcal{L}}{\to} \mathcal{N}(0,1)$ (with a deterministic number of summands), thus enabling further developments. 
For instance, if the underlying data set $\{X_k\}_{k\in\mathbb{N}}$ is iid, then the standard central limit theorem here is refined with the aid of the Berry-Esseen inequality based on higher-order moments (Sections \ref{subsubsection hickernell 2012} and \ref{subsubsection bayer}), to address the limitations of the standard central limit theorem, particularly in cases involving asymmetric and/or heavy-tailed distributions of the underlying random elements.
If $\{X_k\}_{k\in\mathbb{N}}$ is not iid but of martingale difference type (satisfying $\mathbb{E}[X_k|\,\mathcal{F}_{k-1}]=\mu$), then the central limit theorem needs to be a martingale central limit theorem (Section \ref{subsection martingale difference type}).
Let us stress that those developments come at the cost of generating separate sample sets, which may be impractical or prohibitive for expensive experiments.
An exception arises when the simulation length is deterministically designed with the aid of specific problem structures, for instance, those based on the Hoeffding inequality for Bernoulli trials (Section \ref{subsection Stopping rule for Bernoulli trials}).

\color{black}
\section{Recent developments in sequential stopping rules}
\label{section recent developments}

\color{black}
With the fundamental concepts and key components reviewed (Section \ref{section fundamental concepts}), we are now ready to present an updated and structured overview on sequential stopping rules for {standard and lightly generalized Monte Carlo methods (Sections \ref{subsubsection relative error gajek}, \ref{subsection martingale difference type}, \ref{subsection quasi-independent sequences}, and \ref{subsection low discrepancy sequences for quasi-Monte Carlo methods}), employed primarily for estimating an unknown expectation, including binomial proportions,} in the recent literature primarily published after 1991, following the release of the comprehensive monograph \cite{handbook} in that year, acknowledging that what qualifies as ``recent'' since 1991 can be highly debatable.
Since our focus is on clarifying the main ideas behind each development, we refrain, in most cases, from delving into lengthy or non-essential technical details.

\subsection{Asymptotic analyses}
\label{subsection asymptotic analyses}

Building upon the large sample theory summarized in Section \ref{subsection asymptotic validity}, asymptotic analysis remains a central research focus in the study of sequential stopping rules.
{In principle, investigating the limit of the error tolerance $\epsilon \to 0$ to zero is crucial, because only in this limit does the stopping time become large enough for the central limit theorem to provide an accurate approximation and for the estimator to settle into stable behavior.
As the error tolerance $\epsilon$ decreases, the run length necessarily increases, placing the procedure firmly within its asymptotic regime, where confidence sets shrink in a predictable manner and attain the intended coverage.
This also highlights the importance of reliable variance estimation.
The sequential stopping rule tends to depend explicitly on the estimated variance, and any dependence between the point estimator and this variance estimate can lead to premature or distorted stopping unless the variance estimate has nearly converged.
Thus, analyzing the limit $\epsilon \to 0$ is not merely a technical convenience, but rather the regime in which both estimators behave well, ensuring the asymptotic correctness of sequential methods.}

\subsubsection{A general framework for asymptotic analysis}
\label{subsubsection a general framework for asymptotic analysis}

{In the 1992 seminal work of Glynn and Whitt \cite{PW1992}, a general framework for asymptotic analysis is established around a stopping rule based on the Lebesgue volume $m(C(t))$ of a confidence set $C(t)$ formed from the so-called estimation process for the unknown mean $\alpha$, as follows:
\begin{equation}\label{glynn-whitt stopping rule}
 \tau_{\epsilon,\delta}\leftarrow \inf\left\{t\ge 0:\, m(C(t))^{1/d}+a(t) \le \epsilon\right\},
\end{equation}
where $a(t)$ is a strictly positive function decreasing monotonically to zero as $t\to +\infty$, which is put to inflate the volume $m(C(t))^{1/d}$ to get around its undesired underestimation.
Among other asymptotic properties, it is proved, as asymptotic validity of the stopping rule \eqref{glynn-whitt stopping rule}, that $\mathbb{P}(\alpha\in C(\tau_{\epsilon,\delta}))\to 1-\delta$ as $\epsilon\to 0$, with $\delta$ fixed in $(0,1)$.
We note that a similar set of asymptotic analysis is also presented for a relative-precision stopping rule (Section \ref{subsection error tolerance}).}


Various examples illustrate the applicability of the developed framework in diverse scenarios, ranging from standard mean estimation to complex stochastic optimization, provided that suitable consistency conditions are satisfied for the variance estimation: iid random variables, iid random vectors, nonlinear functions of sample means, jackknife estimators, steady-state means of stochastic processes, and stochastic approximation algorithms.

On the contrary, as briefly discussed in Section \ref{subsection asymptotic validity}, such asymptotic properties do not directly extend to the non-asymptotic regime, where the parameter pair $(\epsilon,\delta)$ is fixed (for instance, \cite{D2014}).
A variety of previously unrecognized challenges are identified in \cite{malfunctioning} when applying the established asymptotic theory in the practical non-asymptotic regime, particularly when the empirical variance $s_n^2$ systematically decreases, increases or even diverges \cite{infinitevariance}, certainly without prior knowledge of such behaviors.
These findings highlight the necessity of careful planning and critical evaluation before implementing sequential stopping rules.

\subsubsection{Higher-order asymptotic analysis}
\label{subsubsection asymptotic analysis with refined knowledge}

In \cite{NS1996}, with the aid of second-order expansions of coverage probabilities, the pioneering asymptotic analysis \cite{YH1965} 
is refined as follows (where $\alpha$ denotes the significance level, corresponding to $\delta$ in our notation):
$\mathbb{E}[\tau_{\epsilon,\delta}-\lceil \sigma^2/(\alpha \epsilon^2)\rceil]=\mathbb{E}[R]-m_4/\sigma^4+o(1)$ and
\[
 \mathbb{P}\left(|\mu_{\tau_{\epsilon,\delta}}-\mu|\ge \epsilon\right)\ge 1-\alpha -\frac{\alpha^2\epsilon^2}{\sigma^2}\left(5+6\left(m_3/\sigma^3\right)^2-\mathbb{E}[R]\right)+o(\epsilon^2),
\]
as $\epsilon\to 0$, under the condition $\mathbb{E}[|X_1|^6]<+\infty$, where $m_k:=\mathbb{E}[(X_1-\mu)^k]$ and $R$ is a random variable with probability distribution $(\mathbb{E}[N])^{-1}\mathbb{P}(2N-\sum_{k=1}^N(X_k-\mu)^2/\sigma^2>r)dr$ defined on $(0,+\infty)$, with $N:=\inf\{n\in\mathbb{N}:\,2n-\sum_{k=1}^n(X_k-\mu)^2/\sigma^2>0\}$.

In addition, an accelerated stopping rule is presented as follows.
With $\rho\in (0,1)$ such that $\rho^{-1}$ is an integer, set
\[
 \zeta_{\epsilon,\delta}:=\inf\left\{n\ge \max\left\{2,\left\lceil (\rho^{-1}\alpha\epsilon^2)^{-1/2}\right\rceil+1\right\}:\,n\alpha \epsilon^2\ge \rho \frac{1}{n}\sum_{k=1}^n (X_k-\mu_n)^2\right\},
\]
as an estimate of $\rho \lceil \sigma^2/(\alpha\epsilon^2)\rceil$.
With $\eta_{\epsilon,\delta}:=\zeta_{\epsilon,\delta}/\rho$, more refined results are derived as $\mathbb{E}[\eta_{\epsilon,\delta}-\lceil \sigma^2/(\alpha\epsilon^2)\rceil]=\rho^{-1}(\mathbb{E}[R]-m_4\sigma^{-4})+o(1)$ and 
\[
 \mathbb{P}\left(|\mu_{\eta_{\epsilon,\delta}}-\mu|\ge \epsilon\right)\ge 1-\alpha -\frac{\alpha^2\epsilon^2}{\sigma^2}\left(5+6\left(m_3/\sigma^3\right)^2-\mathbb{E}[R]+\left(\rho^{-1}-1\right)\left(2m_4/\sigma^4-1-\mathbb{E}[R]\right)\right) +o(\epsilon^2),
\]
as $\epsilon\to 0$.
Those theoretical developments are supported through a parameter estimation problem on the exponential distribution.


\subsection{Sequential stopping rules based on higher-order moments}
\label{subsection stopping rules based on higher-order moments}

As discussed in Section \ref{subsection normal approximation}, the performance of stopping rules in typical iid setups can largely depend on the quality of normal approximation \eqref{typical approximation of the criterion}.
{To improve this normal approximation in the non-asymptotic regime (that is, with the error tolerance $\epsilon$ fixed, unlike in the asymptotic analysis (Section \ref{subsection asymptotic analyses})), a natural approach is to leverage more refined results, such as Berry-Esseen inequalities that incorporate higher-order moments beyond the first two moments.}
In this section, we review sequential stopping rules based on such higher-order moments.

\subsubsection{A two-stage algorithm with known bounds on the kurtosis}
\label{subsubsection hickernell 2012}

A two-stage algorithm is proposed in \cite{FLYA2012} for iid Monte Carlo methods under the prior availability of a bound of the kurtosis, that is,  $\mathbb{E}[|X_1-\mu|^4]/({\rm Var}(X_1))^2\le Q_{\max}$, ensuring the reliability of the variance estimation.
The algorithm begins by estimating the variance using the empirical estimate $s^2_{n_{\sigma}}$ based on an initial sample of size $n_\sigma$. 
A conservative estimate is then set as $\widehat{\sigma}^2 = C^2 s^2_{n_{\sigma}}$, where the constant $C(> 1)$ is called a variance inflation factor for {reducing the probability that the variance estimation is below the true value.}
In the second stage, the required sample size $n^*$ is determined using the Chebyshev and Berry-Esseen inequalities, yielding $n^* = \max\{1, \min\{N_{\text{Cheb}}, N_{\text{BE}}\} \}$, where $N_{\text{Cheb}}$ and $N_{\text{BE}}$ denote the sample sizes derived with reference to the Chebyshev and Berry-Esseen inequalities, respectively.

While more computationally intensive than conventional stopping rules, this two-stage approach ensures that the confidence interval meets the desired precision and confidence level by adaptively determining sample sizes based on variance estimates.
As a consequence, it is particularly well-suited for scenarios where the fourth moment of the underlying distribution is bounded, albeit with the bound known (or very accurately estimated) in advance.

Various numerical examples are presented in this work, such as univariate fooling functions designed to mislead common quadrature algorithms, a single hump test integrand, and a geometric-mean Asian option pricing. 
These examples demonstrate that, while the proposed two-stage algorithm is computationally demanding, it ensures guaranteed error bounds, as long as the kurtosis condition is satisfied, unlike conventional automatic quadrature algorithms which can fail despite claiming high accuracy.

 
 
 
 

\subsubsection{A sequential stopping rule based on the Berry-Esseen inequality}
\label{subsubsection bayer}

Sequential stopping rules are developed in \cite{BHST} for iid Monte Carlo methods, based on higher-order moments estimated in a two-stage framework.
It is first demonstrated that a typical second-moment stopping rule, which iteratively doubles the sample size $M_n=2M_{n-1}$ until $2(1 - \Phi(\sqrt{M_n} \epsilon/s_{M_n}))\leq \delta$, may fail in the non-asymptotic regime due to inaccuracies inherent in the central limit theorem approximation.
To address this issue, a refined algorithm is constructed with the aid of third and fourth moments in conjunction with the Berry-Esseen theorem.
There, the simulation is stopped when  
\[
2 \left(1 - \Phi\left(\frac{\sqrt{M_n} \epsilon}{s_{M_n}}\right)\right) + 2 C_{\text{BE}} \left(\frac{\sqrt{M_n} \epsilon}{s_{M_n}}, \beta_{M_n}\right) \frac{1}{\sqrt{M_n}} \leq \delta,
\]
with $C_{\text{BE}}(x, \beta):= \min\{0.3328 (\beta + 0.429),\, (18.1139 \beta)/(1 + |x|^3)\}$, where the argument $\beta$ corresponds to the estimate of the normalized third moment $\mathbb{E}[|X_1|^3]/\sigma^3$.
The effectiveness of this refined stopping rule is demonstrated through numerical experiments on various distributions, including bounded, light-tailed, and heavy-tailed cases, such as the Pareto, normal-inverse Gaussian, uniform, and exponential distributions.

\color{black}
\subsubsection{Stopping rules with known bounds on moments}
\label{subsubsection kunsch 2019}

In \cite{KNR2019}, the complexity of an $(\epsilon,\delta)$-approximation (or, almost equivalently, the solvability as defined in the work) is investigated 
within the cone-shaped class of the underlying distribution $\mathcal{Y}_{p,q,K} := \{Y\in L_q(\Omega):\,\mathbb{E}[|Y-\mathbb{E}[Y]|^q]\leq K\mathbb{E}[|Y-\mathbb{E}[Y]|^p]\}$ for $1\leq p<q\leq +\infty$ and $K>1$.
Note that this class generalizes the case of bounded kurtoses $(p,q)=(2,4)$ examined in \cite{FLYA2012} (Section \ref{subsubsection hickernell 2012}).
In addition, an algorithm is developed for this $(\epsilon,\delta)$-approximation with its expected run length bounded by $C_q K^{pq/(q-p)}(1+(\mathbb{E}[|Y-\mathbb{E}[Y]|]/\epsilon)^{\max\{1+1/(q-1),2\}})\ln(1/\delta)$.
We also note that, in \cite{https://doi-org.utokyo.idm.oclc.org/10.1002/rsa.20839}, a relative-precision stopping rule is built upon a general distribution with a known bound $c$ for the relative standard deviation, that is, $|\sigma/\mu|\le c$.

\subsection{Sequential stopping rules under distributional assumptions}
\label{subsection stopping rules under distributional assumptions}

In the opposite direction to generalizing the underlying random elements, several insightful studies aim to make the non-asymptotic analysis with fixed error tolerance as explicit as possible by instead imposing restrictive assumptions on the underlying random elements.

\subsubsection{Analyses of coverage probabilities under iid normal assumption}
\label{subsubsection coverage probabilities}

In \cite{DL2009, DL2014}, the coverage probability is investigated in depth under the assumption of the random elements being iid normal.
By imposing this assumption, their analyses are made as analytical as possible in the form of so-called coverage functions (Section \ref{subsection coverage functions and related stopping rules}), quantifying the distribution of simulation lengths $\mathbb{P}(\tau_{\epsilon,\delta}=n)$ and the loss in coverage (that is, $(1-\delta)-\mathbb{P}(|\mu_{\tau_{\epsilon,\delta}}-\mu|\le \epsilon)$).
This framework highlights the potential risk of inappropriate early termination of stopping rules, even when the underlying data satisfy all idealized properties.
It is thus rightfully claimed that this loss in coverage can be substantial in the practical scenario where the data are non-normal or exhibit dependence.

\color{black}
\subsubsection{A relative-precision stopping rule for bounded random elements}
\label{subsubsection absolute error}

{The $(\epsilon, \delta)$-approximation of a parameter $\mu$ is defined in \cite{doi:10.1137/S0097539797315306} as a random estimate $\mu_n$, under the assumption that the underlying iid random variables are bounded in $[0,1]$ with no more information about its distribution, satisfying the relative error-based criterion $\mathbb{P}(|(\mu_n - \mu)/\mu| \leq \epsilon) \geq 1 - \delta$, ensuring that the estimate $\mu_n$ lies within a factor of $(1 + \epsilon)$ of the parameter $\mu$ with probability $1-\delta$ at least.
Such approximations are particularly useful when exact computations of expectations are computationally expensive or NP-hard.}


Two stopping rules are proposed and analyzed in this study.
First, with $\epsilon\ll 1$, $\delta\in (0,1)$, $\Upsilon=4(e-2)\lambda\ln(2/\delta)/\epsilon^2$ and $\Upsilon_1=1+(1+\epsilon)\Upsilon$, a stopping rule proposed there is to stop the simulation at $\tau_1:=\inf\{n\in\mathbb{N}:\,X_1+\cdots+X_n\ge \Upsilon_1\}$ and return $\Upsilon_1/\tau_1$ as an estimate of $\mathbb{E}[X_1]$, satisfying $\mathbb{P}(|\Upsilon_1/\tau_1-\mathbb{E}[X_1]|\le \epsilon)>1-\delta$ and $\mathbb{E}[\tau_1]\le \Upsilon_1/\mathbb{E}[X_1]$.

The second stopping rule proposed there consists of two steps.
First, run the first stopping rule with $\min\{1/2,\sqrt{\epsilon}\}$ and $\delta/3$, in place of $\epsilon$ and $\delta$, to generate an estimate $\widehat{\mu}$.
With $\Upsilon_2=2(1+\sqrt{\epsilon})(1+2\sqrt{\epsilon})(1+\ln (3/2)/\ln(2/\delta))\Upsilon$ and $S_n:=2^{-1}\sum_{k=1}^n(X_{2k-1}-X_{2k})^2$ based on realizations independent of the first phase, set $\tau_2=\Upsilon_2 \epsilon/\widehat{\mu}$ and $\rho=\max\{S_{\tau_2}/\tau_2,\epsilon \widehat{\mu}\}$.
Then, with $\tau_3=\Upsilon_2 \rho/\widehat{\mu}^2$, return $\mu_{\tau_3}$ as an estimate for $\mathbb{E}[X_1]$ based on the realizations in the first phase, satisfying $\mathbb{P}(|\mu_{\tau_3}-\mathbb{E}[X_1]|\le \epsilon)>1-\delta$, $\mathbb{P}(\tau_3\ge c \Upsilon\rho/\widehat{\mu}^2)\le \delta$ and $\mathbb{E}[\tau_3]\le c\Upsilon/\mathbb{E}[X_1]$ for some positive $c$.
This algorithm is proved to be robust in the sense that the constant $c$ in the bounds is independent of the underlying distribution, if $\rho$ is instead computed with the theoretical mean and variance, that is, $\rho=\max\{{\rm Var}(X_1),\epsilon \mathbb{E}[X_1]\}$.

\subsubsection{A relative-precision stopping rule for bounded random elements of martingale difference type}
\label{subsubsection relative error gajek}

A relative-precision stopping rule is constructed and investigated in \cite{LWP2013}, under the assumptions (i) $X_n$ is $\mathcal{F}_n$-measurable for a suitable filtration $(\mathcal{F}_n)_{n\in\mathbb{N}_0}$, (ii) $\mathbb{E}[X_n|\mathcal{F}_{n-1}]=\mu$, and (iii) $X_n\in [0,1]$ with probability one, for all $n\in \mathbb{N}$.
We note that the assumptions (i) and (ii) are the same as those imposed in \cite{JWmartingale}.
With the simulation length defined by $n_r:=\{n\in \mathbb{N}:\,n\mu_n \ge r\}$ for $r>0$, 
various upper and lower bounds for the error probability $\mathbb{P}(|(\mu_{n_r}-\mu)/\mu|>\epsilon)$ and the expected simulation length $\mathbb{E}[n_r]$ are derived,
along with numerical results demonstrating a significant reduction in sample size and the accuracy of the derived bounds.
An extension is also explored where the boundedness condition (iii) is suppressed.

\subsubsection{Stopping rules for Bernoulli trials}
\label{subsection Stopping rule for Bernoulli trials}

{One of the most practically important settings involves Bernoulli trials, which have received extensive attention throughout their long history, including various early foundational developments, such as \cite{10.1214/aoms/1177706361, 10.1214/aoms/1177731018, 10.1145/1102586.1102590}.}
{In this context, the unknown parameter of interest is the success probability $p \in (0,1)$, whereas the associated odds $p/(1-p)$ and log odds $\ln (p/(1-p))$ have also been studied (see, for instance, \cite{Bandyopadhyay03072017, 10.1093/biomet/54.1-2.181}).}
In particular, the estimation of the success probability is of utmost importance in structural reliability analysis \cite{SONG2023103479}, where Monte Carlo experiments with appropriately defined stopping criteria are broadly referred to as principled Monte Carlo simulation \cite{ELLINGWOOD2025102474}.

In this context, an absolute-precision stopping rule is examined in \cite{Jiang2014}.
Unlike previous developments relying on asymptotic approximations primarily via the central limit theorem (Section \ref{subsection normal approximation}),
an algorithm is built with the aid of the Hoeffding inequality for bounded random variables $\mathbb{P}(|\mu_n - p| \geq \epsilon) \leq 2e^{-2n\epsilon^2}$, leading to stopping at $\lceil\ln (2 / \delta)/(2 \epsilon^2)\rceil$.
This simple stopping rule is guaranteed to achieve the prescribed error tolerance and confidence level, at a computational cost within a small constant factor (between $3.64$ and $5.09$) of the sample size based on the central limit theorem for typical confidence levels $\delta \in [0.001,0.1]$.
Numerical experiments demonstrate the reliability of the algorithm, though it can be conservative for small values of the success probability $p$.
It is noted that extending the approach to relative error tolerances remains an open challenge, as this would require a lower bound on the unknown value of the success probability $p$.
 
 

In \cite{https://doi-org.utokyo.idm.oclc.org/10.1002/rsa.20654}, a relative-precision stopping rule, referred to as the Gamma Bernoulli approximation scheme (GBAS method, for short), is proposed for Bernoulli trials, with the expected run length bounded by $2\epsilon^{-2}p^{-1}\ln(2\delta^{-1})(1-(4/3)\epsilon)^{-1}$ for general success probability $p\in (0,1)$, or $(1/5)\epsilon^{-2}p^{-1}(1+2\epsilon)(1-\delta)\ln((2-\delta)/\delta)$ if restricted to $p\in (0,1/2]$, as an improvement of the algorithm developed in \cite{doi:10.1137/S0097539797315306} in terms of, for instance, the constant factor and unbiasedness of the output.



{As for fixed-width interval estimation for risk ratios (that is, ratios of two binomial proportions, quantifying the risk of an event under different exposures), a sequential procedure is developed in \cite{Cho2013ApproximateConfidence}, where sampling continues until a stopping rule based on the estimated variance is met, yielding asymptotically valid fixed-width confidence intervals.
In contrast, a two-stage stopping rule is adopted in \cite{cho2019}, in which an initial preliminary sample is used to estimate the variance, after which additional sampling continues only until the fixed-width criterion is satisfied, thereby ensuring that the final sample size meets the prescribed precision without requiring continuous monitoring.}

\subsubsection{Confidence intervals for inverse binomial sampling}
\label{section confidence intervals for inverse binomial sampling}

In contrast to direct binomial sampling, where the sample size is fixed in advance, the so-called inverse binomial sampling has been explored in various studies, in which the simulation continues until a predetermined number of successes is observed.
This stopping rule is based on a negative binomial distribution for the resulting number of trials, for which basics, such as the variance and bounds for the relative mean-square error, are available in early times \cite{10.1093/biomet/61.2.385, 10.1093/biomet/63.1.216, 357b9062-c3ca-3953-ab0e-8de887bd167d}.

Here, we briefly review existing studies on the construction of confidence intervals for the binomial proportion under inverse binomial sampling.
In \cite{LJ2006, 4686254}, the estimator for the success probability $p$ is given as $(N-1)/n$, where the required sample size $n$ is random and determined by the stopping rule of observing a fixed number $N$ of successes.
In this framework, those studies provide a detailed analysis of the statistical guarantees of such estimators, with particular emphasis on the confidence level of relative intervals of the form $[p/\mu_2,p\mu_1]$ for $\mu_1,\mu_2>1$, establishing conditions under which these intervals achieve prescribed coverage probabilities and highlighting the practical reliability of inverse binomial sampling for rare-event probability estimation.
Those studies are advanced in \cite{10.3150/09-BEJ219} by establishing the maximum possible guaranteed confidence for relative interval estimation and by constructing estimators that achieve this optimum for all success probabilities, thereby completing and generalizing the partial confidence-guarantee results of the previous two studies.
Finally, we refer the reader to, for instance, the odds $p/(1-p)$ and the log odds $\ln(p/(1-p))$ \cite{Bandyopadhyay03072017, mendo2025}, and the risk ratio $p_1/p_2$ under two different exposures \cite{https://doi-org.utokyo.idm.oclc.org/10.1002/sim.3158} for respective confidence intervals under inverse binomial sampling.

\color{black}
\subsection{Sequential stopping rules for non-iid sequences}
\label{subsection stopping rules for non-iid sequences}

As discussed in Section \ref{subsection non-iid random elements}, sequential stopping rules for Monte Carlo methods have often been constructed and analyzed for a sequence of non-iid random elements.
These deserve and require separate investigation, as the sample sizes needed to achieve a specified level of accuracy can differ substantially between iid and non-iid sequences.

\subsubsection{Non-iid sequences of martingale difference type}
\label{subsection martingale difference type}

A stopping rule is developed in \cite{JWmartingale} for Monte Carlo methods for non-iid sequences of martingale difference type, satisfying $X_n\in \mathcal{F}_n$ and $\mathbb{E}[X_n|\mathcal{F}_{n-1}]=\mu$ for all $n\in \mathbb{N}$, which is substantially weaker than the usual iid assumption.
With the batch index $t\in\mathbb{N}_0$, fix a sequence of strictly increasing non-negative integers $\{m(t)\}_{t\in\mathbb{N}_0}$ with $m(0)=0$ and $\lim_{t\rightarrow\infty}m(t)=+\infty$, and denote by $M(t):=\{m(t-1)+1,\cdots,m(t)\}$ the set of indices in the $t$th batch.
With the empirical batch mean $\mu(t):=|M(t)|^{-1}\sum_{k\in M(t)} X_k$ and variance $\sigma^{2}(t):=(|M(t)|-1)^{-1} \sum_{k \in M(t)} (X_k-\mu(t))^2$, a stopping rule is built in the usual form $\tau_{\epsilon,\delta}=\inf\{t\in \mathbb{N}:\,2(1-\Phi(\epsilon \sqrt{|M(t)|}/(\sigma(t)+a(t))))\le \delta\}$, as an approximation of $\inf\{t\in\mathbb{N}:\,\mathbb{P}(|\mu(t)-\mu|>\epsilon)\le \delta\}$, where $a(t)$ is a strictly positive and monotonically decreasing function.
Recall that a similar martingale-type assumption has been made in \cite[Assumption 1.1]{LWP2013} in the context of relative-precision stopping rules (Section \ref{subsubsection relative error gajek}), albeit based upon a boundedness condition on the underlying randomness.
{Its effectiveness is demonstrated through numerical experiments conducted on an ARCH model and an adaptive variance reduction method, highlighting a favorable balance between coverage probability and computational complexity.}
It is noteworthy that, unlike the other rules in the non-asymptotic regime, this stopping rule has also been validated in the asymptotic regime explicitly in line with the framework \cite{PW1992}.




\subsubsection{Mixing sequences with independence tests}
\label{subsection quasi-independent sequences}

The sample sizes required for prescribed accuracy can be drastically different between iid and correlated sequences.
In a series of studies \cite{899773, CHEN2003237, Chen01052012}, sequential stopping rules are constructed for the sample mean, where the sequence of random variables $\{X_k\}_{k\in\mathbb{N}}$ is
$\phi$-mixing, that is, $|\mathbb{P}(E_{n,m}|E_n)-\mathbb{P}(E_{n,m})|\le \phi(m)$ for all $n,m\in\mathbb{N}$, where $E_{n,m}$ and $E_n$ denote $\sigma(\{X_k:\,k\ge n+m\})$- and $\sigma(\{X_k:\,k\le n\})$-measurable events and $\phi$ is positive and vanishing.
A typical example is the waiting-time of an M/M/1 delay-in-queue. 
Those studies employ test procedures for checking whether the input data can be regarded as independent so that the simulation may proceed until the desired number of effectively independent random samples is collected, with strongly correlated observations being skipped.
In those studies, the effectiveness of the proposed stopping rules is demonstrated systematically through various time series models and queueing systems.

\subsubsection{Low-discrepancy sequences for quasi-Monte Carlo methods}
\label{subsection low discrepancy sequences for quasi-Monte Carlo methods}

An absolute-precision stopping rule is proposed in \cite{Hickernel2018} for estimating the integral value $\int_{(0,1)^d}f({\bf x})d{\bf x}$ when the random elements $\{X_k\}_{k\in\mathbb{N}}$ are not iid but form a low-discrepancy sequence in $(0,1)^d$.
{This stopping rule is designed with reference to the cubature error obtained from the decay of the discrete Fourier coefficients of the integrand, expressed as $|n^{-1}\sum_{k=1}^n f(X_k)-\mu|=|\sum_{k\in \mathcal{P}}\widehat{f}_k|$ where $\mathcal{P}$ is a suitable infinite set on a Fourier expansion of the integrand $f({\bf x})=\sum_k\widehat{f}_k\phi_k({\bf x})$.
The cubature error is well suited in this context because the quasi-Monte Carlo point sets are tailored to the Fourier structure of the integrand, allowing the error to be estimated from the decay of high-frequency coefficients and thus providing a principled criterion for halting the algorithm.}
Through numerical experiments on a multidimensional integration problem arising in statistical physics, it is reported that this QMC-based stopping rule with randomly scrambled Sobol sequences requires substantially less time and a smaller number of iterations than a stopping rule for iid Monte Carlo methods developed in this study (Section \ref{subsubsection hickernell 2012}).

\subsection{A sequential stopping rule when estimating multiple means}
\label{subsection several means}

A relative-precision stopping rule is constructed in \cite{Luck1993} for Monte Carlo methods for estimating multiple means separately yet simultaneously, resulting in no correlation to be addressed among variates, unlike in the case of multivariate outputs (Section \ref{subsection multivariate outputs}).
When multiple means $\{\theta_k\}_{k\in \{1,\cdots,p\}}$ need to be estimated simultaneously, it is proposed to generalize the relative half-width criterion via 
\[
\mathbb{P}\left(\{|\theta_1 - \widehat{\theta}_1| \leq \epsilon_1|\widehat{\theta}_1|\}\cap \cdots \cap\{|\theta_p - \widehat{\theta}_p| \leq \epsilon_p|\widehat{\theta}_p|\}\right) \geq 1-\delta,
\]
implying that with probability $1-\delta$ at least, the respective relative errors are at most $\epsilon_k/(1-\epsilon_k)$ for all components.
Instead of applying the equally weighted significance level ($\delta_k\equiv \delta/p$) in the Bonferroni inequality, the probabilities $\delta_k$ of exceeding the error bounds are estimated as $\widehat{\delta}_k= 2F_{\nu_k}(-\epsilon_k|\widehat{\theta}_k|/s_{\widehat{\theta}_k})$ (where $F_{\nu_j}$ denotes the cumulative distribution function of $t$-distribution with $\nu_k$ degrees of freedom), with which
the termination occurs when $\sum_{k=1}^p \widehat{\delta}_k \leq \delta$.



{Among several differences, the most notable distinction from the case of multivariate outputs (Section \ref{subsection multivariate outputs}) is that the full covariance matrix need not be estimated.
Estimating the covariance matrix is not only computationally demanding and sensitive but also conceptually central to constructing confidence ellipsoids in the case of multivariate outputs, whereas it plays no role when the components of the mean vector are estimated separately.}

Notably, the experiments conducted on queueing systems result in a 13-40\% reduction in run length when compared to the equal-weights scheme $\delta_k=\delta/p$. 
Furthermore, incorporating one or two additional means increases the run length by 6-83\% (with an average increase of 31.4\%) relative to single-mean estimation.
It is also noted that stopping rules for multivariate outputs based on empirical covariance matrices (as opposed to separate multiple means) may result in overly conservative run lengths when the objective is to control individual relative errors rather than the volume of the joint confidence region.

\subsection{Coverage functions and related stopping rules}
\label{subsection coverage functions and related stopping rules}

The coverage function \cite{doi:10.1287/mnsc.26.1.18} quantifies how well an interval estimator captures the true parameter over all confidence levels.
For an interval estimator producing a random confidence region $R(q,X)$ at confidence coefficient $q$ (corresponding to the confidence level $1-\delta$ in our context), define $q^*:=\inf\{q\in [0,1]:\,\mu \in R(q,X)\}$, the smallest confidence level at which the interval contains the true parameter $\mu$.
The coverage function is then the distribution function $F_{q^*}(q):=\mathbb{P}(q^*\le q)$, which equals the actual coverage probability achieved when the nominal level is $q$.
Under correct model assumptions, the random variable $q^*$ is uniformly distributed on $[0,1]$, that is, $F_{q^*}(q)=q$ for all $q\in [0,1]$.
Departures from uniformity indicate bias, inefficiency, or violated dependence assumptions in the simulation output.

The concept of coverage functions is extended in \cite{1172904}, which introduces a single-criterion framework based on coverage values for comparing fixed-sample confidence interval procedures, emphasizing that validity and closeness to an ideal procedure determine quality.
This idea is formalized in \cite{Yeh02112015} as the VAMP1RE metric, an acronym reflecting the two fundamental drivers of confidence-interval quality and the nature of a unified criterion.
A key motivation for creating a single rating criterion for fixed-sample procedures is that sequential stopping rules generally lack a unique coverage value.
Because sequential methods let sample size vary with the data, an interval may repeatedly cover or miss the true parameter during a simulation.
The VAMP1RE metric converts coverage functions from a graphical diagnostic into a quantitative tool applicable to any fixed-sample confidence interval procedure, regardless of nominal coverage.
Moreover, this metric is further generalized in \cite{Yeh03072025} to include confidence interval procedures with random sample sizes. 



\color{black}
\subsection{Quality assessment of sequential stopping rules}
\label{subsection assessment and selection of sequential stopping rules}

A comprehensive study on the low coverage issue is presented {in the 2014 seminal work of Singham \cite{D2014}} through the quality assessment of a wide range of stopping rules used for iid Monte Carlo methods within the non-asymptotic regime.
Instead of proposing stopping rules, this study focuses on analyzing existing orthodox ones, highlighting the issue of low coverage probabilities by employing a visualization approach, analogous to the graphical method developed in \cite{doi:10.1287/opre.38.3.546}, to illustrate confidence interval coverages.
A key contribution is the introduction of coverage profiles, which is a method for assessing stopping rules by plotting the coverage probability of a stopping rule as a function of different simulation lengths.
Coverage profiles yield useful insights, for instance, indicating how much the initial sample size should be increased to reduce the risk of premature stopping and improve coverage probabilities.
This study delves into the low coverage issue in further detail, examining how the coverage is influenced by various parameters, including error tolerances (Section \ref{subsection error tolerance}), confidence and significance levels (Section \ref{subsection confidence level}), and preliminary and final sample sizes (Section \ref{subsection preliminary and final sample sizes}), through a variety of numerical results on normal, exponential and gamma distributions.

\color{black}
\subsection{Stopping rules outside the conventional framework}
\label{section outside the conventional framework}

We close the present review by very briefly describing three methods for sequentially stopping Monte Carlo sampling that fall outside the scope of the conventional framework \eqref{core requirement for stopping}, in the sense that the first two (Sections \ref{subsubsection the coefficient of variation} and \ref{subsubsection acceptable shifting convergence bands}) do not involve error tolerances (Section \ref{subsection error tolerance}) or confidence levels (Section \ref{subsection confidence level}), while the third one (Section \ref{subsubsection successive changes to the empirical mean}) is based upon successive changes to the empirical mean.
{Finally, we briefly describe the challenges associated with stopping rules for advanced modern simulation techniques, specifically, Markov chain Monte Carlo (Section \ref{subsubsection MCMC}), stochastic programming (Section \ref{subsubsection stochastic programming}), machine learning related techniques (Section \ref{subsubsection machine learning related techniques}), and Bayesian cubature (Section \ref{subsubsection bayesian cubature}), contrasting them with those in the conventional framework and providing motivating and comparative bases for potential future research.}

\color{black}
\subsubsection{Coefficients of variation}
\label{subsubsection the coefficient of variation}

To evaluate the uncertainty associated with the mean estimator $\mu_n$, one often introduces the coefficient of variation (CoV, for short), also referred to as the relative standard deviation, which is conceptually akin to the relative root-mean-square error.
This metric is defined as the ratio of the (empirical) standard deviation to the (empirical) mean $s_n/\mu_n$, or often its magnitude $s_n/|\mu_n|$.
The unitless nature of the coefficient of variation makes it particularly useful for comparing datasets with differing units or substantially different mean values.
It, however, exhibits sensitivity to fluctuations in the mean, especially when the mean is close to zero.
It is clearly unsuitable for directly constructing confidence intervals for the mean, as in \eqref{CoV is useless here}.

In practice, the coefficient of variation has found applications as a stopping criterion across various fields involving Monte Carlo methods.
For instance, in structural reliability analysis, the sampling procedure is often designed to terminate when the CoV reaches a specified threshold (such as $0.05$ \cite{ASDIS, ReliabilitySS}), with the number of iterations and the empirical variance subsequently reported.

\subsubsection{Acceptable shifting convergence bands}
\label{subsubsection acceptable shifting convergence bands}

{The concept of an acceptable shifting convergence band, introduced in \cite{ATA2007237}, provides a dynamic interval within which the empirical mean is required to remain for a specified number of consecutive iterations.
Once the empirical mean stays within this moving band for the prescribed duration, the simulation run is deemed to have effectively converged.}
An associated stopping criterion is proposed based on the number of updates to the acceptable shifting convergence band, which is adjusted only when the sample mean deviates from the most recently established band.
Its reliability and efficiency are empirically validated for some representative underlying distributions including Bernoulli, normal, and (the fifth observation of) a steady-state Markov process.

\color{black}
\subsubsection{Successive changes to the empirical mean}
\label{subsubsection successive changes to the empirical mean}

A stopping rule is developed in \cite{WOO1991179} for iid Monte Carlo methods based upon the successive change of the mean estimates $|\mu_{n+1}-\mu_n|$, in the form of
\begin{equation}\label{successive changes}
 \tau_{\epsilon,\delta}\leftarrow \inf\left\{n\in\mathbb{N}:\,\mathbb{P}(|\mu_{n+1}-\mu_n|>\epsilon)<\delta\right\}.
\end{equation}
As is clear, the stopping criterion here is conceptually different from the one specified in the conventional framework \eqref{core requirement for stopping}.
In \eqref{successive changes}, the error tolerance $\epsilon$ represents the maximum permitted change in the empirical mean $\mu_n$ resulting from an additional sample, as opposed to the conventional framework \eqref{core requirement for stopping}, where the error tolerance $\epsilon$ denotes the maximum allowable deviation of the empirical mean $\mu_n$ from the unknown theoretical mean $\mu$.
The simulation length $\tau_{\epsilon,\delta}$ is then derived based on the criterion \eqref{successive changes} with the aid of a Chebyshev inequality with respect to the $(n+1)$st realization based on the empirical standard deviation $s_n$ \cite{Saw01051984}, as given in the form of $\mathbb{P}(|X_{n+1}-\mu_n|>\lambda \sqrt{(n+1)/n}s_n)\le \lfloor (n+1)(n^2-1+n\lambda^2)/(n^2\lambda^2)\rfloor/(n+1)$ for $\lambda\ge 1$ and $n\in \mathbb{N}$.
The relative precision (Section \ref{subsection error tolerance}) is also addressed by setting $\epsilon=c \mu_n$ with positive constant $c$, where the simulation length is characterized in terms of the coefficient of variation (Section \ref{subsubsection the coefficient of variation}). 
Numerical results are presented for radiological risk assessment involving the release of the radionuclide Ra from a hypothetical shallow waste disposal site.
We note that a multivariate variant could potentially be developed with the aid of the multivariate version of the Chebyshev inequality with estimated mean and variance \cite{Stellato03042017}.
Note that such successive changes have been applied in various problem-specific forms, both similarly and differently (such as the relative change $|(\mu_{n+1}-\mu_n)/\mu_n|$), in the literature (for instance, in Monte Carlo criticality analysis in nuclear science \cite{Shim01102007, Ueki01012005}).

\subsubsection{Stopping rules for Markov chain Monte Carlo}
\label{subsubsection MCMC}

A central challenge in developing stopping rules for MCMC is the lack of universally reliable or easily verifiable criteria for determining when a chain has run “long enough.” Stopping too early risks halting before the chain reaches stationarity, producing biased estimates, while running too long wastes computational resources. This balance is especially difficult in high-dimensional or strongly correlated posterior spaces, where convergence is hard to diagnose and available diagnostics capture only partial and sometimes conflicting aspects of chain behavior.

Many widely used criteria rely on heuristics, assume ideal mixing, or depend on user-chosen thresholds lacking firm theoretical grounding. Practical complications also arise because posterior shapes vary widely across models, leading to inconsistent performance, and because multi-chain diagnostics can be influenced by initialization and between-chain dispersion, while single-chain precision-based rules require stable Monte Carlo error estimates that may be noisy. Consequently, robust stopping rules that generalize across models remain elusive, and practitioners often combine multiple diagnostics, visual inspection, and domain expertise to judge termination.
In the specific context of latent variable models used in psychometrics, \cite{https://doi.org/10.1111/bmsp.12357} provides a systematic simulation comparison of four stopping rules, the Gelman-Rubin diagnostic, Geweke's diagnostic, the effective sample size, and the relative fixed-width stopping rule, showing that single-chain precision-based approaches tend to outperform multi-chain convergence diagnostics for item parameter recovery, while the choice of stopping rule has little effect on person parameter estimates. 
We refer the reader to \cite{https://doi.org/10.1111/bmsp.12357, doi:https://doi-org.utokyo.idm.oclc.org/10.1002/9781118445112.stat08283} for recent reviews and empirical studies.

\subsubsection{Stopping rules for stochastic programming}
\label{subsubsection stochastic programming}

Constructing effective stopping rules for stochastic programming is difficult because such methods must balance statistical uncertainty, computational cost, and model complexity, often in the absence of solid theoretical guarantees.
Many stochastic algorithms, especially those relying on sampling, Monte Carlo simulation, or multi-start procedures, produce dependent or poorly understood iterate sequences, complicating probabilistic modeling or reliable estimation of optimality gaps.
Even when independence assumptions are imposed to facilitate analytical stopping rules, they rarely hold in practice, particularly in adaptive algorithms that blend global exploration with local refinement.
As a result, stopping rules based on simplified stochastic models can be overly conservative, requiring excessive computation, or overly optimistic, offering little meaningful error control.

Practical difficulties further arise because essential quantities, such as the distribution of objective values, the structure of attractive regions in the search space, or measures of residual uncertainty, are unknown and hard to approximate.
Practitioners therefore rely heavily on heuristics whose sensitivity to algorithm design and problem structure can be unpredictable.
Ultimately, the interplay between complex algorithmic dynamics and intrinsic randomness makes it challenging to develop stopping rules that are both efficient and theoretically sound, leaving many users dependent on heuristic thresholds, time limits, or ad-hoc diagnostics.
For further accounts, we refer the reader to, for instance, \cite{df3b10f8-26b4-38aa-9eb3-41173d4233f0, Schoen2009}.

\subsubsection{Stopping rules for machine learning related techniques}
\label{subsubsection machine learning related techniques}

Designing stopping rules for machine learning is challenging because learning trajectories in models like deep neural networks are highly non-convex and noisy, making it difficult to determine whether improvements reflect true progress or transient fluctuations.
Validation-based criteria can be misleading when data are scarce, imbalanced, or not fully representative of deployment conditions.
Moreover, modern training pipelines often involve stochastic optimization, data augmentation, and regularization, all of which introduce noise that obscures reliable signals for convergence.

In reinforcement learning, stopping criteria are even harder to define because performance can oscillate as the agent explores, policies evolve, or environments change, preventing stable convergence indicators from emerging.
The stochasticity of reward signals and the delayed effect of actions make it difficult to assess at any given point whether further training will improve policy quality.
As a result, practitioners frequently rely on heuristics such as fixed training budgets, moving-average performance thresholds, or domain-specific cues rather than theoretically grounded, general-purpose stopping rules.

Research on stopping rules for machine-learning methods remains highly active, driven by the increasing complexity, scale, and stochastic nature of modern learning systems.
As models grow larger and training dynamics become more intricate, researchers continue to develop theory-informed and data-adaptive stopping criteria that more accurately capture real training behavior rather than relying on idealized assumptions.
This line of work spans deep learning, reinforcement learning, and numerous related subfields, and is expected to remain vibrant for years as practitioners seek more reliable, efficient, and theoretically grounded methods for deciding when to halt training.
For further accounts, we refer the reader to, for instance, \cite{ Hardt2016TrainFaster, HuLei2022JMLR_EarlyStoppingIterativeReg,Prechelt2012EarlyStopping, JMLR:v24:23-0646}, among many past and ongoing studies.

\subsubsection{Stopping rules for Bayesian cubature}
\label{subsubsection bayesian cubature}

Bayesian cubature, also known as Bayesian quadrature, is an implementable framework recasting numerical integration as a problem of statistical inference.
Instead of treating the integrand as an unknown but fixed function, Bayesian cubature places a prior (typically a Gaussian process) over the space of possible integrands.
Because integration is a linear functional, this prior can be pushed through the integral to produce a posterior distribution over the value of the integral.
The resulting estimator is therefore not just a point estimate, but a full probabilistic characterization of the uncertainty arising from approximating the integral using a finite set of function evaluations.
This perspective belongs to the broader field of probabilistic numerics, which seeks to quantify numerical error in a principled statistical manner.

Although Bayesian cubature may employ Monte Carlo, Markov chain Monte Carlo, or quasi-Monte Carlo point sets to select sample locations, it differs fundamentally from classical Monte Carlo integration.
Standard Monte Carlo methods rely on the central limit theorem (Section \ref{subsection normal approximation}), which guarantees asymptotic normality only in the limit of large sample sizes with their canonical convergence rate limited to $\mathcal{O}(n^{-1/2})$ regardless of smoothness.
In contrast, Bayesian cubature explicitly models the smoothness of the integrand through the Gaussian process kernel, allowing it to achieve substantially faster rates when the integrand is sufficiently regular.
Importantly, whereas Monte Carlo methods obtain error estimates only through asymptotic approximations, Bayesian cubature provides a finite-sample posterior variance that quantifies uncertainty even with only a handful of function evaluations.
This makes it particularly attractive in scientific computation where each function evaluation is computationally expensive.
For a comprehensive survey of this approach, we refer the reader to \cite{mahsereci2026bayesianquadraturegaussianprocesses}.

Despite their advantages and practical relevance, systematic stopping rules remain underdeveloped for Bayesian cubature, with only a few notable exceptions such as \cite{jagadeeswaran_hickernell_2019}.
In many scenarios, the sampling budget is constrained but not rigidly fixed, allowing practitioners to allocate a few additional evaluations when these are expected to substantially reduce uncertainty.
A principled stopping rule would help determine when posterior uncertainty is small enough relative to a target tolerance, or when further sampling is unlikely to yield meaningful improvement, and may additionally signal whether the Gaussian-process prior is adequately informed by the data or whether kernel choices and hyperparameters require refinement.
Accordingly, even with the variety of techniques reviewed in the present article, developing robust stopping criteria is an important and still largely open direction in Bayesian cubature, crucial not only for computational efficiency but also for trustworthy uncertainty quantification.

\section{Practical guidance and comparative evaluation}
\label{section practical guide and comparison}

We have so far reviewed a wide spectrum of sequential stopping rules, each carrying its own theoretical assumptions and guarantees.
Practitioners, however, are often faced with a more concrete question: how do these rules behave relative to one another under representative simulation settings?
\textcolor{black}{To address this question, this section provides a structured comparison of representative stopping rules together with practical guidance for their use. We present a selection checklist, a qualitative overview of method families, and illustrative numerical studies to highlight key trade-offs across different settings. These components are intended to offer heuristic and interpretive guidance rather than a complete algorithmic framework, an exhaustive benchmark, or definitive performance rankings.}



\subsection{Choosing a right stopping rule}
\label{subsection selection checklist}

Let us begin with a compact checklist for selecting stopping rules in the settings emphasized in this review.
The goal is to indicate which methods are commonly employed under which assumptions, and where reliable tools are currently limited.

\begin{tcolorbox}[title={Checklist for selection},enhanced,breakable=true]
\begin{itemize}[leftmargin=*, itemsep=0em]
  \item \textbf{Step 1. The target guarantee:}
Most stopping rules in this review target a fixed-width (absolute or relative) accuracy statement of the form $\mathbb{P}(|\mu_{\tau}-\mu|\le \epsilon)\ge 1-\delta$ (or its relative variants).
If the goal is not this fixed-width probabilistic guarantee, but instead a stopping rule based on the coefficient of variation, acceptable shifting convergence bands, or successive changes in the empirical mean, we direct the reader to Section \ref{section outside the conventional framework}.

\item \textbf{Step 2. Identify the sampling mechanism:}
\vspace{-0.5em}
\begin{itemize}[leftmargin=10pt,itemsep=0cm]
    \item \textbf{Standard iid sampling:}
    If the samples are iid with a finite variance, classical fixed-width methods based on normal approximation and variance estimation are natural starting points (Section \ref{subsection normal approximation}).
    When several scalar means must be estimated simultaneously, the extension to multiple means (Section \ref{subsection several means}) avoids full covariance estimation.
    
\item  \textbf{Non-iid sampling with structural dependence:}
When dependency is present among the samples, constructing a stopping rule requires an explicit mechanism to account for correlation.
For sequences of martingale difference type, the martingale central limit theorem is applied (Section \ref{subsection martingale difference type}).
For mixing sequences, stopping rules often rely on testing for approximate independence: strongly correlated observations are skipped so that fixed-width criteria can be applied to the remaining approximately independent samples (Section \ref{subsection quasi-independent sequences}).    
    \item \textbf{Quasi-Monte Carlo:}
    When the integrand admits sufficient smoothness and the point set is a low-discrepancy sequence, 
    a dedicated stopping rule based on cubature error estimation is available (Section \ref{subsection low discrepancy sequences for quasi-Monte Carlo methods}).
\end{itemize}

\item \textbf{Step 3. Check what moment/tail information you can justify:}
Without additional information beyond finite mean and variance, uniformly valid fixed-width guarantees are impossible in general (Section \ref{subsection preliminary and final sample sizes}).
\vspace{-0.5em}
\begin{itemize}[leftmargin=10pt,itemsep=0cm]
    \item \textbf{Finite higher-order moments / kurtosis control:} Apply two-stage variance estimation plus Berry-Esseen-type refinements (Sections \ref{subsubsection hickernell 2012}, \ref{subsubsection bayer}, and \ref{subsubsection kunsch 2019}).

    \item \textbf{Boundedness:} If, for instance, $X\in[0,1]$ or Bernoulli, then employ concentration-based or distribution-specific guarantees.
    For relative precision, a multi-phase algorithm provides distribution-free guarantees (Section \ref{subsubsection absolute error}) and boundedness comes with the martingale-difference structure (Section \ref{subsubsection relative error gajek}).
    For Bernoulli trials, Hoeffding-type rules yield simple deterministic run lengths (Section \ref{subsection Stopping rule for Bernoulli trials}), while inverse binomial sampling fixes the number of successes and achieves optimally guaranteed confidence for relative intervals (Section \ref{section confidence intervals for inverse binomial sampling}).
\end{itemize}

\item \textbf{Step 4. Decide on single-stage vs two-stage structure:}
\vspace{-0.5em}
\begin{itemize}[leftmargin=10pt,itemsep=0cm]
\item \textbf{Single-stage (continuous monitoring):}
Conceptually simple but vulnerable to early underestimation of variance and premature stopping.
The resulting loss of coverage in the non-asymptotic regime has been documented in detail, motivating various adjustments, such as larger preliminary sample sizes or conservative variance estimation in specific constructions
(Sections \ref{subsubsection hickernell 2012},
\ref{subsection martingale difference type}, and
\ref{subsection assessment and selection of sequential stopping rules}).
\item \textbf{Two-stage (pilot + final run):}
Often preferred when one can justify moment/tail conditions and wants non-asymptotic control; requires choosing a pilot size and possibly variance-inflation parameters (Sections \ref{subsubsection hickernell 2012}, \ref{subsubsection bayer}, and \ref{subsubsection kunsch 2019}).
\end{itemize}

\item \textbf{Step 5. When reliable fixed-width tools are currently limited:}
The literature reviewed here provides strong tools in several structured regimes, whereas reliable fixed-width guarantees remain limited when
\vspace{-0.5em}
\begin{itemize}[leftmargin=10pt,itemsep=0cm]
    \item only a finite-variance assumption is defensible, with no further information about variance, tail, or boundedness available (Section \ref{subsection preliminary and final sample sizes});
\item the practitioner seeks a fixed-width guarantee as in {\bf Step 1} under dependent sampling, while the dependence structure cannot be adequately characterized to support the martingale or mixing-type conditions as in {\bf Step 2}.
\end{itemize}
\end{itemize}
\end{tcolorbox}

\textcolor{black}{To translate the checklist into concrete starting recommendations, Table~\ref{table:decision} condenses, for each sampling regime emphasized in this review, a reasonable default rule, its main caveat or fallback, and a cross-reference to the relevant section. The final row indicates the situations in which we would advise against relying on existing fixed-width rules without further diagnostics.
We stress that these entries are defaults rather than prescriptions, and should be read together with the assumptions and coverage behavior detailed in Table \ref{method_comparison}.}


\small\color{black}
\begin{longtable}{p{4.9cm} p{4.3cm} p{5.1cm} p{1.9cm}}
\caption{Recommended default stopping-rule choices by sampling regime, under a fixed-width $(\epsilon,\delta)$ target. Entries are reasonable starting points rather than prescriptions and should be read together with Table \ref{method_comparison}.}
\label{table:decision}\\

\hline\hline
\rowcolor{gray!10}
\textbf{Setting (what can justified)} & \textbf{Default rule} & \textbf{Main caveat / fallback} & \textbf{Section} \\
\hline\hline
iid, light-tailed, finite variance & CLT / normal-approximation fixed-width rule (single- or two-stage) & May undercover in small samples when the variance estimate is unstable; enlarge the pilot $n_0$ or switch to a two-stage form & Sections \ref{subsection normal approximation} and \ref{subsection single- or multi-stage procedures} \\ \hline
iid, heavy-tailed but with finite higher moments (kurtosis bound) & Two-stage kurtosis-bounded rule or Berry-Esseen refinement & More conservative and costlier as skewness/kurtosis grows; requires a justified kurtosis bound or estimable third/fourth moments & Sections \ref{subsubsection hickernell 2012}, \ref{subsubsection bayer}, and \ref{subsubsection kunsch 2019} \\ \hline
Bounded support / Bernoulli ($X\in[0,1]$) & Hoeffding-type (absolute); multi-phase $(\epsilon,\delta)$-approximation or GBAS (relative); inverse binomial sampling (relative, rare events) & Distribution-free but conservative, especially for small success probability $p$ & Sections \ref{subsubsection absolute error}, \ref{subsection Stopping rule for Bernoulli trials}, and \ref{section confidence intervals for inverse binomial sampling} \\ \hline
Dependence, martingale-difference type ($\mathbb{E}[X_n|\mathcal{F}_{n-1}]=\mu$) & Batched, variance-estimated martingale-CLT rule & Conservative when dependence inflates the effective variance; performance depends on the batch schedule & Section \ref{subsection martingale difference type} \\ \hline
Dependence, mixing / autocorrelated & Independence-screening rule (skip strongly correlated observations) & Run length can grow sharply as dependence strengthens & Section \ref{subsection quasi-independent sequences} \\ \hline
Quasi-Monte Carlo (smooth integrand, low-discrepancy points) & Cubature-error-based stopping rule & Requires sufficient smoothness / Fourier decay and a low-discrepancy point set & Section \ref{subsection low discrepancy sequences for quasi-Monte Carlo methods} \\
\hline\hline
Only finite variance defensible (no tail/moment/boundedness information), or dependence not characterizable as martingale-difference or mixing type & \emph{No reliable fixed-width rule}; do not rely on existing rules without further diagnostics & Run diagnostics first (e.g., coverage profiles, variance/tail and dependence checks) and enlarge the pilot; treat any fixed-width output as heuristic & Sections \ref{subsection preliminary and final sample sizes} and \ref{subsection assessment and selection of sequential stopping rules} \\
\hline\hline
\end{longtable}
\normalsize

\textcolor{black}{Let us note that two further cases fall outside this table.
When several means are estimated simultaneously, the multiple-means rule (Section \ref{subsection several means}) controls componentwise relative errors without estimating the full covariance matrix.
When the target is not a fixed-width guarantee (for instance, a coefficient-of-variation threshold, an acceptable shifting convergence band, or a successive-change criterion), we refer the reader to Section \ref{section outside the conventional framework}.}

\color{black}
\subsection{Qualitative overview}
\label{subsection qualitative comparison table}

In this subsection, we discuss a structured qualitative overview of the major families of sequential stopping rules and closely related methodological and diagnostic tools along Table \ref{method_comparison}, complementing the selection checklist and the recommended default stopping-rule choices provided in Section \ref{subsection selection checklist}.
The purpose of this overview is not to rank methods, but rather to bring together in a single table the assumptions they rely on, the types of failure modes they address, and the forms of conservatism they typically exhibit.  
Accordingly, the table highlights three key dimensions across method families: (i) modeling assumptions and moment or dependence requirements;  
(ii) sensitivity to user-specified tuning parameters; and  
(iii) typical coverage behavior in the non-asymptotic regime.

To help interpret Table \ref{method_comparison}, we briefly summarize the structure and connections among method families.  
The rows are organized into six groups reflecting progressively different assumptions and objectives:  
(A) iid, CLT-based methods, including single-stage and higher-moment refinements;  
(B) methods exploiting bounded support or distributional structure (for instance, Bernoulli or bounded martingale-difference settings);  
(C) methods addressing dependence, including martingale-difference and mixing-based approaches, as well as deterministic low-discrepancy sequences;  
(D) methods for simultaneous estimation of multiple means;  
(E) diagnostic frameworks, such as coverage functions and VAMP1RE-type metrics, which assess interval quality but do not implement stopping rules; and  
(F)~criteria falling outside the conventional $(\epsilon,\delta)$ framework, such as coefficient-of-variation thresholds or acceptable shifting convergence bands.

\small
\begin{longtable}{p{3.0cm} p{4.0cm} p{3.4cm} p{3.4cm} p{2.0cm}}
\caption{Qualitative overview of representative stopping rules and related methodological and diagnostic tools.}
\label{method_comparison}\\

\hline\hline
\textbf{Method class} &
\textbf{Assumptions required} &
\textbf{Parameter sensitivity} &
\textbf{Coverage behavior} &
\textbf{Typical context} \\
\hline
\endfirsthead

\hline\hline
\textbf{Method class} &
\textbf{Assumptions required} &
\textbf{Parameter sensitivity} &
\textbf{Coverage behavior} &
\textbf{Typical context} \\
\hline
\endhead

\hline
\multicolumn{5}{r}{Continued on next page}\\
\endfoot

\hline
\endlastfoot

\hline\hline
\rowcolor{gray!10}
\multicolumn{5}{l}{\textbf{Group A: iid, CLT-based methods}} \\
\hline\hline
Variance-based fixed-width stopping rules (CLT / normal approximation)
 &
iid; finite variance; asymptotic normality effective at stopping time &
Sensitive to early variance underestimation and stopping-time variability &
Tends to undercover in small samples when variance estimates are unstable; no built-in conservatism &
Sections \ref{subsection normal approximation} and \ref{subsection single- or multi-stage procedures} \\ \hline 
Two-stage with kurtosis bound (variance inflation + B-E/Chebyshev) &
iid; finite variance; known/justified kurtosis bound or related moment control &
Pilot size $n_0$ and inflation factor $C$ materially affect validity/efficiency &
Often conservative (designed to protect against underestimation) &
Section \ref{subsubsection hickernell 2012} \\ \hline
Berry-Esseen refined stopping (two-stage) &
iid; finite variance; third/fourth moments estimable or controlled &
Moment estimation in pilot stage; tuning of B-E constants; doubling schedule &
Typically more conservative than CLT but more reliable in non-asymptotic regime &
Section \ref{subsubsection bayer} \\ \hline
Moment-cone / bounded moment ratio classes &
Distribution class restricted by known $(p,q)$-moment relationship &
Choice of class parameters $(p,q,K)$ governs feasibility and cost &
Conservative when constraints are loose (designed for worst-case control) &
Section \ref{subsubsection kunsch 2019} \\

\hline\hline
\rowcolor{gray!10}
\multicolumn{5}{l}{\textbf{Group B: bounded support / distributional structure}} \\
\hline\hline
iid relative precision (multi-phase) &
iid; $X\in[0,1]$; no distributional information beyond boundedness &
Internal thresholds ($\Upsilon$, $\Upsilon_1$, $\Upsilon_2$) set by $(\epsilon,\delta)$; two-phase variant adapts to estimated variance &
Conservative due to worst-case concentration bounds; constants independent of the unknown distribution &
Section \ref{subsubsection absolute error} \\ \hline
Bounded martingale-difference relative precision &
Adaptedness and $\mathbb{E}[X_n|\mathcal{F}_{n-1}]=\mu$; $X_n\in[0,1]$ a.s.; no CLT invoked &
Stopping threshold parameter (e.g., $r$) governs run length and error bounds &
Performance characterized via derived upper and lower bounds for error probability and run length &
Section \ref{subsubsection relative error gajek} \\ \hline
Bounded/Bernoulli (Hoeffding-type) absolute precision &
$X\in[0,1]$; Bernoulli$(p)$ as a primary case &
Stopping length determined directly by $(\epsilon,\delta)$ via concentration bounds &
Often conservative due to worst-case variance bounds, particularly for small $p$ &
Section \ref{subsection Stopping rule for Bernoulli trials} \\ \hline
Inverse binomial sampling &
Bernoulli with fixed number $N$ of successes; negative-binomial run length &
Choice of required successes $N$ governs precision; no tuning beyond $N$ and $(\epsilon,\delta)$ &
Achieves optimal guaranteed confidence for relative intervals; can be conservative for very small $p$ &
Section \ref{section confidence intervals for inverse binomial sampling} \\ 

\hline\hline\hline
\rowcolor{gray!10}
\multicolumn{5}{l}{\textbf{Group C: non-iid / dependence}} \\
\hline
Martingale difference (batched, variance-estimated) &
$\mathbb{E}[X_n|\mathcal{F}_{n-1}]=\mu$; martingale CLT conditions; no boundedness required; stable batch variance estimation &
Batch schedule $M(t)$ and inflation term $a(t)$; dependence impacts variance stability &
Can be conservative when dependence inflates effective variance; protects against early stopping &
Section \ref{subsection martingale difference type} \\ \hline
Mixing sequences with independence screening &
$\phi$-mixing (or related mixing) assumptions; testing procedures to identify approximate independence &
Choice of independence test and criteria for skipping correlated observations affect effective sample size &
Run length can increase substantially when strong dependence requires discarding many observations &
Section \ref{subsection quasi-independent sequences} \\ \hline
Low-discrepancy (quasi-Monte Carlo) stopping &
Deterministic low-discrepancy sequences; structural assumptions on integrand (e.g., Fourier decay) &
Stopping depends on discretization schedule and error estimator tied to frequency decay &
Performance depends on smoothness assumptions; may outperform iid Monte Carlo when structure is favorable &
Section \ref{subsection low discrepancy sequences for quasi-Monte Carlo methods} \\

\hline\hline
\rowcolor{gray!10}
\multicolumn{5}{l}{\textbf{Group D: multiple means}} \\
\hline\hline
Multiple-means relative precision &
iid sampling; separate control of componentwise relative errors &
Allocation of significance levels across components affects termination &
More conservative than single-mean estimation; avoids covariance estimation but increases run length &
Section \ref{subsection several means} \\

\hline\hline
\rowcolor{gray!10}
\multicolumn{5}{l}{\textbf{Group E: diagnostic frameworks}} \\
\hline\hline
Coverage-function / quality-assessment frameworks (diagnostic) &
Often assumes iid normality for analytic tractability; requires ability to compute stopping-time distribution &
No tuning for stopping; evaluates coverage distortion induced by stopping mechanisms via coverage functions,
VAMP1RE metrics,
and coverage profiles & 
Reveals structural coverage loss and informs conservative design adjustments; does not propose stopping rules itself &
Sections \ref{subsubsection coverage probabilities},
\ref{subsection coverage functions and related stopping rules}, and
\ref{subsection assessment and selection of sequential stopping rules} \\

\hline\hline
\rowcolor{gray!10}
\multicolumn{5}{l}{\textbf{Group F: non-conventional criteria}} \\
\hline\hline
Non-conventional stopping criteria (CoV, convergence bands, successive changes) &
Varies: CoV requires only finite variance; convergence bands require stationarity; successive-change rule assumes iid with finite variance &
CoV threshold is user-specified (such as $0.05$); band width and duration are tuning parameters; successive-change rule depends on Chebyshev-based tolerance &
CoV and convergence bands lack formal $(\epsilon,\delta)$ coverage guarantees; successive-change rule targets stability of the empirical mean rather than proximity to $\mu$, so may stop prematurely or late relative to conventional criteria &
Section \ref{section outside the conventional framework} \\
\hline
\end{longtable}

\normalsize
Many of the algorithmic developments reviewed in this article can be understood as direct responses to limitations of earlier approaches.  
Variance-based fixed-width rules (Group~A) depend heavily on asymptotic normality and can underperform in small samples;  
two-stage and Berry-Esseen refinements (Sections \ref{subsubsection hickernell 2012} and \ref{subsubsection bayer}) directly address this by incorporating higher-order moment information.  
Boundedness-based methods (Sections \ref{subsubsection absolute error}, \ref{subsubsection relative error gajek}, and~\ref{subsection Stopping rule for Bernoulli trials}, Group B) circumvent the impossibility phenomena of Section \ref{subsection preliminary and final sample sizes} by restricting the distribution class;  
inverse binomial sampling (Group B) represents a structurally distinct approach by fixing the number of successes rather than the total number of trials, yielding optimal relative-interval guarantees.  
Dependence-handling methods (Sections \ref{subsection martingale difference type} and \ref{subsection quasi-independent sequences}, Group C) extend fixed-width ideas beyond iid settings but require additional structural conditions.  
Finally, low-discrepancy methods (Section \ref{subsection low discrepancy sequences for quasi-Monte Carlo methods}, Group C) replace random sampling with deterministic sequences, shifting the error analysis from probabilistic concentration to Fourier-analytic decay.

Note that not all entries in the table correspond to operational stopping rules.  
In particular, the coverage-function and related quality-assessment frameworks (Group E) do not prescribe how or when to terminate simulation runs; rather, they quantify how random stopping distorts coverage, even under idealized assumptions.  
These diagnostic tools play a foundational role by isolating the intrinsic effects of random stopping (separate from variance-estimation error or tail behavior) and thereby motivate more conservative or structured designs.  
Similarly, the non-conventional criteria collected in Group F do not target fixed-width guarantees of the form $\mathbb{P}(|\mu_\tau - \mu| \le \epsilon) \ge 1-\delta$ and therefore lack formal coverage guarantees, though they remain widely used in practice.

\subsection{Empirical studies}\label{sec:numerical}

To build on the practice-oriented resources presented in Sections \ref{subsection selection checklist} and \ref{subsection qualitative comparison table}, we present a collection of illustrative numerical examples across four canonical scenarios: light-tailed iid sampling (Section \ref{Exponential sampling}), heavy-tailed iid sampling (Section \ref{Lognormal sampling}), probability estimation via Bernoulli trials (Section \ref{Bernoulli rare-event estimation}), and serially dependent time series of martingale difference and mixing types (Section \ref{ARCH(1) dependence}).
For each scenario, we examine empirical coverage, the behavior of termination times, and robustness to variance instability, with the primary aim of highlighting qualitative differences among the representative method families summarized in Table \ref{method_comparison}.
\textcolor{black}{While our primary focus is on fixed-width stopping rules, we also include selected non-fixed-width criteria (such as coefficient-of-variation and successive-change rules) to illustrate how commonly used alternatives behave when evaluated against this fixed-width benchmark.
All the following experiments are intended solely to illustrate typical trade-offs among the method families; they are not designed as a systematic benchmark or as definitive performance rankings, and the four scenarios are chosen to span representative sampling regimes rather than to constitute an exhaustive or standardized test suite.}

\textcolor{black}{Throughout this section, the \emph{nominal} level refers to the target confidence $1-\delta$ (here $95\%$), and for each rule we report two performance metrics: the empirical coverage and the mean $\mathbb{E}[\tau]$ of the stopping time $\tau$.
We do not report the interval length as a separate metric, because the rules considered here are of fixed-width type, for which the half-width is fixed at $\epsilon$ by construction.
The cost of any conservatism is therefore reflected entirely in the stopping time $\tau$, equivalently, in its mean $\mathbb{E}[\tau]$ which we report, rather than in the (fixed) interval width.
We stress that high coverage is not, by itself, a desirable goal.
Coverage close to $100\%$ typically reflects conservative over-sampling (a larger $\mathbb{E}[\tau]$) or, for procedures that do not fix the width, wider intervals, and, in the limit, an interval of unbounded length attains full coverage trivially.
Accordingly, we describe coverage as \emph{near-nominal} only when it is close to the $95\%$ target, as \emph{conservative} or \emph{over-covering} when it lies appreciably above it, and as \emph{under-covering} when it falls below; for the fixed-width rules considered here, coverage should always be read jointly with $\mathbb{E}[\tau]$.}

\subsubsection{Exponential sampling}
\label{Exponential sampling}

Consider iid sampling from the Exponential$(1)$ distribution ($\mu = 1$, $\sigma^2 = 1$, skewness $= 2$, and kurtosis $= 9$).
Although this setting is too simple, the light-tailed but moderately skewed distribution provides a natural first test case for illustrative purposes.
Since all moments are finite, every method in Group A in Table \ref{method_comparison} is, in principle, applicable.
At the same time, a skewness of $2$ is large enough to produce noticeable departures from normality in the distribution of the sample mean at small to moderate sample sizes, making this a useful stress test for the normal approximation in the non-asymptotic regime.
Here, we fix $\epsilon=0.1$, $\delta=0.05$ (nominal $95\%$ coverage) and the preliminary sample size $n_0=30$, and compare the following five rules:
\begin{itemize}[leftmargin=20pt,itemsep=0cm]
\setlength{\parskip}{0cm}
\item[(a)] the normal-approximation stopping rule
(Section \ref{subsection normal approximation});
\item[(b)] the two-stage procedure (Section \ref{subsubsection hickernell 2012}) with kurtosis bound $Q_{\max}=9$ and inflation factor $C=1.1$;
\item[(c)] the Berry-Esseen refined rule with doubling schedule
(Section \ref{subsubsection bayer});
\item[(d)] the coefficient-of-variation criterion with a threshold of $0.05$ (Section \ref{subsubsection the coefficient of variation});
\item[(e)] the successive-change stopping rule (Section \ref{subsubsection successive changes to the empirical mean}).
\end{itemize}

For each rule, we record in Table \ref{table:exp41} the empirical coverage probability $\mathbb{P}(|\mu_{\tau}-\mu|\le \epsilon)$, the mean and standard deviation of the stopping time $\tau$, the median stopping time ${\rm Med}(\tau)$, and the mean absolute error $\mathbb{E}[|\mu_{\tau}-\mu|]$, all estimated based on $10{,}000$ macro-replications.

\begin{table}[ht]
\centering
\caption{Five rules under Exponential$(1)$ sampling with $\epsilon=0.1$, $\delta=0.05$, and $n_0=30$ over $10{,}000$ macro-replications.}
\label{table:exp41}
\begin{tabular}{lccccc}
\hline
rule & $\mathbb{P}(|\mu_{\tau}-\mu|\le \epsilon)$ & $\mathbb{E}[\tau]$ & $\sqrt{{\rm Var}(\tau)}$ & ${\rm Med}(\tau)$ & $\mathbb{E}[|\mu_{\tau}-\mu|]$ \\
\hline
(a) CLT & $0.957$ & $431$ & $62$ & $430$ & $0.039$ \\
(b) Kurtosis two-stage & $0.999$ & $2{,}194$ & $797$ & $2{,}137$ & $0.018$ \\
(c) Berry-Esseen & $1.000$ & $1{,}919$ & $24$ & $1{,}920$ & $0.018$ \\
(d) CoV & $0.964$ & $446$ & $46$ & $430$ & $0.038$ \\
(e) Successive-change & $0.743$ & $129$ & $11$ & $130$ & $0.073$ \\
\hline
\end{tabular}
\end{table}

The CLT rule results in $\mathbb{E}[\tau]=431$ and an empirical coverage of $95.7\%$, achieving the nominal $95\%$ target, as expected in this light-tailed setting where the asymptotic normal approximation is deemed adequate.
The kurtosis two-stage and Berry-Esseen rules deliver coverage well above the nominal $95\%$, but at four to five times that of the CLT rule ($2{,}194$ and $1{,}919$ versus $431$).
This illustrates the cost of non-asymptotic conservatism: the kurtosis two-stage and Berry-Esseen rules exchange a substantial increase in sample size for finite-sample coverage guarantees, which is an expense that may be unnecessary when the CLT approximation is already adequate.

Next, the CoV criterion yields an empirical coverage of $96.4\%$ with $\mathbb{E}[\tau]=446$, which is fairly comparable to the CLT rule.
For the Exponential$(1)$ distribution, the CoV threshold effectively imposes a relative standard-error bound that nearly coincides with the CLT half-width condition when the mean equals the standard deviation.
However, its satisfactory performance in this case should not be extrapolated to other distributional settings, since the CoV criterion provides no formal $(\epsilon, \delta)$ guarantee.

The successive-change rule performs poorly relative to fixed-width criteria in this setting, achieving only $74.3\%$ coverage with $\mathbb{E}[\tau]$ of only $129$.
Because the incremental change decays at rate $\mathcal{O}(1/n)$ regardless of whether $\mu_n$ has approached $\mu$, the stopping rule can terminate far too early.
Since the Exponential$(1)$ distribution has unit variance ($\sigma^2= 1$), the theoretical standard error of $\mu_n$ at $n = 129$ is $\sigma/\sqrt{n} = 1/\sqrt{129} \approx 0.088$.
The half-width of the corresponding $95\%$ confidence interval is therefore $z_{0.025} \times 0.088 \approx 1.96 \times 0.088 \approx 0.173$, which still exceeds the target precision $\epsilon = 0.1$.

\begin{figure}[ht]
\centering
\includegraphics[width=0.5\textwidth]{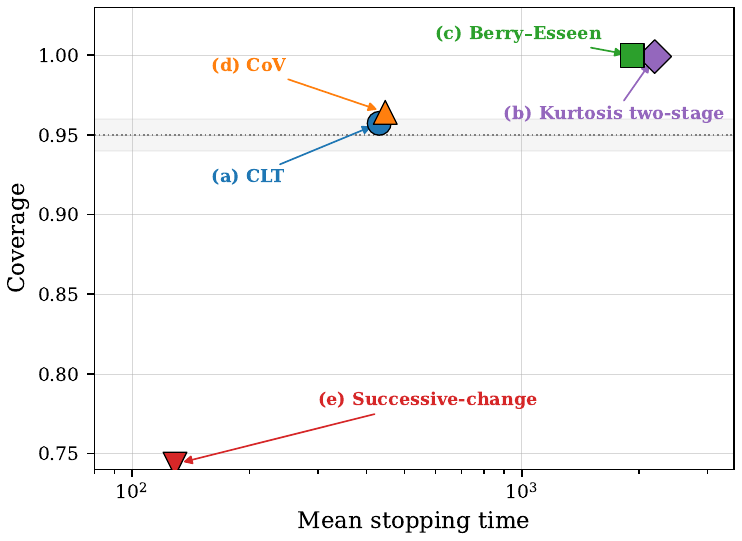}
\caption{Exponential$(1)$ sampling: coverage versus the mean of the stopping time $\tau$ for the five rules summarized in Table \ref{table:exp41}.
The gray band indicates the nominal $95\%$ coverage.
Rules positioned above and to the left of the band are generally preferable.}
\label{fig:exp41_scatter}
\end{figure}

The coverage-efficiency tradeoff is illustrated in Figure \ref{fig:exp41_scatter}, where each rule is represented by a single point.
The CLT and CoV rules lie close to the efficient frontier, achieving early stopping while maintaining adequate coverage.
In contrast, the kurtosis two-stage and Berry-Esseen rules occupy the conservative region, characterized by late stopping and \textcolor{black}{coverage close to $100\%$}.
The successive-change rule appears isolated in the lower-left corner, indicating that although it stops quickly, it fails to provide reliable coverage.

\subsubsection{Lognormal sampling}
\label{Lognormal sampling}

Next, consider iid sampling from the lognormal$(0,\sigma^2)$ distribution with $\sigma^2\in\{1,4,9\}$, corresponding to means given by $\mu=\exp[\sigma^2/2]\in\{1.65, 7.39, 90.02\}$ and kurtoses that grow super-exponentially: approximately $33$, $1.3\times 10^7$, and $4.3\times 10^{15}$, respectively. 
Since all moments are finite, every method in Group A (Table \ref{method_comparison}) can, in principle, once again be applied.
The main theme here is whether these stopping rules remain practically useful as the kurtosis increases.

To maintain comparable difficulty across the three values of $\sigma^2$, we set the absolute error tolerance proportional to the mean, $\epsilon=0.1\mu$, so that the stopping criterion $\mathbb{P}(|\mu_{\tau}-\mu|\le \epsilon)\ge 1-\delta$ amounts to a $10\%$ relative precision requirement.
With $\delta=0.05$ fixed throughout, we compare the following three rules:
\begin{itemize}[leftmargin=20pt,itemsep=0cm]
\setlength{\parskip}{0cm}
\item[(a)] the normal-approximation stopping rule 
(Section \ref{subsection normal approximation}), with 
preliminary sample sizes $n_0 \in \{20, 100, 500, 1000\}$;
\item[(b)] the two-stage procedure (Section \ref{subsubsection hickernell 2012}) with kurtosis bound $Q_{\max}$ set to the true kurtosis of the lognormal$(0,\sigma^2)$ distribution and inflation factor $C=1.1$;
\item[(c)] the Berry-Esseen refined rule with doubling schedule 
(Section \ref{subsubsection bayer}).
\end{itemize}
For rules (a), we report the empirical coverage probability $\mathbb{P}(|\mu_{\tau}-\mu|\le \epsilon)$ and the mean $\mathbb{E}[\tau]$ of the stopping time $\tau$ as a function of the preliminary sample size $n_0$ in Table \ref{tab:lognormal_CLT}.
In contrast, for rules (b) and (c), we report in Table \ref{table:exp42_be_kurt} the coverage probability $\mathbb{P}(|\mu_{\tau}-\mu|\le \epsilon)$, the mean and standard deviation of the stopping time $\tau$, the median stopping time ${\rm Med}(\tau)$, and the mean absolute error $\mathbb{E}[|\mu_{\tau}-\mu|]$, all estimated based on $10{,}000$ macro-replications.
Note that results are shown only for $\sigma^2 \in \{1,4\}$ because, for $\sigma^2 = 9$ (kurtosis $\approx 4.3 \times 10^{15}$), all three rules exceeded our computational budget of four hours per configuration, with even the CLT rule completing fewer than $12\%$ of replications.

\begin{table}[ht]
\centering
\caption{CLT rule under lognormal$(0, \sigma^2)$ sampling: coverage and the mean of the stopping time $\tau$ as functions of the preliminary sample size $n_0$, with $\epsilon = 0.1\mu$ and $\delta = 0.05$.}
\label{tab:lognormal_CLT}
\begin{tabular}{ccccc}
\hline
 & \multicolumn{2}{c}{$\sigma^2=1$} & \multicolumn{2}{c}{$\sigma^2=4$} \\
\cline{2-3} \cline{4-5}
$n_0$ & $\mathbb{P}(|\mu_{\tau}-\mu|\le \epsilon)$ & $\mathbb{E}[\tau]$ & $\mathbb{P}(|\mu_{\tau}-\mu|\le \epsilon)$ & $\mathbb{E}[\tau]$ \\
\hline
$20$   & $0.933$ & $667$      & $0.828$ & $11{,}847$ \\
$100$  & $0.935$ & $667$      & $0.844$ & $12{,}079$ \\
$500$  & $0.956$ & $681$      & $0.846$ & $12{,}097$ \\
$1000$ & $0.987$ & $1{,}011$  & $0.846$ & $12{,}104$ \\
\hline
\end{tabular}
\end{table}

\begin{table}[ht]
\centering
\caption{Kurtosis two-stage and Berry-Esseen rules under lognormal$(0,\sigma^2)$ sampling with $\delta=0.05$.
Results for the Berry-Esseen rule are reported at the preliminary sample size $n_0 = 100$, while those for the kurtosis rule are reported at $n_0 = 30$.}
\label{table:exp42_be_kurt}
\begin{tabular}{llcccc}
\hline
rule & $\sigma^2$ & $\mathbb{P}(|\mu_{\tau}-\mu|\le \epsilon)$ & $\mathbb{E}[\tau]$ & $\sqrt{{\rm Var}(\tau)}$ & $\mathbb{E}[|\mu_{\tau}-\mu|]$ \\
\hline
(b) Kurtosis two-stage & $1$ & $0.992$ & $4{,}044$ & $5{,}486$ & $0.037$ \\
(b) Kurtosis two-stage & $4$ & $0.743$ & $70{,}893$ & $477{,}511$ & $0.612$ \\
(c) Berry-Esseen & $1$ & $1.000$ & $4{,}129$ & $1{,}588$ & $0.028$ \\
(c) Berry-Esseen & $4$ & $1.000$ & $168{,}911$ & $141{,}593$ & $0.116$ \\ \hline
\end{tabular}
\end{table}

The anticipated effects are evident in the concrete magnitudes reported in Tables \ref{tab:lognormal_CLT} and \ref{table:exp42_be_kurt}.
Under $\sigma^2 = 1$, the CLT-based rule undercovers (for instance, $93.3\%$ when $n_0 = 20$) but improves to $98.7\%$ once the preliminary sample size is increased to $n_0 = 1{,}000$, indicating that a sufficiently large pilot can restore the accuracy of the normal approximation under moderate skewness.
In contrast, when $\sigma^2 = 4$, coverage saturates at roughly $84\%$ regardless of the choice of the preliminary sample size $n_0$.
Even a pilot as large as $n_0 = 1{,}000$ yields no meaningful improvement, as the extreme skewness of the $\mathrm{lognormal}(0,4)$ distribution makes the CLT approximation fundamentally unreliable.
Second, the Berry-Esseen rule maintains \textcolor{black}{coverage close to $100\%$} at both values of $\sigma^2$, confirming its non-asymptotic robustness, but at a steep cost: $\mathbb{E}[\tau]=4{,}129$ at $\sigma^2=1$ and $\mathbb{E}[\tau]=168{,}911$ at $\sigma^2=4$, reflecting the growth of the normalized third moment as $\sigma^2$ increases.
Third, the kurtosis two-stage rule attains $99.2\%$ coverage at $\sigma^2 = 1$ with $\mathbb{E}[\tau]=4{,}044$, comparable to the Berry-Esseen rule, but collapses to $74.3\%$ coverage at $\sigma^2 = 4$, accompanied by enormous variability in the stopping time (mean $70{,}893$, standard deviation $477{,}511$).
This failure arises because the kurtosis of the $\mathrm{lognormal}(0,4)$ distribution is on the order of $10^7$, making the variance estimate at the preliminary stage far too noisy to yield a reliable stopping rule.

Taken together, these results reveal a widening gap between theoretical validity and practical feasibility as the kurtosis increases.
Although all methods in Group A are, in principle, guaranteed to work for the lognormal distribution since all moments are finite, the sample sizes needed to ensure these guarantees can become prohibitively large.
The case of $\sigma^2 = 9$, where no rule completed within four hours, illustrates this limitation most starkly.
Developing stopping rules that remain both valid and practically implementable in such heavy-tailed settings remains an open challenge.


\subsubsection{Probability estimation}
\label{Bernoulli rare-event estimation}

Next, consider iid Bernoulli$(p)$ trials with success probabilities $p\in\{0.5, 0.01, 0.001\}$, representing common, moderately rare, and rare events.
This setting provides a natural testbed for comparing rules that exploit bounded support against those that do not.
We examine two precision criteria:
\begin{itemize}[leftmargin=20pt,itemsep=0cm]
\setlength{\parskip}{0cm}
\item \textbf{Absolute precision} with $\epsilon=0.01$ and $\delta=0.05$: 
we compare the classical normal approximation rule 
(Section \ref{subsection normal approximation}) against the 
Hoeffding-based rule with deterministic stopping at 
$\lceil\ln(2/\delta)/(2\epsilon^2)\rceil$ 
(Section \ref{subsection Stopping rule for Bernoulli trials}).
\item \textbf{Relative precision} with $\epsilon=0.1$ and $\delta=0.05$:
we compare the $(\epsilon,\delta)$-approximation algorithm $\mathcal{AA}$ (Section \ref{subsubsection absolute error}), the GBAS method
(Section \ref{subsection Stopping rule for Bernoulli trials}), and inverse binomial sampling with required successes of $N\in\{50,100\}$ (Section \ref{section confidence intervals for inverse
binomial sampling}).
\end{itemize}

Here, we aim to demonstrate that concentration-based rules guarantee nominal coverage uniformly in $p$, but at a cost that becomes increasingly conservative as $p$ decreases.
In contrast, CLT-based rules adapt naturally to the value of $p$ but risk undercoverage when $p$ is small and the normal approximation breaks down.
In addition, inverse binomial sampling provides a natural form of adaptation to unknown $p$ in the relative-precision setting by fixing the number of observed successes rather than the sample size.
For each rule and success probability $p$, we report in Tables \ref{table:exp43_abs} and \ref{table:exp43_rel} below the coverage probability and the mean of the stopping time $\tau$ for the absolute-precision and relative-precision settings, respectively, each based on $10{,}000$ macro-replications.

\begin{table}[ht]
\centering
\caption{Absolute-precision setting for probability estimation with $\epsilon=0.01$ and $\delta=0.05$.}
\label{table:exp43_abs}
\begin{tabular}{l cc cc cc}
\hline
& \multicolumn{2}{c}{$p=0.5$} & \multicolumn{2}{c}{$p=0.01$} & \multicolumn{2}{c}{$p=0.001$} \\
rule & $\mathbb{P}(|\mu_{\tau}-\mu|\le \epsilon)$ & $\mathbb{E}[\tau]$ & $\mathbb{P}(|\mu_{\tau}-\mu|\le \epsilon)$ & $\mathbb{E}[\tau]$ & $\mathbb{P}(|\mu_{\tau}-\mu|\le \epsilon)$ & $\mathbb{E}[\tau]$ \\
\hline
CLT & $0.948$ & $9{,}630$ & $0.998$ & $147$ & $1.000$ & $37$ \\
Hoeffding & $0.994$ & $18{,}445$ & $1.000$ & $18{,}445$ & $1.000$ & $18{,}445$ \\
\hline
\end{tabular}
\end{table}

\begin{table}[ht]
\centering
\caption{Relative-precision setting for probability estimation with $\epsilon=0.1$ and $\delta=0.05$.}
\label{table:exp43_rel}
\begin{tabular}{l cc cc cc}
\hline
& \multicolumn{2}{c}{$p=0.5$} & \multicolumn{2}{c}{$p=0.01$} & \multicolumn{2}{c}{$p=0.001$} \\
rule & $\mathbb{P}(|\mu_{\tau}-\mu|\le \epsilon)$ & $\mathbb{E}[\tau]$ & $\mathbb{P}(|\mu_{\tau}-\mu|\le \epsilon)$ & $\mathbb{E}[\tau]$ & $\mathbb{P}(|\mu_{\tau}-\mu|\le \epsilon)$ & $\mathbb{E}[\tau]$ \\
\hline
$(\epsilon,\delta)$-approximation $\mathcal{AA}$ & $1.000$ & $44{,}680$ & $1.000$ & $660{,}166$ & $1.000$ & $6{,}354{,}191$ \\
GBAS & $0.994$ & $778$ & $0.949$ & $38{,}900$ & $0.683$ & $9{,}997$ \\
Inv.\ binom.\ ($N\!=\!50$) & $0.654$ & $100$ & $0.522$ & $5{,}001$ & $0.514$ & $50{,}014$ \\
Inv.\ binom.\ ($N\!=\!100$) & $0.831$ & $200$ & $0.683$ & $9{,}997$ & $0.677$ & $99{,}987$ \\
\hline
\end{tabular}
\end{table}

The ratio of the mean of the stopping time $\tau$ to the CLT-based benchmark ($z_{\delta/2}^2 p(1-p)/\epsilon^2$ in the absolute-precision case, and $z_{\delta/2}^2(1-p)/(\epsilon^2 p)$ in the relative-precision case) measures the excess conservatism relative to asymptotic efficiency.
That is, ratios near $1$ indicate that the rule operates close to the minimum sample size predicted by the CLT rule, whereas larger ratios reflect the additional cost required to achieve finite-sample guarantees.
In the absolute-precision setting, the CLT rule at $p=0.5$ achieves $94.8\%$ coverage with $\mathbb{E}[\tau]=9{,}630$, closely matching the CLT-based benchmark $z_{\delta/2}^2 p(1-p)/\epsilon^2 \approx 9{,}604$ (ratio $\approx 1.00$).
The Hoeffding rule, which is based on the worst-case bound $p(1-p)\le 1/4$, requires a deterministic sample size of $n=\lceil\ln(2/\delta)/(2\epsilon^2)\rceil=18{,}445$, regardless of the unknown probability $p$.
This is roughly twice the cost of the CLT rule at $p=0.5$ (ratio $\approx 1.92$, relative to the CLT-based benchmark), but with stronger coverage of $99.4\%$.
At smaller $p$, however, the CLT rule exhibits anomalous early stopping driven by a zero-variance artifact, even though the CLT benchmark $z_{\delta/2}^2 p(1-p)/\epsilon^2$ itself remains moderate (approximately $380$ at $p=0.01$ and $38$ at $p=0.001$, both above the preliminary sample size $n_0=30$).
For instance, when $p=0.01$, the probability of observing no success in the first $n_0=30$ Bernoulli trials is $0.99^{30}\approx 0.74$.
In this event, the empirical variance is zero ($s_{n_0}^2=0$), which in turn yields the zero CLT half-width $z_{\delta/2}\,s_{n_0}/\sqrt{n_0}=0$.
The stopping criterion is therefore immediately triggered at $n_0$, producing a zero-width confidence interval centered at $\mu_{n_0}=0$.
This zero-variance event occurs even more likely at $p=0.001$, where $0.999^{30}\approx 0.97$.
The resulting high coverage probabilities ($99.8\%$ at $p=0.01$ and $100\%$ at $p=0.001$) are thus purely artifacts of this degeneracy: since the tolerance $\epsilon=0.01$ is equal to or exceeds the true mean $p$ itself, even the trivial estimate $\mu_n=0$ satisfies the condition $|\mu_n - p|\le \epsilon$.

In the relative precision setting, the GBAS method \textcolor{black}{over-covers at $p=0.5$ ($99.4\%$, well above the $95\%$ nominal level) and achieves near-nominal coverage at $p=0.01$ ($94.9\%$)}, while its coverage deteriorates sharply to $68.3\%$ at $p=0.001$.
The inverse binomial sampling with a target of $N=50$ successes exhibits severe undercoverage across all three values of the success probability $p$ ($51.4\%$-$65.4\%$).
Even doubling the requirement to $N=100$ raises coverage only to $67.7\%$-$83.1\%$.
In both cases, the mean of the stopping time $\tau$ scales as $N/p$, consistent with the geometric waiting time for each success.
In constrast, the $(\epsilon,\delta)$-approximation algorithm $\mathcal{AA}$ guarantees $100\%$ coverage uniformly in the success probability $p$, but at the price of rapidly increasing sample sizes: $\mathbb{E}[\tau]\approx 4.5\times 10^4$ at $p=0.5$, $\approx 6.6\times 10^5$ at $p=0.01$, and $\approx 6.4\times 10^6$ at $p=0.001$.
This sharp growth highlights the inherent tension between distribution-free coverage guarantees and sample efficiency, a trade-off that becomes especially pronounced in the rare-event regime.

\subsubsection{Serially dependent time series}
\label{ARCH(1) dependence}

The preceding settings all assume iid sampling.
In practice, however, Monte Carlo estimators frequently arise from serially dependent sequences, for instance, in MCMC output, adaptive importance sampling, or simulations of queueing and financial systems.
When an iid stopping rule is applied to such dependent data without accounting for the underlying dependence structure, the variance of the sample mean is typically underestimated, leading to premature termination and inadequate coverage.
In this section, we aim to demonstrate that those issues can be addressed with stopping rules reviewed in Sections \ref{subsection martingale difference type} and \ref{subsection quasi-independent sequences}, if appropriately employed.

We examine a stationary ARCH(1) process defined by 
$X_t = \sigma_t Z_t$ with $X_0=0$, $\sigma_t^2 = 0.1 + \alpha\,X_{t-1}^2$, and $Z_t\stackrel{\text{iid}}{\sim}\mathcal{N}(0,1)$. 
We fix $\alpha=0.8$, so that the stationarity condition $\alpha<1$ is met.
Here, suppose one is concerned with estimating the unconditional stationary mean $\mathbb{E}[X_t]=0(=\mu)$.
Although the process is uncorrelated (that is, $\text{Cov}(X_t,X_s)=0$ for all $t\ne s$), its squared observations exhibit strong time dependence (that is,  $\operatorname{Cov}(X_t^2, X_s^2) \neq 0$ for $t \neq s$), reflecting the fact that the conditional variance depends on past values.
This dependence in the second-order structure generates the characteristic volatility clustering of ARCH processes.
The process also satisfies $\mathbb{E}[X_t\,|\,\mathcal{F}_{t-1}]=0(=\mu=\mathbb{E}[X_t])$, placing it squarely within the framework of martingale difference type, reviewed in Section \ref{subsection martingale difference type}.

We additionally consider an AR(1) process defined by $Y_t = \rho\,Y_{t-1} + \eta_t$ with $Y_0=0$, $\eta_t\stackrel{\text{iid}}{\sim}\mathcal{N}(0,1)$ and $\rho\in\{0.3, 0.7, 0.95\}$, estimating the unconditional stationary mean $\mathbb{E}[Y_t]=0(=\mu)$.
Unlike the ARCH(1) model, which forms a martingale difference sequence with zero autocorrelation, the AR(1) process has non-zero autocorrelation $\text{Corr}(Y_t, Y_{t+k})=\rho^k(\ne 0)$ and satisfies $\phi$-mixing conditions, making it the natural setting for the independence-screening framework reviewed in Section \ref{subsection quasi-independent sequences}.

Throughout, we fix $\epsilon=0.1$ and $\delta=0.05$, and compare three stopping rules over $5{,}000$ macro-replications:
\begin{itemize}[leftmargin=20pt,itemsep=0cm]
\setlength{\parskip}{0cm}
\item[(a)] the normal-approximation stopping rule (Section \ref{subsection normal approximation}), applied as though the sequence were iid based on the standard sample variance $s_n^2$;
\item[(b)] the martingale-difference stopping rule (Section \ref{subsection martingale difference type}) with linearly growing batch sizes $|M(t)|=100\,t$ and inflation 
function $a(t)=t^{-1/2}$;
\item[(c)] the independence-screening stopping rule 
(Section \ref{subsection quasi-independent sequences}), which applies a test identifying approximately independent subsequences and constructs a fixed-width interval from the retained observations.
\end{itemize}
For illustrative purposes, we apply all three rules to both models, regardless of whether the standing assumptions underlying each stopping rule are satisfied.
Below in Tables \ref{table:exp44_arch} and \ref{table:exp44_ar}, we report numerical results for ARCH(1) and AR(1) models, respectively.

\begin{table}[ht]
\centering
\caption{ARCH(1) model with $\epsilon=0.1$ and $\delta=0.05$.}
\label{table:exp44_arch}
\begin{tabular}{lccccc}
\hline
rule & $\mathbb{P}(|\mu_{\tau}-\mu|\le \epsilon)$ & $\mathbb{E}[\tau]$ & $\sqrt{{\rm Var}(\tau)}$ & ${\rm Med}(\tau)$ & $\mathbb{E}[|\mu_{\tau}-\mu|]$ \\
\hline
(a) CLT-as-iid & $0.970$ & $195$ & $111$ & $130$ & $0.035$ \\
(b) martingale difference & $1.000$ & $1{,}567$ & $373$ & $1{,}500$ & $0.014$ \\
(c) independence screening & $1.000$ & $4{,}212$ & $1{,}013$ & $4{,}000$ & $0.009$ \\
\hline
\end{tabular}
\end{table}

\begin{table}[ht]
\centering
\caption{AR(1) model with $\epsilon=0.1$ and $\delta=0.05$.}
\label{table:exp44_ar}
\begin{tabular}{l cc cc cc}
\hline
& \multicolumn{2}{c}{$\rho=0.3$} & \multicolumn{2}{c}{$\rho=0.7$} & \multicolumn{2}{c}{$\rho=0.95$} \\
rule & $\mathbb{P}(|\mu_{\tau}-\mu|\le \epsilon)$ & $\mathbb{E}[\tau]$ & $\mathbb{P}(|\mu_{\tau}-\mu|\le \epsilon)$ & $\mathbb{E}[\tau]$ & $\mathbb{P}(|\mu_{\tau}-\mu|\le \epsilon)$ & $\mathbb{E}[\tau]$ \\
\hline
(a) CLT-as-iid & $0.865$ & $473$ & $0.596$ & $802$ & $0.248$ & $3{,}972$ \\
(b) martingale difference & $1.000$ & $3{,}665$ & $0.990$ & $7{,}238$ & $0.842$ & $83{,}663$ \\
(c) independence screening & $1.000$ & $16{,}837$ & $1.000$ & $55{,}142$ & $0.997$ & $415{,}898$ \\
\hline
\end{tabular}
\end{table}

For the setting of ARCH(1), the CLT-as-iid rule achieves $97.0\%$ coverage, slightly above the nominal level.
This is not surprising.
Since the process is uncorrelated despite the volatility clustering, the sample variance is a consistent (though not fully efficient) variance estimator of the true variance, making the CLT rule reasonably valid.
The martingale difference and independence screening rules both achieve $100\%$ coverage, but at costs that are approximately $8\times$ and $22\times$ higher, respectively.

\begin{figure}[ht]
\centering
\includegraphics[width=0.8\textwidth]{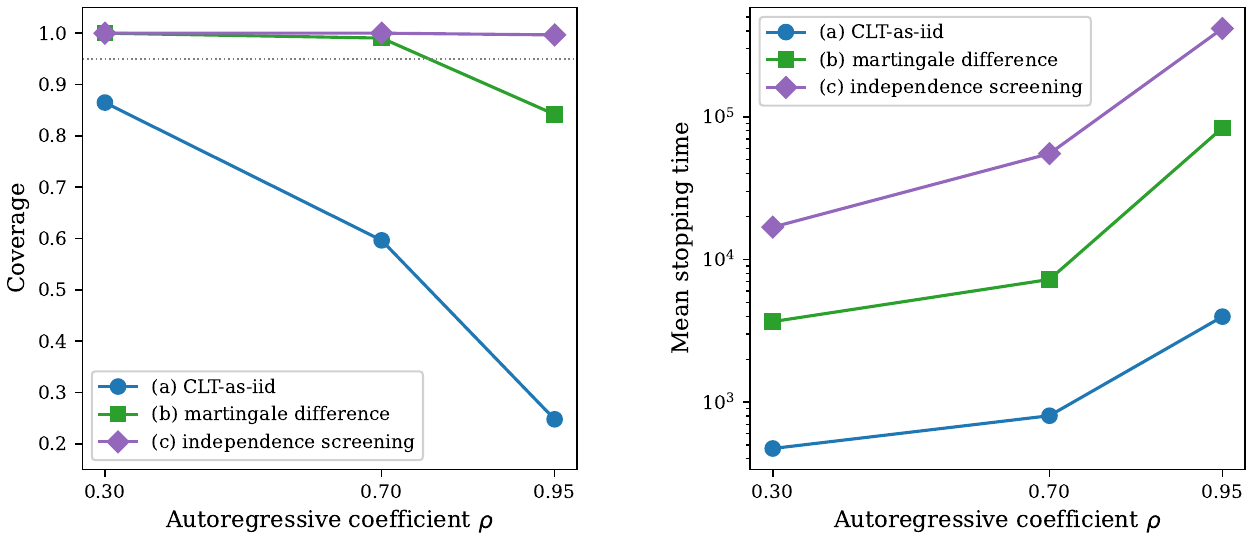}
\caption{AR(1) model: coverage and the mean of the stopping time $\tau$ as functions of the autocorrelation coefficient~$\rho$. The nominal $95\%$ coverage level is marked by the dotted line.}
\label{fig:exp44_ar}
\end{figure}

The results for the AR(1) model tell a strikingly different story. 
The coverage of CLT-as-iid degrades monotonically as the autoregressive coefficient $\rho$ increases: $86.5\%$ at $\rho=0.3$, $59.6\%$ at $\rho=0.7$, and only $24.8\%$ at $\rho=0.95$.
The cause is well understood: the iid variance estimator $s_n^2$ estimates the marginal variance $\text{Var}(Y_t) = \sigma_\eta^2/(1-\rho^2)$, 
obtained by solving $\text{Var}(Y_t)=\rho^2 \text{Var}(Y_{t-1})+\sigma_\eta^2$ under stationarity, rather than the long-run variance $\sigma_{\text{LR}}^2=\sum_{t=-\infty}^{\infty}\text{Cov}(Y_0,Y_t) = \sigma_\eta^2/(1-\rho)^2$, with the ratio $\sigma_{\text{LR}}^2/\text{Var}(Y_1)=(1+\rho)/(1-\rho)$, which grows without bound as $\rho$ approaches 1, leading to substantial undercoverage.
The martingale difference rule is applied here outside its theoretical scope, since the AR(1) process violates its standing assumption: $\mathbb{E}[Y_t|\,\mathcal{F}_{t-1}] = \rho\,Y_{t-1} \ne \mu(=0)$ unless $\rho=0$.
Nonetheless, the stopping rule performs remarkably well.
It maintains $100\%$ coverage at $\rho=0.3$ and $99.0\%$ at $\rho=0.7$, and achieves $84.2\%$ coverage even at the extreme autocorrelation $\rho=0.95$, which is a threefold improvement over $24.8\%$ of the CLT rule.
This robustness can be attributed to the growing batch sizes (as $|M(t)|=100\,t$), which allow within-batch statistics to average over progressively longer stretches of the dependent sequence, partially capturing the long-run variance even though the within-batch estimator is not designed for correlated data.
The independence screening rule, which is specifically designed for correlated sequences via an independence-screening runs test, achieves \textcolor{black}{coverage at or near $100\%$} at all three autocorrelation levels ($100\%$, $100\%$, and $99.7\%$), but at a cost that scales sharply with $\rho$: from $\mathbb{E}[\tau]= 16{,}837$ at $\rho=0.3$ to $\mathbb{E}[\tau]= 415{,}898$ at $\rho=0.95$.
For illustrative purposes, we plot in Figure \ref{fig:exp44_ar} the coverage-cost tradeoff as the autoregressive coefficient varies in the AR(1) model.

\subsection{Key insights}
\label{subsection key insights}

A central and recurring theme across the literature, and one that is confirmed by our empirical studies in Section \ref{sec:numerical}, is that many stopping rules can substantially undercover when 
asymptotic approximations are not reliable (Sections \ref{subsection asymptotic validity} and \ref{subsection assessment and selection of sequential stopping rules}). 
Such coverage loss has been demonstrated even analytically under the idealized iid normal model (Sections \ref{subsubsection coverage probabilities} and \ref{subsection assessment and selection of sequential stopping rules}), indicating that the distortion arises from the random stopping mechanism itself rather than from distributional misspecification.

Drawing on the coverage-analysis literature and the numerical experiments in Section \ref{sec:numerical}, several practices help mitigate this risk:
(i) use a sufficiently large preliminary sample size $n_0$, as coverage profiles show that increasing $n_0$ can substantially reduce premature stopping (Section \ref{subsection assessment and selection of sequential stopping rules}; see also Table \ref{tab:lognormal_CLT} for an empirical illustration, where pilot sizes on the order of several hundreds improve coverage for moderately skewed distributions, with the adequate size depending on tail behavior);
(ii) apply a variance inflation factor $C>1$ to guard against early underestimation of variability (Section \ref{subsubsection hickernell 2012});
(iii) when higher-order moments are available or can be estimated, prefer Berry-Esseen or kurtosis-bounded refinements over plain CLT-based rules, as they provide more reliable non-asymptotic coverage at the cost of larger sample sizes (Sections \ref{subsubsection hickernell 2012} and \ref{subsubsection bayer});
(iv) employ coverage functions and coverage profiles as diagnostics before committing to production runs (Sections \ref{subsection coverage functions and related stopping rules} and \ref{subsection assessment and selection of sequential stopping rules}); and
(v) when the normal approximation is practically inadequate, even with a finite variance (as in the heavy-tailed settings of Section \ref{Lognormal sampling}, where increasing the pilot size fails to restore coverage), switch to a different method family (for instance, Groups B or C in Table \ref{method_comparison}) rather than attempting to patch the CLT-based rule.

The four representative problem settings examined in Sections \ref{Exponential sampling}, \ref{Lognormal sampling}, \ref{Bernoulli rare-event estimation}, and \ref{ARCH(1) dependence} span five of the six method families summarized in Table \ref{method_comparison}.  
In favorable situations (iid sampling with light tails), almost any well-designed stopping rule seems to attain acceptable empirical coverage.  
In contrast, when these are violated, whether through heavy-tailed distributions, bounded support, or serial dependence, the choice of stopping rule becomes crucial, and a mismatched rule can lead to pronounced undercoverage.  
Let us now summarize the key insights that emerge from this comparison and translate them into concrete guidance.

\paragraph{Key insight 1. CLT-based rules are efficient but fragile:}
When the underlying assumptions hold (iid, finite variance, and a sufficiently large sample size), the CLT stopping rule (Group A) is the most economical option, achieving high coverage with small samples.
Its performance, however, degrades in two distinct ways.
Under heavy tails (such as lognormal with $\sigma^2=4$), coverage plateaus below the nominal regardless of the pilot size, because the normal approximation itself breaks down.
Under the serial dependence of the AR(1) model, coverage drops as the autoregressive coefficient increases, 
because the iid variance estimator fails to capture the long-run variance.
The kurtosis- and Berry-Esseen-based refinements (also Group A) recover valid coverage for heavy-tailed iid data but require roughly $2$-$5\times$ simulation lengths.

\paragraph{Key insight 2. Distribution-free guarantees come at a steep price:}
If the support of the underlying distribution is known to be bounded, then the $(\epsilon,\delta)$-approximation algorithm $\mathcal{AA}$, Hoeffding, and GBAS (Group B) may provide coverage guarantees without further distributional assumptions.
The case of probability estimation (Section \ref{Bernoulli rare-event estimation}) illustrates the associated cost: for absolute precision at $p=0.5$, the Hoeffding rule requires roughly twice the sample size of the CLT-based rule, albeit delivering stronger coverage.
For relative precision, the $(\epsilon,\delta)$-approximation algorithm $\mathcal{AA}$ achieves $100\%$ coverage uniformly in $p$, but at sample sizes that grow prohibitively in practice.
The GBAS method provides a more practical alternative for moderate $p$, 
though its coverage degrades sharply in the rare-event regime.

\paragraph{Key insight 3. Dependence structure determines the appropriate method:}
The experiment on ARCH and AR sequences (Section \ref{ARCH(1) dependence}) is particularly informative for practitioners working with serially dependent data.
For sequences of martingale difference type (Group C), where the conditional mean is unbiased but the variance exhibits nontrivial dynamics, the batched rule delivers reliable coverage.
For correlated processes such as AR(1), where the martingale-difference condition fails, the independence-screening rule (also in Group C) attains robust coverage, albeit at the cost of substantially larger sample sizes.
Notably, the martingale difference rule still outperforms the na\"ive CLT-as-iid approach even outside its theoretical validity, illustrating a meaningful degree of practical robustness.

\paragraph{Key insight 4. Non-conventional criteria require caution:}
The results for the CoV and successive-change rules (Group F), tested in the experiment on the Exponential(1) distribution (Section \ref{Exponential sampling}), imply that formal reasoning grounded in $(\epsilon,\delta)$ guarantees is essential.
The CoV rule attains reasonable coverage for the specific parameter settings tested, but is unlikely to do so uniformly, especially when the mean is near zero, where it becomes extremely sensitive to fluctuations in the sample mean.
The successive-change rule, which tracks the stability of the running mean rather than its proximity to the true mean, exhibits coverage behavior that depends strongly on the chosen tolerance.

\vspace{1em}
We note that the substance of the points above is familiar rather than novel, each of which can be traced, in one form or another, to earlier works on sequential analysis.
However, they tend to appear in a scattered manner across the literature, often embedded within specific methodological developments rather than stated as explicit and consolidated guidance.
One contribution of the present comparison is therefore to bring these insights together in a unified empirical setting, making them directly actionable for practitioners.
For broader treatments of the underlying theory and comprehensive assessments of stopping rules, we refer the reader to, for instance, \cite{Mukhopadhyaybook, D1985, D2014}.

\color{black}
\section{Concluding remarks}
\label{section concluding remarks} 

Sequential stopping rules have been, and will highly likely continue to be, a focal point of interest in Monte Carlo methods due to their fundamental importance.
As highlighted throughout the present review, these rules, along with their variants and integrations with other techniques, play a central role in reducing computational costs while achieving the desired accuracy.
Looking ahead, further developments are certainly expected, whether through innovative combinations of existing methods or entirely novel breakthroughs, with potential applications in expanding fields.

As such, the ultimate aim of the present review is to promote further research on sequential stopping rules for more general and modern Monte Carlo methods by providing \textcolor{black}{an up-to-date methodological guidebook, with a deliberate focus on standard and lightly generalized Monte Carlo methods for estimating an unknown mean.
In this spirit, we provide a synthesis of existing results, along with comparative insights and illustrative examples.
While this article does not introduce new methodological developments, it clarifies the relationships among existing approaches and supports their informed use in practice.}

To achieve this goal, we have started the present article with an overview of the essential concepts and main components (Section \ref{section fundamental concepts}), specifically, error tolerances (Section \ref{subsection error tolerance}), confidence and significance levels (Section \ref{subsection confidence level}), multivariate outputs or multiple means (Section \ref{subsection multivariate outputs}), non-iid random elements (Section \ref{subsection non-iid random elements}), preliminary and final sample sizes (Section \ref{subsection preliminary and final sample sizes}), large sample theory (Section \ref{subsection asymptotic validity}), normal approximation (Section \ref{subsection normal approximation}) and procedures for mean estimation (Section \ref{subsection single- or multi-stage procedures}).
The overview on these fundamentals is then followed by a review of the recent developments (Section \ref{section recent developments}) on sequential stopping rules for standard and lightly generalized Monte Carlo methods.
Given the somewhat sporadic attention this area has received in the literature, we have tried to collectively include discussions and comparisons of the main assumptions and key convergence properties, along with summaries of the primary advantages and disadvantages.

{To link theory with practice, we have then provided a practice-oriented guide for selecting stopping rules across different contexts (Section \ref{subsection selection checklist}), along with a qualitative overview of major method families (Section \ref{subsection qualitative comparison table}).
These are further supported by a series of numerical illustrations (Section \ref{sec:numerical}), in which representative stopping rules are evaluated across four canonical sampling regimes: iid light-tailed (Section \ref{Exponential sampling}), iid heavy-tailed (Section \ref{Lognormal sampling}), probability estimation (Section \ref{Bernoulli rare-event estimation}), and serially dependent time series (Section \ref{ARCH(1) dependence}).
The resulting key insights translate the underlying theoretical distinctions into actionable guidance (Section \ref{subsection key insights}), highlighting in particular the fragility of CLT-based rules under heavy tails and serial dependence, as well as the substantial sample-size demands required for distribution-free guarantees.}

{Despite their flexibility and potential, sequential stopping rules seem largely unexploited in practical applications of Monte Carlo methods, often appearing only in an ad hoc manner when employed at all.
A crucial reason is the limited research focus on their direct implementation in real-world problems, making it challenging for practitioners to adapt these methods effectively.
Practical studies addressing this gap would undoubtedly stimulate further research in diverse directions.
To support further studies, the four scenarios examined in Section \ref{sec:numerical} \textcolor{black}{may serve as illustrative starting points for future comparative evaluations of stopping rules, in that they span the major methodological families and highlight the most practically relevant failure modes.
Nevertheless, we did not aim to propose them as a standardized benchmark suite, which would require a more systematic protocol and a fuller justification of representativeness than is within the scope of the present review.}
In addition, implementing sequential stopping rules presents significant challenges, particularly when addressing large-scale and high-dimensional problems in emerging fields.}

{There remain significant room for further developments, driven by the necessity of making these methods more accessible and practical.
In particular, bridging the gap between asymptotic validity and non-asymptotic performance remains the central unresolved tension.
As demonstrated via the numerical experiments (Section \ref{sec:numerical}), coverage can fall well below the nominal level or that computational costs become prohibitive under heavy tails and strong dependence.
Developing stopping rules for serially dependent sequences under realistic conditions without requiring strong structural assumptions or incurring very large sample sizes is another important target, given the prevalence of dependent outputs in MCMC, queueing simulations, and financial modeling.}

{Moreover, external techniques, such as variance reduction and machine learning, have yet to be systematically tailored to sequential stopping rules for performance enhancements, presenting promising avenues for future research.
For variance reduction, the central barrier is that techniques such as control variates modify the effective variance of the estimator adaptively, so that the empirical variance supplied to the stopping criterion is no longer a simple sample variance.
The interaction between such adaptive variance reduction and sequential stopping criteria is not well understood and warrants dedicated investigation.
For machine learning techniques, the fundamental difficulty is that the estimand itself evolves during training due to non-convex optimization and stochastic gradients.
As a consequence, the fixed-mean framework underpinning the stopping rules reviewed here does not directly apply.
Looking further ahead, modern Monte Carlo paradigms (such as Markov chain Monte Carlo, stochastic programming, and Bayesian cubature) represent natural yet largely unexplored territories for sequential stopping rules, where achieving theory-guaranteed stopping criteria demands significantly deeper and more technical analysis than in the standard Monte Carlo setting reviewed here.}

{In conclusion, we hope this survey article serves as a valuable guide, inspiring further exploration and innovation in the design and enhancement of sequential stopping rules for a broader range of Monte Carlo methods, and moreover, as a motivating reference point for establishing rigorously justified stopping rules in these advanced methodologies.}

\color{black}
\bibliographystyle{abbrv}
\small\bibliography{ref.bib}

@article{jagadeeswaran_hickernell_2019,
	author = {Jagadeeswaran, R. and Hickernell, Fred J.},
	journal = {Statistics and Computing},
	number = {6},
	pages = {1215-1229},
	title = {Fast automatic {Bayesian} cubature using lattice sampling},
	volume = {29},
	year = {2019}}

@misc{mahsereci2026bayesianquadraturegaussianprocesses,
      title={Bayesian Quadrature: Gaussian Processes for Integration}, 
      author={Maren Mahsereci and Toni Karvonen},
      year={2026},
      eprint={2602.16218},
      archivePrefix={arXiv},
      primaryClass={cs.LG},
      url={https://arxiv.org/abs/2602.16218}, 
      howpublished={arXiv:2602.16218}
}

@incollection{Prechelt2012EarlyStopping,
  author    = {Lutz Prechelt},
  title     = {Early Stopping — But When?},
  booktitle = {Neural Networks: Tricks of the Trade},
  series    = {Lecture Notes in Computer Science},
  volume    = {7700},
  pages     = {53--67},
  publisher = {Springer},
  year      = {2012},
  url       = {https://link.springer.com/chapter/10.1007/978-3-642-35289-8_5},
}

@article{JMLR:v24:23-0646,
  author  = {Eric Xia and Koulik Khamaru and Martin J. Wainwright and Michael I. Jordan},
  title   = {Instance-Dependent Confidence and Early Stopping for Reinforcement Learning},
  journal = {Journal of Machine Learning Research},
  year    = {2023},
  volume  = {24},
  number  = {392},
  pages   = {1--43},
  url     = {http://jmlr.org/papers/v24/23-0646.html}
}

@article{HuLei2022JMLR_EarlyStoppingIterativeReg,
  author  = {Ting Hu and Yunwen Lei},
  title   = {Early Stopping for Iterative Regularization with General Loss Functions},
  journal = {Journal of Machine Learning Research},
  volume  = {23},
  pages   = {1--36},
  year    = {2022},
  url     = {https://jmlr.org/papers/volume23/21-0983/21-0983.pdf}
}

@article{df3b10f8-26b4-38aa-9eb3-41173d4233f0,
 ISSN = {0030364X, 15265463},
 URL = {http://www.jstor.org/stable/223014},
 author = {David P. Morton},
 journal = {Operations Research},
 number = {5},
 pages = {710--718},
 publisher = {INFORMS},
 title = {Stopping Rules for a Class of Sampling-Based Stochastic Programming Algorithms},
 urldate = {2026-02-13},
 volume = {46},
 year = {1998}
}

@article{https://doi.org/10.1111/bmsp.12357,
author = {Kwon, Sunbeom and Zhang, Susu and K\"ohn, Hans Friedrich and Zhang, Bo},
title = {{MCMC} stopping rules in latent variable modelling},
journal = {British Journal of Mathematical and Statistical Psychology},
volume = {78},
number = {1},
pages = {225-257},
doi = {https://doi.org/10.1111/bmsp.12357},
url = {https://bpspsychub.onlinelibrary.wiley.com/doi/abs/10.1111/bmsp.12357},
eprint = {https://bpspsychub.onlinelibrary.wiley.com/doi/pdf/10.1111/bmsp.12357},
year = {2025}
}

@article{10.1214/aoms/1177731018,
author = {M. A. Girshick and Frederick Mosteller and L. J. Savage},
title = {Unbiased Estimates for Certain Binomial Sampling Problems with Applications},
volume = {17},
journal = {The Annals of Mathematical Statistics},
number = {1},
publisher = {Institute of Mathematical Statistics},
pages = {13 -- 23},
year = {1946},
doi = {10.1214/aoms/1177731018},
URL = {https://doi.org/10.1214/aoms/1177731018}
}

@article{10.1214/aoms/1177706361,
author = {Morris H. DeGroot},
title = {Unbiased Sequential Estimation for Binomial Populations},
volume = {30},
journal = {The Annals of Mathematical Statistics},
number = {1},
publisher = {Institute of Mathematical Statistics},
pages = {80 -- 101},
year = {1959},
doi = {10.1214/aoms/1177706361},
URL = {https://doi.org/10.1214/aoms/1177706361}
}

@article{10.1093/biomet/61.2.385,
    author = {Best, D. J.},
    title = {The variance of the inverse binomial estimator},
    journal = {Biometrika},
    volume = {61},
    number = {2},
    pages = {385-386},
    year = {1974},
    month = {08},
    issn = {0006-3444},
    doi = {10.1093/biomet/61.2.385},
    url = {https://doi.org/10.1093/biomet/61.2.385},
    eprint = {https://academic.oup.com/biomet/article-pdf/61/2/385/746002/61-2-385.pdf},
}

@article{357b9062-c3ca-3953-ab0e-8de887bd167d,
 ISSN = {05815738},
 URL = {http://www.jstor.org/stable/25052355},
 author = {P. K. Pathak and Y. S. Sathe},
 journal = {Sankhyā: The Indian Journal of Statistics, Series B (1960-2002)},
 number = {3},
 pages = {301--305},
 publisher = {Springer},
 title = {A New Variance Formula for Unbiased Estimation in Inverse Sampling},
 urldate = {2026-01-31},
 volume = {46},
 year = {1984}
}

@article{10.1093/biomet/63.1.216,
    author = {Mikulski, PIOTR W. and Smith, PAUL J.},
    title = {A variance bound for unbiased estimation in inverse sampling},
    journal = {Biometrika},
    volume = {63},
    number = {1},
    pages = {216-217},
    year = {1976},
    month = {04},
    issn = {0006-3444},
    doi = {10.1093/biomet/63.1.216},
    url = {https://doi.org/10.1093/biomet/63.1.216},
    eprint = {https://academic.oup.com/biomet/article-pdf/63/1/216/6689483/63-1-216.pdf},
}

@article{10.3150/09-BEJ219,
author = {Luis Mendo and Jos{\'e} M. Hernando},
title = {Estimation of a probability with optimum guaranteed confidence in inverse binomial sampling},
volume = {16},
journal = {Bernoulli},
number = {2},
publisher = {Bernoulli Society for Mathematical Statistics and Probability},
pages = {493-513},
year = {2010},
doi = {10.3150/09-BEJ219},
URL = {https://doi.org/10.3150/09-BEJ219}
}

@article{cho2019,
	author = {Cho, Hokwon},
	journal = {Methodology and Computing in Applied Probability},
	number = {3},
	pages = {721--733},
	title = {Two-Stage Procedure of Fixed-Width Confidence Intervals for the Risk Ratio},
	volume = {21},
	year = {2019}}

@article{Cho2013ApproximateConfidence,
  author    = {Hokwon Cho},
  title     = {Approximate Confidence Limits for the Ratio of Two Binomial Variates with Unequal Sample Sizes},
  journal   = {Communications for Statistical Applications and Methods},
  volume    = {20},
  number    = {5},
  pages     = {347--356},
  year      = {2013},
  doi       = {10.5351/CSAM.2013.20.5.347},
  publisher = {The Korean Statistical Society}
}

@article{https://doi-org.utokyo.idm.oclc.org/10.1002/sim.3158,
author = {Tian, M. and Tang, M. L. and Ng, H. K. T. and Chan, P. S.},
title = {Confidence intervals for the risk ratio under inverse sampling},
journal = {Statistics in Medicine},
volume = {27},
number = {17},
pages = {3301-3324},
doi = {https://doi-org.utokyo.idm.oclc.org/10.1002/sim.3158},
url = {https://onlinelibrary-wiley-com.utokyo.idm.oclc.org/doi/abs/10.1002/sim.3158},
eprint = {https://onlinelibrary-wiley-com.utokyo.idm.oclc.org/doi/pdf/10.1002/sim.3158},
year = {2008}
}

@article{10.1214/aoms/1177730439,
author = {J. Wolfowitz},
title = {The Efficiency of Sequential Estimates and {Wald's} Equation for Sequential Processes},
volume = {18},
journal = {The Annals of Mathematical Statistics},
number = {2},
publisher = {Institute of Mathematical Statistics},
pages = {215-230},
year = {1947},
doi = {10.1214/aoms/1177730439},
URL = {https://doi.org/10.1214/aoms/1177730439}
}

@article{mendo2025,
	author = {Mendo, Luis},
	journal = {Statistical Papers},
	number = {1},
	pages = {26},
	title = {Estimating odds and log odds with guaranteed accuracy},
	volume = {66},
	year = {2025}}

@article{Bandyopadhyay03072017,
author = {Uttam Bandyopadhyay and Suman Sarkar and Atanu Biswas},
title = {Fixed-width confidence interval of log odds ratio for joint binomial and inverse binomial sampling},
journal = {Sequential Analysis},
volume = {36},
number = {3},
pages = {345--354},
year = {2017},
publisher = {Taylor \& Francis},
doi = {10.1080/07474946.2017.1360087},


URL = { 
    
        https://doi.org/10.1080/07474946.2017.1360087
    
    

},
eprint = { 
    
        https://doi.org/10.1080/07474946.2017.1360087
    
    

}

}

@article{10.1093/biomet/54.1-2.181,
    author = {Gart, JOHN J. and Zweifel, JAMES R.},
    title = {On the bias of various estimators of the logit and its variance with application to quantal bioassay},
    journal = {Biometrika},
    volume = {54},
    number = {1-2},
    pages = {181-187},
    year = {1967},
    month = {06},
    issn = {0006-3444},
    doi = {10.1093/biomet/54.1-2.181},
    url = {https://doi.org/10.1093/biomet/54.1-2.181},
    eprint = {https://academic.oup.com/biomet/article-pdf/54/1-2/181/947312/54-1-2-181.pdf},
}

@INPROCEEDINGS{1172904,
  author={Schmeiser, B. and Yingchieh Yeh},
  booktitle={Proceedings of the Winter Simulation Conference}, 
  title={On choosing a single criterion for confidence-interval procedures}, 
  year={2002},
  volume={1},
  number={},
  pages={345-352 vol.1},
  doi={10.1109/WSC.2002.1172904}}

@article{Yeh02112015,
author = {Yingchieh Yeh and Bruce W. Schmeiser},
title = {{VAMP1RE}: a single criterion for rating and ranking confidence-interval procedures},
journal = {IIE Transactions},
volume = {47},
number = {11},
pages = {1203--1216},
year = {2015},
publisher = {Taylor \& Francis},
doi = {10.1080/0740817X.2015.1047068},


URL = { 
    
        https://doi.org/10.1080/0740817X.2015.1047068
    
    

},
eprint = { 
    
        https://doi.org/10.1080/0740817X.2015.1047068
    
    

}

}

@article{Yeh03072025,
author = {Yingchieh Yeh and Shing Chih Tsai},
title = {Single criterion for rating confidence interval procedures: a generalization to stochastic stopping rules},
journal = {Journal of the Operational Research Society},
volume = {76},
number = {7},
pages = {1481--1489},
year = {2025},
publisher = {Taylor \& Francis},
doi = {10.1080/01605682.2024.2441227},


URL = { 
    
        https://doi.org/10.1080/01605682.2024.2441227
    
    

},
eprint = { 
    
        https://doi.org/10.1080/01605682.2024.2441227
    
    

}

}

@article{10.1145/1102586.1102590,
author = {Solomon, Susan L.},
title = {A simplified stopping rule for simulations},
year = {1972},
issue_date = {October 1972},
publisher = {Association for Computing Machinery},
address = {New York, NY, USA},
volume = {4},
number = {1},
issn = {0163-6103},
url = {https://doi.org/10.1145/1102586.1102590},
doi = {10.1145/1102586.1102590},
journal = {ACM SIGSIM Simulation Digest},
pages = {27–28},
numpages = {2}
}

@article{ADLAKHA1982379,
title = {Starting and stopping rules for simulations using a priori information},
journal = {European Journal of Operational Research},
volume = {10},
number = {4},
pages = {379-394},
year = {1982},
issn = {0377-2217},
doi = {https://doi.org/10.1016/0377-2217(82)90089-3},
url = {https://www.sciencedirect.com/science/article/pii/0377221782900893},
author = {Veena G. Adlakha and George S. Fishman}
}

@article{5078cc27-f2a4-395f-864d-471f6693e06d,
 ISSN = {0030364X, 15265463},
 author = {Irwin W. Kabak},
 journal = {Operations Research},
 number = {2},
 pages = {431--437},
 publisher = {INFORMS},
 title = {Stopping rules for queuing simulations},
 urldate = {2025-04-12},
 volume = {16},
 year = {1968}
}

@INPROCEEDINGS{7843498,
  author={Chiew, Yeong Shiong and Shaw, Geoffrey M and Docherty, Paul and Dickson, Jennifer and Pretty, Christopher and Chase, J. Geoffrey},
  booktitle={2016 IEEE EMBS Conference on Biomedical Engineering and Sciences (IECBES)}, 
  title={Early clinical trial termination: Simulation-based design of a robust stopping rule using difference in interventional effect on mortality}, 
  year={2016},
  volume={},
  number={},
  pages={484-489},
  doi={10.1109/IECBES.2016.7843498}}

@article{10.1145/2567907,
author = {Gupta, Vivek and Andrad\'{o}ttir, Sigr\'{u}n and Goldsman, David},
title = {Variance estimation and sequential stopping in steady-state simulations using linear regression},
year = {2014},
issue_date = {February 2014},
publisher = {Association for Computing Machinery},
address = {New York, NY, USA},
volume = {24},
number = {2},
issn = {1049-3301},
url = {https://doi.org/10.1145/2567907},
doi = {10.1145/2567907},
journal = {ACM Transactions on Modeling and Computer Simulation},
articleno = {7},
numpages = {25}
}

@book{10.5555/554952,
author = {Law, Averill M. and Kelton, David M.},
title = {Simulation Modeling and Analysis},
year = {1999},
isbn = {0070592926},
publisher = {McGraw-Hill Higher Education},
edition = {3rd}
}

@article{infinitevariance,
author = {Reiichiro Kawai},
title = {{Monte Carlo methods with infinite variances}},
volume = {23},
journal = {Probability Surveys},
publisher = {Institute of Mathematical Statistics and Bernoulli Society},
pages = {1-42},
year = {2026},
doi = {10.1214/25-PS32},
URL = {https://doi.org/10.1214/25-PS32}
}

@article{ELLINGWOOD2025102474,
title = {Development of methods of structural reliability},
journal = {Structural Safety},
volume = {113},
pages = {102474},
year = {2025},
issn = {0167-4730},
doi = {https://doi.org/10.1016/j.strusafe.2024.102474},
url = {https://www.sciencedirect.com/science/article/pii/S0167473024000456},
author = {Bruce Ellingwood and Marc Maes and F. {Michael Bartlett} and Andre T. Beck and Colin Caprani and Armen {Der Kiureghian} and Leonardo Dueñas-Osorio and Neryvaldo Galvão and Robert Gilbert and Jie Li and Jose Matos and Yasuhiro Mori and Iason Papaioannou and Roger Parades and Daniel Straub and Bruno Sudret}
}

@inproceedings{dammertz09wscg,
  author =       {Holger Dammertz and Johannes Hanika and Alexander Keller and Hendrik Lensch},
  title =        {A hierarchical automatic stopping condition for {Monte} {Carlo} global illumination},
  booktitle =    {Proceedings of the WSCG 2010},
  pages =        {159--164},
  year =         2010
}

@article{Shim01102007,
author = {Hyung Jin Shim and Chang Hyo Kim},
title = {Stopping criteria of inactive cycle {Monte} {Carlo} calculations},
journal = {Nuclear Science and Engineering},
volume = {157},
number = {2},
pages = {132--141},
year = {2007},
publisher = {Taylor \& Francis},
doi = {10.13182/NSE06-33},


URL = { 
    
        https://doi.org/10.13182/NSE06-33
    
    

}
}

@article{Ueki01012005,
author = {Taro Ueki and Forrest B. Brown},
title = {Stationarity modeling and informatics-based diagnostics in {Monte} {Carlo} criticality calculations},
journal = {Nuclear Science and Engineering},
volume = {149},
number = {1},
pages = {38--50},
year = {2005},
publisher = {Taylor \& Francis},
doi = {10.13182/NSE04-15},


URL = { 
    
        https://doi.org/10.13182/NSE04-15
    
    

}
}

@article{frick1998,
	author = {Frick, Robert W. },
	journal = {Behavior Research Methods, Instruments, \& Computers},
	number = {4},
	pages = {690--697},
	title = {A better stopping rule for conventional statistical tests},
	volume = {30},
	year = {1998}}

@article{Costanza01121979,
author = {Michael C. Costanza and A. A. Afifi},
title = {Comparison of stopping rules in forward stepwise discriminant analysis},
journal = {Journal of the American Statistical Association},
volume = {74},
number = {368},
pages = {777--785},
year = {1979},
publisher = {ASA Website},
doi = {10.1080/01621459.1979.10481030},


URL = { 
    
    
        https://www.tandfonline.com/doi/abs/10.1080/01621459.1979.10481030
    

}
}

@article{doi:10.1287/opre.21.6.1309,
author = {Randolph, Paul H. and Swinson, Gary and Ellingsen, Carl},
title = {Technical note - Stopping rules for sequencing problems},
journal = {Operations Research},
volume = {21},
number = {6},
pages = {1309-1315},
year = {1973},
doi = {10.1287/opre.21.6.1309},

URL = { 
    
        https://doi.org/10.1287/opre.21.6.1309
    
    

}
}

@book{billingsley1995probability,
  title={Probability and Measure},
  author={Billingsley, P.},
  isbn={9780471007104},
  lccn={gb95051456},
  series={Wiley Series in Probability and Statistics},
  year={1995},
  publisher={Wiley}
}

@article{Stellato03042017,
author = {Bartolomeo Stellato and Bart P. G. Van Parys and Paul J. Goulart},
title = {Multivariate {Chebyshev} inequality with estimated mean and variance},
journal = {The American Statistician},
volume = {71},
number = {2},
pages = {123--127},
year = {2017},
publisher = {ASA Website},
doi = {10.1080/00031305.2016.1186559},


URL = { 
    
        https://doi.org/10.1080/00031305.2016.1186559
    
    

}

}

@article{Saw01051984,
author = {John G. Saw and Mark C.K. Yang and Tse Chin Mo},
title = {Chebyshev inequality with estimated mean and variance},
journal = {The American Statistician},
volume = {38},
number = {2},
pages = {130--132},
year = {1984},
publisher = {ASA Website},
doi = {10.1080/00031305.1984.10483182},


URL = { 
    
    
        https://www.tandfonline.com/doi/abs/10.1080/00031305.1984.10483182
    

}
}

@article{WOO1991179,
title = {A quitting rule for {M}onte {C}arlo simulation of extreme risks},
journal = {Reliability Engineering \& System Safety},
volume = {31},
number = {2},
pages = {179-189},
year = {1991},
issn = {0951-8320},
doi = {https://doi.org/10.1016/0951-8320(91)90117-P},
url = {https://www.sciencedirect.com/science/article/pii/095183209190117P},
author = {Gordon Woo}
}

@article{landers_rogge_1976,
	author = {Landers, D. and Rogge, L.},
	journal = {Zeitschrift f{\"u}r Wahrscheinlichkeitstheorie und Verwandte Gebiete},
	number = {4},
	pages = {269--283},
	title = {The exact approximation order in the central-limit-theorem for random summation},
	volume = {36},
	year = {1976}}

@article{10.1093/jrsssb/qkad009,
    author = {Waudby-Smith, Ian and Ramdas, Aaditya},
    title = {Estimating means of bounded random variables by betting},
    journal = {Journal of the Royal Statistical Society Series B: Statistical Methodology},
    volume = {86},
    number = {1},
    pages = {1-27},
    year = {2023},
    month = {02},
    issn = {1369-7412},
    doi = {10.1093/jrsssb/qkad009},
    url = {https://doi.org/10.1093/jrsssb/qkad009}
}

@article{10.1093/biomet/33.3.222,
    author = {Haldane, J. B. S.},
    title = {On a method of estimating frequencies},
    journal = {Biometrika},
    volume = {33},
    number = {3},
    pages = {222-225},
    year = {1945},
    month = {11},
    issn = {0006-3444},
    doi = {10.1093/biomet/33.3.222},
    url = {https://doi.org/10.1093/biomet/33.3.222}
}

@article{10.1214/aoms/1177729586,
author = {Herbert Robbins and Sutton Monro},
title = {A stochastic approximation method},
volume = {22},
journal = {The Annals of Mathematical Statistics},
number = {3},
publisher = {Institute of Mathematical Statistics},
pages = {400-407},
year = {1951},
doi = {10.1214/aoms/1177729586},
URL = {https://doi.org/10.1214/aoms/1177729586}
}

@article{https://doi.org/10.1002/j.1538-7305.1930.tb00373.x,
author = {Shewhart, W. A.},
title = {Economic quality control of manufactured product},
journal = {Bell System Technical Journal},
volume = {9},
number = {2},
pages = {364-389},
doi = {https://doi.org/10.1002/j.1538-7305.1930.tb00373.x},
url = {https://onlinelibrary.wiley.com/doi/abs/10.1002/j.1538-7305.1930.tb00373.x},
year = {1930}
}

@book{wald1947,
title={Sequential Analysis},
author={Abraham Wald},
publisher={John Wiley and Sons, New York},
year={1947}
}

@article{1737d468-f05a-3760-96b3-0b4b960811fd,
 ISSN = {10170405, 19968507},
 URL = {http://www.jstor.org/stable/24306697},
 author = {Chin-Shan Chuang and Tze Leung Lai},
 journal = {Statistica Sinica},
 number = {1},
 pages = {1--33},
 publisher = {Institute of Statistical Science, Academia Sinica},
 title = {Hybrid resampling methods for confidence intervals},
 urldate = {2025-01-01},
 volume = {10},
 year = {2000}
}

@book{jennison1999group,
  title={Group Sequential Methods with Applications to Clinical Trials},
  author={Jennison, C. and Turnbull, B.W.},
  isbn={9781584888581},
  lccn={99043218},
  series={Chapman \& Hall/CRC Interdisciplinary Statistics},
  year={1999},
  publisher={CRC Press}
}

@article{WOODROOFE198670,
title = {Asymptotic optimality in sequential interval estimation},
journal = {Advances in Applied Mathematics},
volume = {7},
number = {1},
pages = {70-79},
year = {1986},
issn = {0196-8858},
doi = {https://doi.org/10.1016/0196-8858(86)90007-2},
url = {https://www.sciencedirect.com/science/article/pii/0196885886900072},
author = {Michael Woodroofe}
}

@article{10.1214/aos/1176345639,
author = {Peter Hall},
title = {Asymptotic theory of triple sampling for sequential estimation of a mean},
volume = {9},
journal = {The Annals of Statistics},
number = {6},
publisher = {Institute of Mathematical Statistics},
pages = {1229-1238},
year = {1981},
doi = {10.1214/aos/1176345639},
URL = {https://doi.org/10.1214/aos/1176345639}
}

@article{4a785a9d-f0f6-3ad7-b6ff-a8b0190a3f52,
 ISSN = {00357596, 19453795},
 URL = {http://www.jstor.org/stable/44236218},
 author = {L.H. Koopmans and CLIFFORD Qualls},
 journal = {The Rocky Mountain Journal of Mathematics},
 number = {4},
 pages = {587--602},
 publisher = {Rocky Mountain Mathematics Consortium},
 title = {Fixed length confidence intervals for parameters of the normal distribution based on two-stage sampling procedures},
 urldate = {2024-12-29},
 volume = {1},
 year = {1971}
}

@incollection{doi:https://doi-org.utokyo.idm.oclc.org/10.1002/9781118445112.stat08283,
author = {Vats, Dootika and Flegal, James M. and Jones, Galin L.},
publisher = {John Wiley \& Sons, Ltd},
isbn = {9781118445112},
title = {Monte {C}arlo Simulation: Are We There Yet?},
booktitle = {Wiley StatsRef: Statistics Reference Online},
chapter = {},
pages = {1-15},
doi = {https://doi-org.utokyo.idm.oclc.org/10.1002/9781118445112.stat08283},
year = {2021}
}

@article{Law01011981,
author = {Averill M. Law and W. David Kelton and Lloyd W Koenig},
title = {Relative width sequential confidence intervals for the mean},
journal = {Communications in Statistics - Simulation and Computation},
volume = {10},
number = {1},
pages = {29--39},
year = {1981},
publisher = {Taylor \& Francis},
doi = {10.1080/03610918108812190},


URL = { 
    
        https://doi.org/10.1080/03610918108812190
    
    

}

}

@Incollection{Schoen2009,
author="Schoen, Fabio",
editor="Floudas, Christodoulos A.
and Pardalos, Panos M.",
title="Stochastic global optimization: Stopping rules",
bookTitle="Encyclopedia of Optimization",
year="2009",
publisher="Springer US",
address="Boston, MA",
pages="3743--3746",
isbn="978-0-387-74759-0",
doi="10.1007/978-0-387-74759-0_647",
url="https://doi.org/10.1007/978-0-387-74759-0_647"
}

@article{e1db1290-dea4-3684-b116-06f625239089,
 ISSN = {00063444, 14643510},
 URL = {http://www.jstor.org/stable/2334018},
 author = {D. R. Cox},
 journal = {Biometrika},
 number = {3/4},
 pages = {217--227},
 publisher = {[Oxford University Press, Biometrika Trust]},
 title = {Estimation by Double Sampling},
 urldate = {2024-12-24},
 volume = {39},
 year = {1952}
}

@article{dantzig1940,
	author = {George B.  Dantzig},
	journal = {The Annals of Mathematical Statistics},
	number = {2},
	pages = {186--192 },
	title = {On the non-existence of tests of ``{S}tudent's'' hypothesis having power functions independent of $\sigma$},
	volume = {11},
	year = {1940}}

@article{10.1214/aoms/1177731088,
author = {Charles Stein},
title = {A two-sample test for a linear hypothesis whose power is independent of the variance},
volume = {16},
journal = {The Annals of Mathematical Statistics},
number = {3},
publisher = {Institute of Mathematical Statistics},
pages = {243-258},
year = {1945},
doi = {10.1214/aoms/1177731088},
URL = {https://doi.org/10.1214/aoms/1177731088}
}

@article{mukhopadhyay1989,
	author = {Mukhopadhyay, Nitis and Sen, Pranab Kumar and Sinha, Bikas Kumar},
	journal = {Annals of the Institute of Statistical Mathematics},
	number = {1},
	pages = {121--138},
	title = {Stopping rules, permutation invariance and sufficiency principle},
	volume = {41},
	year = {1989}}

@article{OHKUBO2021,
  title={Revisiting the two predominant statistical problems: The stopping-rule problem and the catch-all hypothesis problem},
  author={Yusaku Ohkubo},
  journal={Annals of the Japan Association for Philosophy of Science},
  volume={30},
  number={ },
  pages={23-41},
  year={2021},
  doi={10.4288/jafpos.30.0_23}
}

@article{fletcher2024,
	author = {Fletcher, Samuel C. },
	journal = {Journal for General Philosophy of Science},
	number = {1},
	pages = {1--28},
	title = {The stopping rule principle and confirmational reliability},
	volume = {55},
	year = {2024}}

@article{takada,
	author = {Takada, Yoshikazu},
	journal = {Annals of the Institute of Statistical Mathematics},
	number = {2},
	pages = {325--335},
	title = {The nonexistence of procedures with bounded performance characteristics in certain parametric inference problems},
	volume = {50},
	year = {1998}}

@article{10.1214/aoms/1177700157,
author = {Leon J. Gleser},
title = {On the asymptotic theory of fixed-size sequential confidence bounds for linear regression parameters},
volume = {36},
journal = {The Annals of Mathematical Statistics},
number = {2},
publisher = {Institute of Mathematical Statistics},
pages = {463-467},
year = {1965},
doi = {10.1214/aoms/1177700157},
URL = {https://doi.org/10.1214/aoms/1177700157}
}

@book{d2019monte,
  title={Monte Carlo Methods for Medical Physics: A Practical Introduction},
  author={Schuemann, J. and Jia, X. and Paganetti, H.},
  isbn={9781498736718},
  year={2019},
  publisher={Taylor \& Francis Group}
}

@article{SONG2023103479,
title = {{M}onte {C}arlo and variance reduction methods for structural reliability analysis: A comprehensive review},
journal = {Probabilistic Engineering Mechanics},
volume = {73},
pages = {103479},
year = {2023},
issn = {0266-8920},
doi = {https://doi.org/10.1016/j.probengmech.2023.103479},
url = {https://www.sciencedirect.com/science/article/pii/S0266892023000681},
author = {Chenxiao Song and Reiichiro Kawai}
}

@book{nightingale1998quantum,
  title={Quantum Monte Carlo Methods in Physics and Chemistry},
  author={Nightingale, M.P. and Umrigar, C.J.},
  isbn={9780792355519},
  lccn={98051578},
  series={NATO Science Series C},
  year={1998},
  publisher={Springer Netherlands}
}

@article{HE20231,
title = {The {G}erber-{S}hiu discounted penalty function: A review from practical perspectives},
journal = {Insurance: Mathematics and Economics},
volume = {109},
pages = {1-28},
year = {2023},
issn = {0167-6687},
doi = {https://doi.org/10.1016/j.insmatheco.2022.12.003},
url = {https://www.sciencedirect.com/science/article/pii/S0167668722001263},
author = {Yue He and Reiichiro Kawai and Yasutaka Shimizu and Kazutoshi Yamazaki}
}

@article{doi:10.1287/opre.38.3.546,
author = {Kang, Keebom and Schmeiser, Bruce},
title = {Graphical methods for evaluating and comparing confidence-interval procedures},
journal = {Operations Research},
volume = {38},
number = {3},
pages = {546-553},
year = {1990},
doi = {10.1287/opre.38.3.546},

URL = { 
    
        https://doi.org/10.1287/opre.38.3.546
    
    

}
}

@article{https://doi-org.utokyo.idm.oclc.org/10.1002/rsa.20839,
author = {Huber, Mark},
title = {An optimal $(\epsilon,\delta)$-randomized approximation scheme for the mean of random variables with bounded relative variance},
journal = {Random Structures \& Algorithms},
volume = {55},
number = {2},
pages = {356-370},
doi = {https://doi-org.utokyo.idm.oclc.org/10.1002/rsa.20839},
url = {https://onlinelibrary-wiley-com.utokyo.idm.oclc.org/doi/abs/10.1002/rsa.20839},
year = {2019}
}

@article{https://doi-org.utokyo.idm.oclc.org/10.1002/rsa.20654,
author = {Huber, Mark},
title = {A {B}ernoulli mean estimate with known relative error distribution},
journal = {Random Structures \& Algorithms},
volume = {50},
number = {2},
pages = {173-182},
doi = {https://doi-org.utokyo.idm.oclc.org/10.1002/rsa.20654},
url = {https://onlinelibrary-wiley-com.utokyo.idm.oclc.org/doi/abs/10.1002/rsa.20654},
year = {2017}
}

@INPROCEEDINGS{899773,
  author={Chen, E.J. and Kelton, W.D.},
  booktitle={2000 Winter Simulation Conference Proceedings}, 
  title={A stopping procedure based on phi-mixing conditions}, 
  year={2000},
  volume={},
  number={},
  pages={617-626},
  doi={10.1109/WSC.2000.899773}}

@article{doi:10.1137/S0097539797315306,
author = {Dagum, Paul and Karp, Richard and Luby, Michael and Ross, Sheldon},
title = {An optimal algorithm for {Monte} {Carlo} estimation},
journal = {SIAM Journal on Computing},
volume = {29},
number = {5},
pages = {1484-1496},
year = {2000},
doi = {10.1137/S0097539797315306},

URL = { 
    
        https://doi.org/10.1137/S0097539797315306
    
    

}
}

@article{CHEN2003237,
title = {Determining simulation run length with the runs test},
journal = {Simulation Modelling Practice and Theory},
volume = {11},
number = {3},
pages = {237-250},
year = {2003},
issn = {1569-190X},
doi = {https://doi.org/10.1016/S1569-190X(03)00048-0},
url = {https://www.sciencedirect.com/science/article/pii/S1569190X03000480},
author = {E.Jack Chen and W.David Kelton}
}

@article{ATA2007237,
title = {A convergence criterion for the {Monte Carlo} estimates},
journal = {Simulation Modelling Practice and Theory},
volume = {15},
number = {3},
pages = {237-246},
year = {2007},
issn = {1569-190X},
doi = {https://doi.org/10.1016/j.simpat.2006.12.002},
url = {https://www.sciencedirect.com/science/article/pii/S1569190X06001146},
author = {Mustafa Y. Ata}
}

@article{doi:10.1287/mnsc.26.1.18,
author = {Schruben, Lee W.},
title = {A coverage function for interval estimators of simulation response},
journal = {Management Science},
volume = {26},
number = {1},
pages = {18-27},
year = {1980},
doi = {10.1287/mnsc.26.1.18},

URL = { 
    
        https://doi.org/10.1287/mnsc.26.1.18
    
    

}
}

@article{Chen01052012,
author = {E J Chen},
title = {A stopping rule using the quasi-independent sequence},
journal = {Journal of Simulation},
volume = {6},
number = {2},
pages = {71--80},
year = {2012},
publisher = {Taylor \& Francis},
doi = {10.1057/jos.2011.23},


URL = { 
    
        https://doi.org/10.1057/jos.2011.23
    
    

},

}

@article{doi:10.1287/opre.40.2.279,
author = {Whitt, Ward},
title = {Asymptotic formulas for {Markov} processes with applications to simulation},
journal = {Operations Research},
volume = {40},
number = {2},
pages = {279-291},
year = {1992},
doi = {10.1287/opre.40.2.279},

URL = { 
    
        https://doi.org/10.1287/opre.40.2.279
    
    

}
}

@article{doi:10.1287/mnsc.35.11.1341,
author = {Whitt, Ward},
title = {Planning queueing simulations},
journal = {Management Science},
volume = {35},
number = {11},
pages = {1341-1366},
year = {1989},
doi = {10.1287/mnsc.35.11.1341},

URL = { 
    
        https://doi.org/10.1287/mnsc.35.11.1341
    
    

}
}

@article{10.1214/aos/1176350049,
author = {Michael Woodroofe},
title = {Very weak expansions for sequential confidence levels},
volume = {14},
journal = {The Annals of Statistics},
number = {3},
publisher = {Institute of Mathematical Statistics},
pages = {1049-1067},
year = {1986},
doi = {10.1214/aos/1176350049},
URL = {https://doi.org/10.1214/aos/1176350049}
}

@article{doi:10.1080/03610928208828311,
author = {J.W.H. Swanepoel and J.W.J. van Wyk},
title = {Fixed width confidence intervals for the location parameter of an exponential distribution},
journal = {Communications in Statistics - Theory and Methods},
volume = {11},
number = {11},
pages = {1279--1289},
year = {1982},
publisher = {Taylor \& Francis},
doi = {10.1080/03610928208828311},


URL = { 
    
        https://doi.org/10.1080/03610928208828311
    
    

}
}

@article{2274e8f1-41d9-3449-bdd4-d9796a099dc0,
 ISSN = {0581572X},
 URL = {http://www.jstor.org/stable/25050270},
 author = {H. Callaert and P. Janssen},
 journal = {Sankhyā: The Indian Journal of Statistics, Series A (1961-2002)},
 number = {2},
 pages = {211--219},
 publisher = {Springer},
 title = {The convergence rate of fixed-width sequential confidence intervals for the mean},
 urldate = {2024-11-29},
 volume = {43},
 year = {1981}
}

@article{doi:10.1080/07474949108836228,
author = {N. Mukhopadhyay and T.K.S. Solanky},
title = {Second order properties of accelerated stopping times with applications in sequential estimation},
journal = {Sequential Analysis},
volume = {10},
number = {1-2},
pages = {99--123},
year = {1991},
publisher = {Taylor \& Francis},
doi = {10.1080/07474949108836228},


URL = { 
    
        https://doi.org/10.1080/07474949108836228
    
    

}
}

@article{hirose_isogai_uni_1997,
	author = {Hirose, Keiichi and Isogai, Eiichi and Uno, Chikara},
	journal = {Annals of the Institute of Statistical Mathematics},
	number = {2},
	pages = {199--209},
	title = {The Convergence Rate of Fixed-Width Sequential Confidence Intervals for a Parameter of an Exponential Distribution},
	volume = {49},
	year = {1997}}

@article{https://doi-org.utokyo.idm.oclc.org/10.1111/j.1467-842X.1986.tb00587.x,
author = {Aerts, M. and Callaert, H.},
title = {The convergence rate of sequential fixed-width confidence intervals for regular functionals},
journal = {Australian Journal of Statistics},
volume = {28},
number = {1},
pages = {97-106},
doi = {https://doi-org.utokyo.idm.oclc.org/10.1111/j.1467-842X.1986.tb00587.x},
year = {1986}
}

@article{10.1111/j.2517-6161.1983.tb01243.x,
    author = {Hall, Peter},
    title = {Sequential estimation saving sampling operations},
    journal = {Journal of the Royal Statistical Society: Series B (Methodological)},
    volume = {45},
    number = {2},
    pages = {219-223},
    year = {1983},
    month = {12},
    issn = {0035-9246},
    doi = {10.1111/j.2517-6161.1983.tb01243.x},
    url = {https://doi.org/10.1111/j.2517-6161.1983.tb01243.x},
}

@article{e0ef43b4-f485-3a19-b019-5fb86fa01603,
 ISSN = {10170405, 19968507},
 URL = {http://www.jstor.org/stable/24306854},
 author = {Tze Leung Lai},
 journal = {Statistica Sinica},
 number = {2},
 pages = {303--351},
 publisher = {Institute of Statistical Science, Academia Sinica},
 title = {Sequential analysis: Some classical problems and new challenges},
 urldate = {2024-10-23},
 volume = {11},
 year = {2001}
}

@article{10.1214/aos/1051027870,
author = {David Siegmund},
title = {{Herbert Robbins and sequential analysis: invited paper}},
volume = {31},
journal = {The Annals of Statistics},
number = {2},
publisher = {Institute of Mathematical Statistics},
pages = {349-365},
year = {2003},
doi = {10.1214/aos/1051027870},
URL = {https://doi.org/10.1214/aos/1051027870}
}

@book{Mukhopadhyaybook,
title={Sequential Methods and Their Applications},
author={Nitis Mukhopadhyay and Basil M. de Silva},
year={2008},
edition={1st},
publisher={Chapman and Hall/CRC, New York}
}

@article{bicher,
title={Review on {Monte} {Carlo} simulation stopping rules: How many samples are really enough?},
journal={Simulation Notes Europe SNE},
volume={32},
number={1},
pages={1-8},
year={2022},
doi={https://doi.org/10.11128/sne.32.on.10591},
author={Martin Bicher and Matthias Wastian and Dominik Brunmeir and Niki Popper}
}

@article{ALMEIDA2023108797,
title = {Stopping criterion for {M}onte {C}arlo method-based simulations of the lightning performance of transmission lines},
journal = {Electric Power Systems Research},
volume = {214},
pages = {108797},
year = {2023},
issn = {0378-7796},
doi = {https://doi.org/10.1016/j.epsr.2022.108797},
url = {https://www.sciencedirect.com/science/article/pii/S0378779622008513},
author = {Frederico S. Almeida and Fernando H. Silveira and Silvério Visacro}
}

@misc{malfunctioning,
author={Reiichiro Kawai},
title={Sequential stopping rules malfunctioning},
note={preprint available \href{https://www.researchgate.net/publication/396847522_Sequential_stopping_rules_malfunctioning}{here}},
year={2025}}

@article{JWmartingale,
author={Jiezhong Wu and Reiichiro Kawai},
title={Stopping rules for {Monte} {Carlo} methods of martingale difference type},
journal={SIAM Journal on Scientific Computing},
volume={48},
year={2026}}

@article{ASDIS,
title = {Adaptive radial importance sampling under directional stratification},
journal = {Probabilistic Engineering Mechanics},
volume = {72},
pages = {103443},
year = {2023},
issn = {0266-8920},
doi = {https://doi.org/10.1016/j.probengmech.2023.103443},
url = {https://www.sciencedirect.com/science/article/pii/S0266892023000322},
author = {Chenxiao Song and Reiichiro Kawai}
}

@article{ReliabilitySS,
title = {Adaptive stratified sampling for structural reliability analysis},
journal = {Structural Safety},
volume = {101},
pages = {102292},
year = {2023},
issn = {0167-4730},
doi = {https://doi.org/10.1016/j.strusafe.2022.102292},
url = {https://www.sciencedirect.com/science/article/pii/S0167473022000996},
author = {Chenxiao Song and Reiichiro Kawai}
}

@article{BHST,
author = {Bayer, Christian and Hoel, Håkon and von Schwerin, Erik and Tempone, Raúl},
title = {On nonasymptotic optimal stopping criteria in {Monte} {Carlo} simulations},
journal = {SIAM Journal on Scientific Computing},
volume = {36},
number = {2},
pages = {A869-A885},
year = {2014},
doi = {10.1137/130911433},

URL = { 
        https://doi.org/10.1137/130911433
    
}
}

@article{GS1,
author = {Fishman, George S.},
title = {Achieving specific accuracy in simulation output analysis},
year = {1977},
issue_date = {May 1977},
publisher = {Association for Computing Machinery},
address = {New York, NY, USA},
volume = {20},
number = {5},
issn = {0001-0782},
url = {https://doi.org/10.1145/359581.359589},
doi = {10.1145/359581.359589},
journal = {Communications of the ACM},
pages = {310–315},
numpages = {6}
}

@article{AJ1,
 ISSN = {0030364X, 15265463},
 URL = {http://www.jstor.org/stable/170064},
 author = {Averill M. Law and John S. Carson},
 journal = {Operations Research},
 number = {5},
 pages = {1011-1025},
 publisher = {INFORMS},
 title = {A sequential procedure for determining the length of a steady-state simulation},
 volume = {27},
 year = {1979}
}

@article{YH1965,
author = {Y. S. Chow and Herbert Robbins},
title = {On the asymptotic theory of fixed-width sequential confidence intervals for the mean},
volume = {36},
journal = {The Annals of Mathematical Statistics},
number = {2},
publisher = {Institute of Mathematical Statistics},
pages = {457-462},
year = {1965},
doi = {10.1214/aoms/1177700156},
URL = {https://doi.org/10.1214/aoms/1177700156}
}

@article{NS1996,
author = {N. Mukhopadhyay and S. Datta},
year = {1996},
month = {},
pages = {497-507},
title = {On sequential fixed-width confidence intervals for the mean and second-order expansions of the associated coverage probabilities},
volume = {48},
number={3},
journal = {Annals of the Institute of Statistical Mathematics},
doi = {10}
}

@article{RL1956,
author = {R. R. Bahadur and Leonard J. Savage},
title = {The nonexistence of certain statistical procedures in nonparametric problems},
volume = {27},
journal = {The Annals of Mathematical Statistics},
number = {4},
publisher = {Institute of Mathematical Statistics},
pages = {1115-1122},
year = {1956},
doi = {10.1214/aoms/1177728077}
}

@inproceedings{M1968,
author = {Gilman, Michael J.},
title = {A brief survey of stopping rules in {Monte} {Carlo} simulations},
year = {1968},
publisher = {Winter Simulation Conference},
booktitle = {Proceedings of the Second Conference on Applications of Simulations},
pages = {16–20},
numpages = {5},
location = {New York, New York, USA}
}

@article{O1994,
	author = {Balci, Osman},
	journal = {Annals of Operations Research},
	number = {1},
	pages = {121--173},
	title = {Validation, verification, and testing techniques throughout the life cycle of a simulation study},
	volume = {53},
	year = {1994}}

@article{RD1976,
author = { R. J.   Serfling  and  D. D.   Wackerly },
title = {Asymptotic theory of sequential fixed-width confidence interval procedures},
journal = {Journal of the American Statistical Association},
volume = {71},
number = {356},
pages = {949-955},
year  = {1976},
publisher = {Taylor & Francis},
doi = {10.1080/01621459.1976.10480975},

URL = { 
        https://www.tandfonline.com/doi/abs/10.1080/01621459.1976.10480975
    
}
}

@article{PW1992,
author = {Peter W. Glynn and Ward Whitt},
title = {The asymptotic validity of sequential stopping rules for stochastic simulations},
volume = {2},
journal = {The Annals of Applied Probability},
number = {1},
publisher = {Institute of Mathematical Statistics},
pages = {180-198},
year = {1992},
doi = {10.1214/aoap/1177005777},
URL = {https://doi.org/10.1214/aoap/1177005777}
}

@article{D2014,
author = {Singham, Dashi I.},
title = {Selecting stopping rules for confidence interval procedures},
year = {2014},
issue_date = {May 2014},
publisher = {Association for Computing Machinery},
address = {New York, NY, USA},
volume = {24},
number = {3},
issn = {1049-3301},
url = {https://doi.org/10.1145/2627734},
doi = {10.1145/2627734},
journal = {ACM Transactions on Modeling and Computer Simulation},
articleno = {18},
numpages = {18}
}

@INPROCEEDINGS{DL2009,
  author={Singham, Dashi I. and Schruben, Lee W.},
  booktitle={Proceedings of the 2009 Winter Simulation Conference (WSC)}, 
  title={Analysis of sequential stopping rules}, 
  year={2009},
  volume={},
  number={},
  pages={723-730},
  doi={10.1109/WSC.2009.5429686}}

@article{JA2006,
author = {Xu, Jin and Gupta, Arjun K.},
title = {Improved confidence regions for a mean vector under general conditions},
year = {2006},
issue_date = {November, 2006},
publisher = {Elsevier Science Publishers B. V.},
address = {NLD},
volume = {51},
number = {2},
issn = {0167-9473},
url = {https://doi.org/10.1016/j.csda.2005.10.011},
doi = {10.1016/j.csda.2005.10.011},
journal = {Computational Statistics \& Data Analysis},
pages = {1051–1062},
numpages = {12}
}

@ARTICLE{4686254,
  author={Mendo, Luis and Hernando, Jose M.},
  journal={IEEE Transactions on Communications}, 
  title={Improved Sequential Stopping Rule for {Monte} {Carlo} Simulation}, 
  year={2008},
  volume={56},
  number={11},
  pages={1761-1764},
  doi={10.1109/TCOMM.2008.070015}}

@ARTICLE{LJ2006,
  author={Mendo, L. and Hernando, J.M.},
  journal={IEEE Transactions on Communications}, 
  title={A simple sequential stopping rule for {Monte} {Carlo} simulation}, 
  year={2006},
  volume={54},
  number={2},
  pages={231-241},
  doi={10.1109/TCOMM.2005.863780}}

@article{AJ1982,
 ISSN = {00251909, 15265501},
 URL = {http://www.jstor.org/stable/2631324},
 author = {Averill M. Law and W. David Kelton},
 journal = {Management Science},
 number = {5},
 pages = {550-562},
 publisher = {INFORMS},
 title = {Confidence intervals for steady-state simulations, {II}: A survey of sequential procedures},
 volume = {28},
 year = {1982}
}

@ARTICLE{SC1977,
  author={Lavenberg, S. S. and Sauer, C. H.},
  journal={IBM Journal of Research and Development}, 
  title={Sequential stopping rules for the regenerative method of simulation}, 
  year={1977},
  volume={21},
  number={6},
  pages={545-558},
  doi={10.1147/rd.216.0545}}

@ARTICLE{PP1981b,
  author={Heidelberger, Philip and Welch, Peter D.},
  journal={IBM Journal of Research and Development}, 
  title={Adaptive spectral methods for simulation output analysis}, 
  year={1981},
  volume={25},
  number={6},
  pages={860-876},
  doi={10.1147/rd.256.0860}}

@article{Luck1993,
author = {Raatikainen, Kimmo E. E.},
title = {A sequential procedure for simultaneous estimation of several means},
year = {1993},
issue_date = {April 1993},
publisher = {Association for Computing Machinery},
address = {New York, NY, USA},
volume = {3},
number = {2},
issn = {1049-3301},
url = {https://doi.org/10.1145/169702.169690},
doi = {10.1145/169702.169690},
journal = {ACM Transactions on Modeling and Computer Simulation},
pages = {108–133},
numpages = {26}
}

@article{LWP2013,
title = {Optimal {M}onte {C}arlo integration with fixed relative precision},
journal = {Journal of Complexity},
volume = {29},
number = {1},
pages = {4-26},
year = {2013},
issn = {0885-064X},
doi = {https://doi.org/10.1016/j.jco.2012.09.001},
url = {https://www.sciencedirect.com/science/article/pii/S0885064X12000805},
author = {Lesław Gajek and Wojciech Niemiro and Piotr Pokarowski}
}

@article{A1969,
author = {Arthur Nadas},
title = {An extension of a theorem of {Chow} and {Robbins} on sequential confidence intervals for the mean},
volume = {40},
journal = {The Annals of Mathematical Statistics},
number = {2},
publisher = {Institute of Mathematical Statistics},
pages = {667-671},
year = {1969},
doi = {10.1214/aoms/1177697737},
URL = {https://doi.org/10.1214/aoms/1177697737}
}

@article{N1966,
author = {Norman Starr},
title = {The performance of a sequential procedure for the fixed-width interval estimation of the mean},
volume = {37},
journal = {The Annals of Mathematical Statistics},
number = {1},
publisher = {Institute of Mathematical Statistics},
pages = {36-50},
year = {1966},
doi = {10.1214/aoms/1177699596},
URL = {https://doi.org/10.1214/aoms/1177699596}
}

@book{D1985,
author = {D. Siegmund},
year = {1985},
month = {},
pages = {},
title = {Sequential Analysis},
publisher = {Springer},
address={New York},
doi = {10.1007/978-1-4757-1862-1}
}

@book{GK1986,
author = {G. B. Wetherill and K. D. Glazebrook},
year = {1986},
month = {},
pages = {},
title = {Sequential {M}ethods in {S}tatistics},
publisher = {3rd ed. Chapman and Hall},
address={London},
doi = {10}
}

@book{Z1987,
author = {Z. Govindarajulu},
year = {1987},
month = {},
pages = {},
title = {The Sequential Statistical Analysis of Hypothesis Testing, Point and Interval Estimation, and Decision Theory},
publisher = {2nd ed. American Science Press},
address={Columbus, OH},
doi = {10}
}

@article{PP1981a,
author = {Heidelberger, Philip and Welch, Peter D.},
title = {A spectral method for confidence interval generation and run length control in simulations},
year = {1981},
issue_date = {April 1981},
publisher = {Association for Computing Machinery},
address = {New York, NY, USA},
volume = {24},
number = {4},
issn = {0001-0782},
url = {https://doi.org/10.1145/358598.358630},
doi = {10.1145/358598.358630},
journal = {Communications of the ACM},
month = apr,
pages = {233–245},
numpages = {13}
}

@article{PP1983,
author = {Heidelberger, Philip and Welch, Peter D.},
title = {Simulation run length control in the presence of an initial transient},
journal = {Operations Research},
volume = {31},
number = {6},
pages = {1109-1144},
year = {1983},
doi = {10.1287/opre.31.6.1109},

URL = { 
        https://doi.org/10.1287/opre.31.6.1109
    
}
}

@book{PBL1987,
author = {P. Bratley and B. L. Fox and L. E. Schrage},
year = {1987},
month = {},
pages = {},
title = {A Guide to Simulation},
edition={2nd},
publisher = {Springer},
address={New York},
doi = {10.1007/978-1-4419-8724-2}
}

@article{c1980,
author = { Attila   Csenki },
title = {On the convergence rate of fixed-width sequential confidence intervals},
journal = {Scandinavian Actuarial Journal},
volume = {1980},
number = {2},
pages = {107-111},
year  = {1980},
publisher = {Taylor & Francis},
doi = {10.1080/03461238.1980.10408644},

URL = { 
        https://doi.org/10.1080/03461238.1980.10408644
    
}
}

@article{W1977,
author = {Michael Woodroofe},
title = {Second order approximations for sequential point and interval estimation},
volume = {5},
journal = {The Annals of Statistics},
number = {5},
publisher = {Institute of Mathematical Statistics},
pages = {984-995},
year = {1977},
doi = {10.1214/aos/1176343953},
URL = {https://doi.org/10.1214/aos/1176343953}
}

@article{DL2014,
author = {Singham, Dashi I. and Schruben, Lee W.},
title = {Finite-sample performance of absolute precision stopping rules},
year = {2012},
issue_date = {Fall 2012},
publisher = {INFORMS},
address = {Linthicum, MD, USA},
volume = {24},
number = {4},
issn = {1526-5528},
url = {https://doi.org/10.1287/ijoc.1110.0471},
doi = {10.1287/ijoc.1110.0471},
journal = {INFORMS Journal on Computing},
pages = {624–635},
numpages = {12}
}

@article{DL1976,
author = {M.   Aerts  and    H. Callaert },
title = {The exact approximation order in the central limit theorem for randam u-statistics},
journal = {Sequential Analysis},
volume = {5},
number = {1},
pages = {19-35},
year  = {1986},
publisher = {Taylor & Francis},
doi = {10.1080/07474948608836089},

URL = { 
        https://doi.org/10.1080/07474948608836089
    
}
}

@article{survey_SteadyStateI,
author = {Law, Averill M. and Kelton, W. David},
title = {Confidence intervals for steady-state simulations: I. {A} survey of fixed sample size procedures},
journal = {Operations Research},
volume = {32},
number = {6},
pages = {1221-1239},
year = {1984},
doi = {10.1287/opre.32.6.1221},

URL = { 
        https://doi.org/10.1287/opre.32.6.1221
    
}
}

@article{KNR2019,
title = {Solvable integration problems and optimal sample size selection},
journal = {Journal of Complexity},
volume = {53},
pages = {40-67},
year = {2019},
issn = {0885-064X},
doi = {https://doi.org/10.1016/j.jco.2018.10.007},
url = {https://www.sciencedirect.com/science/article/pii/S0885064X18300839},
author = {Robert J. Kunsch and Erich Novak and Daniel Rudolf}
}

@misc{Jiang2014,
      title={Guaranteed {Monte} {Carlo} methods for {Bernoulli} random variables}, 
      author={Lan Jiang and Fred J. Hickernell},
      year={2014},
      archivePrefix={arXiv},
      primaryClass={math.NA},
howpublished={arXiv:1411.1151}
}

@incollection{Hickernel2018,
author = {Hickernell, Fred J. and Choi, Sou-Cheng T. and Jiang, Lan and Jim\'enez Rugama, Llu{\'i}s Antoni},
publisher = {American Cancer Society},
isbn = {9781118445112},
title = {Monte {Carlo} simulation, automatic stopping criteria for},
booktitle = {Wiley StatsRef: Statistics Reference Online},
chapter = {},
pages = {1-7},
doi = {https://doi.org/10.1002/9781118445112.stat08035},
url = {https://onlinelibrary.wiley.com/doi/abs/10.1002/9781118445112.stat08035},
year = {2018}
}

@InProceedings{FLYA2012,
author="Hickernell, Fred J.
and Jiang, Lan
and Liu, Yuewei
and Owen, Art B.",
editor="Dick, Josef
and Kuo, Frances Y.
and Peters, Gareth W.
and Sloan, Ian H.",
title="Guaranteed Conservative Fixed Width Confidence Intervals via {M}onte {C}arlo Sampling",
booktitle="Monte Carlo and Quasi-Monte Carlo Methods 2012",
year="2013",
publisher="Springer Berlin Heidelberg",
pages="105-128",
isbn="978-3-642-41095-6"
}

@article{Stein_Wald_1947,
 ISSN = {00034851},
 URL = {http://www.jstor.org/stable/2235738},
 author = {Charles Stein and Abraham Wald},
 journal = {The Annals of Mathematical Statistics},
 number = {3},
 pages = {427-433},
 publisher = {Institute of Mathematical Statistics},
 title = {Sequential confidence intervals for the mean of a normal distribution with known variance},
 volume = {18},
 year = {1947}
}

@article{Anscombe_1949,
 ISSN = {00063444},
 author = {F. J. Anscombe},
 journal = {Biometrika},
 number = {3/4},
 pages = {455-458},
 publisher = {[Oxford University Press, Biometrika Trust]},
 title = {Large-sample theory of sequential estimation},
 volume = {36},
 year = {1949}
}

@article{Anscombe_1952, title={Large-sample theory of sequential estimation}, volume={48}, DOI={10.1017/S0305004100076386}, number={4}, journal={Mathematical Proceedings of the Cambridge Philosophical Society}, publisher={Cambridge University Press}, author={Anscombe, F. J.}, year={1952}, pages={600–607}}

@article{Anscombe_1953,
 ISSN = {00359246},
 author = {F. J. Anscombe},
 journal = {Journal of the Royal Statistical Society. Series B (Methodological)},
 number = {1},
 pages = {1-29},
 publisher = {[Royal Statistical Society, Wiley]},
 title = {Sequential estimation},
 volume = {15},
 year = {1953}
}

@article{Anscombe_1954,
 ISSN = {0006341X, 15410420},
 author = {F. J. Anscombe},
 journal = {Biometrics},
 number = {1},
 pages = {89-100},
 publisher = {[Wiley, International Biometric Society]},
 title = {Fixed-sample-size analysis of sequential observations},
 volume = {10},
 year = {1954}
}

@article{Ray_1957,
 ISSN = {00359246},
 author = {W. D. Ray},
 journal = {Journal of the Royal Statistical Society. Series B (Methodological)},
 number = {1},
 pages = {133-143},
 publisher = {[Royal Statistical Society, Wiley]},
 title = {Sequential confidence intervals for the mean of a normal population with unknown variance},
 volume = {19},
 year = {1957}
}

@book{handbook,
  title={Handbook of Sequential Analysis},
  author={Ghosh, B.K. and Sen, P.K.},
  isbn={9780824784089},
  lccn={91011594},
  year={1991},
  publisher={CRC Press},
address={New York}
}

@inproceedings{Hardt2016TrainFaster,
  author    = {Moritz Hardt and Ben Recht and Yoram Singer},
  title     = {Train Faster, Generalize Better: Stability of Stochastic Gradient Descent},
  booktitle = {Proceedings of the 33rd International Conference on Machine Learning (ICML)},
  series    = {JMLR Workshop and Conference Proceedings},
  volume    = {48},
  pages     = {1225--1234},
  year      = {2016}
}
\end{document}